\documentclass[prb,twocolumn,showpacs,superscriptaddress,eqsecnum]{revtex4-1}

\usepackage{graphicx}
\usepackage{amsmath,amssymb,bm}
\usepackage{color}

\begin{document}

\title{
Higher-order Fermi-liquid corrections for an Anderson impurity away from half-filling  II: 
Equilibrium properties
}

 \author{Akira Oguri}
 \affiliation{
 Department of Physics, Osaka City University, Sumiyoshi-ku, 
 Osaka 558-8585, Japan }

 \author{A.\ C.\ Hewson}
 \affiliation{
 Department of Mathematics, Imperial College London, London SW7 2AZ, 
 United Kingdom }

\date{\today}

\begin{abstract}
We study the low-energy behavior of the vertex  function 
of a single Anderson impurity away from half-filling for finite magnetic fields,   
using the Ward identities with careful consideration 
of the  anti-symmetry  and analytic properties.
The asymptotic form  of  the vertex function 
 $\Gamma_{\sigma\sigma';\sigma'\sigma}^{}(i\omega,i\omega';i\omega',i\omega)$  
is determined  
up to terms of  linear order with respect to 
the two frequencies  $\omega$ and $\omega'$,  
as well as the  $\omega^2$ contribution for  anti-parallel spins 
 $\sigma'\neq \sigma$ at $\omega'=0$. 
From these results,  
we also obtain a series of the Fermi-liquid relations beyond those 
of Yamada-Yosida [Prog.\ Theor.\ Phys.\, {\bf  54}, 316 (1975)].  
The $\omega^2$ real part 
of the self-energy $\Sigma_{\sigma}^{}(i\omega)$ 
 is shown to be expressed  in terms of 
 the double derivative 
 $\partial^2\Sigma_{\sigma}^{}(0)/\partial \epsilon_{d\sigma}^{2}$  
with respect to the impurity energy level  $\epsilon_{d\sigma}^{}$, 
and   agrees with the formula obtained recently 
by  Filippone,  Moca, von Delft, and Mora (FMvDM) 
in the Nozi\`{e}res phenomenological Fermi-liquid theory  
[Phys.\ Rev.\ B {\bf  95}, 165404  (2017)].
We also calculate the $T^2$ correction of the self-energy,   
and find that the real part can be expressed in terms of  
the three-body correlation function 
$\partial \chi_{\uparrow\downarrow}/\partial \epsilon_{d,-\sigma}^{}$,  
where  $ \chi_{\uparrow\downarrow}$  is 
the static susceptibility between anti-parallel spins.
We also provide an alternative derivation of the asymptotic form of the vertex function.
Specifically,  we calculate  the skeleton diagrams for the vertex function 
 $\Gamma_{\sigma\sigma;\sigma\sigma}^{}(i\omega,0;0,i\omega)$ 
 for parallel spins up to order $U^4$ in  the Coulomb repulsion $U$.
It directly clarifies the fact that the analytic components of order $\omega$ vanish 
as a result of the cancellation of four related Feynman diagrams  
which are related to each other through the anti-symmetry operation.

\end{abstract}

\pacs{71.10.Ay, 71.27.+a, 72.15.Qm}

\maketitle

\section{Introduction}

The  Anderson impurity  model has been studied 
 as a model for the Kondo effect in dilute magnetic alloys\cite{AndersonModel}
and quantum dots.\cite{Hershfield1,MeirWingreen}
The  low-energy behavior of the model  as a Fermi liquid 
 has successfully  been explained by the  
Nozi\`{e}res' phenomenological description\cite{NozieresFermiLiquid} 
and  Yamada-Yosida's  microscopic  formulation.\cite{YamadaYosida2,YamadaYosida4,ShibaKorringa,Yoshimori} 
In particular,  the three leading-order parameters, i.e.,  
the scattering phase shift  $\delta$,  
the  Kondo energy scale $T_K$,  and the Wilson ratio $R_W^{}$,  
determine  the universal behavior and 
 explain the low-lying excited states obtained with    
the Wilson numerical normalization 
group (NRG).\cite{WilsonRMP,KWW1,KWW2,HewsonOguriMeyer}

However,  the next-leading Fermi-liquid corrections,  
such as  the low-frequency  $\omega^2$ and  low-temperature $T^2$ corrections, 
had not been fully understood away from half-filling over the years  
despite their importance.  
What made the problem difficult was 
the real part of the self-energy which also shows the $\omega^2$ and $T^2$ dependences  
 away from half-filling.\cite{Yoshimori,HorvaticZlatic2}  
Recently, a significant breakthrough has been achieved by
Mora, Moca, von Delft, and Zar\'{a}nd (MMvDZ),\cite{MoraMocaVonDelftZarand}
and  Filippone, Moca, von Delft and Mora  (FMvDM).\cite{FilipponeMocaVonDelftMora}
They have provided  an explicit  way to clarify  
the next-leading corrections away from half-filling 
by extending Nozi\`{e}res' phenomenological theory.
Specifically,  FMvDM have determined the coefficients for 
 the quadratic  $\omega^2$, $T^2$ and $(eV)^2$ terms 
of the real part of the self-energy  for a non-equilibrium steady state 
driven by a bias voltage $V$.

In the present work,  we have constructed  the  microscopic theory 
for the next-leading corrections away from half-filling, 
extending  the seminal works of Yamada-Yosida, Shiba, 
and Yoshimori.\cite{YamadaYosida4,ShibaKorringa,Yoshimori}
Our microscopic formulation of the higher-order Fermi-liquid 
relations  is applicable to a wide class of  impurity correlation functions  
in various situations.
It is hard to give a comprehensive description in a single account, 
 so  we present our work  
in a series of three separate papers.  
The first report,  referred to as {\it paper I},\cite{ao2017_1}
is a short, less technical, report which concisely describes the 
formulation and  main results of the whole series.
The second one is the present paper,  referred to as {\it paper II}, 
where we mainly describe equilibrium properties, 
using the Matsubara imaginary-frequency Green's function. 
The third one, referred to as {\it paper III},\cite{ao2017_3_PRB}
 describes the microscopic theory for nonlinear transport through 
quantum dots  away from half-filling 
 and also thermoelectric transport in dilute magnetic alloys,    
using the Keldysh Green's function.

The main purpose of the present paper is to give a complete derivation 
of  the higher-order corrections  away from half-filling at finite magnetic fields.
In the first half of the paper,
we show that a series of the higher-order Fermi-liquid relations 
can be deduced  from the analytic and anti-symmetry properties of  the vertex function 
  $\Gamma_{\sigma\sigma';\sigma'\sigma} (i\omega, i\omega'; i\omega', i\omega)$, 
which we obtain explicitly up to linear-order of 
the two frequency arguments  $\omega$ and $\omega'$ 
using the Ward identities.  
The higher-order Fermi-liquid correction involves  
the static $n$-body correlation function, 
$\chi_{\sigma_1\sigma_2\sigma_3\cdots}^{[n]}$, 
for the spin-resolved impurity occupation  $n_{d\sigma}^{}$.  
Specifically, the three-body fluctuations for $n=3$ contribute 
to the $\omega^2$, $T^2$, and $(eV)^2$ corrections away from half-filling.
In the second half of the paper, we perturbatively examine 
the low-frequency behavior of the vertex function 
in order to give an alternative derivation of the higher-order corrections. 
To this end,   we calculate  the skeleton-diagrams for  the vertex function 
 for the parallel spins up to order $U^4$ 
with respect  to the Coulomb interaction $U$.
It explicitly demonstrates  that the analytic $\omega$-linear part of 
 $\Gamma_{\sigma\sigma;\sigma\sigma} (i\omega, 0; 0, i\omega)$ 
vanishes  as a result of the anti-symmetry.
The calculations are systematically carried out 
by introducing an operator that extracts 
 the next-leading contributions from 
 a singular particle-hole-pair propagator.

The first part, Sec.\ \ref{sec:model}--Sec.\ \ref{sec:T2_self_energy},  
 is devoted to the general description based on the Ward identities and 
the  analytic and antisymmetry  properties of  the vertex function. 
The second part,  
Sec.\ \ref{sec:green_function_product_expansion}--Sec.\  
\ref{sec:skeleton_diagram_expansion_for_vertex}, 
is devoted to perturbative calculations;   
details of the order $U^4$ contributions 
are provided  in Supplemental Material.
\footnote{Calculations of order $U^4$ skeleton-diagrams are given in Supplemental Material.}

\section{Model \&  Formulation}
\label{sec:model}

In this section,  
we describe  the renormalization factors 
and the  nonlinear susceptibilities  
that we use throughout the present work.
We start with  the single Anderson impurity, defined by  
\begin{align}
\mathcal{H} =&  
 \sum_{\sigma}
 \epsilon_{d\sigma}^{}\, n_{d\sigma}^{} 
 + U\,n_{d\uparrow}\,n_{d\downarrow} + 
\sum_{\sigma} 
\int_{-D}^D  \!\! d\epsilon\,  \epsilon\, 
 c^{\dagger}_{\epsilon  \sigma} c_{\epsilon  \sigma}^{}
 \nonumber \\
 & 
+  
\sum_{\sigma}  v  \left( \psi_{\sigma}^\dag d_{\sigma}^{} + 
  d_{\sigma}^{\dag} \psi_{\sigma}^{} \right) \;.
 \label{Hami_seri_part}
\end{align}
Here, 
 $d^{\dag}_{\sigma}$ creates an impurity electron 
with spin $\sigma$ 
 in the impurity level of energy $\epsilon _{d\sigma}$, and  
 $n_{d\sigma}^{} = d^{\dag}_{\sigma} d^{}_{\sigma}$. 
 $U$ is the Coulomb interaction between  electrons occupying the impurity level. 
Electrons in the leads obey  the  anti-commutation relation 
$
\{ c^{\phantom{\dagger}}_{\epsilon\sigma}, 
c^{\dagger}_{\epsilon'\sigma'}
\} = 
\delta_{\sigma\sigma'}   
\delta(\epsilon-\epsilon')$. 
The linear combination of the conduction electrons, 
$\psi^{}_{\sigma} \equiv  \int_{-D}^D d\epsilon \sqrt{\rho_c^{}} 
\, c^{\phantom{\dagger}}_{\epsilon\sigma}$  with $\rho_c^{}=1/(2D)$, 
 couples to the impurity level, the  bare width of which  
 is given by 
 $\Delta = \pi \rho_c^{} v_{}^2$.  
We consider the parameter region, where 
 the  half band-width 
 $D$ is much greater than the other energy scales,    
$D \gg \max( U, \Delta, |\epsilon_{d\sigma}^{}|, |\omega|, T)$.   
For  finite magnetic fields $h$, the impurity energy takes the form
 $\epsilon_{d\sigma}^{} = \epsilon_{d}^{} - \sigma h$, 
 where  $\sigma = +1$ (-1) for $\uparrow$ ($\downarrow$) spin.
The relation between the differentiations is 
\begin{align}
&
\frac{\partial}{\partial \epsilon_{d}^{}} 
= 
\frac{\partial}{\partial \epsilon_{d\uparrow}^{}} 
+
\frac{\partial}{\partial \epsilon_{d\downarrow}^{}} 
,  
\qquad 
\frac{\partial}{\partial h} 
= 
-\frac{\partial}{\partial \epsilon_{d\uparrow}^{}} 
+
\frac{\partial}{\partial \epsilon_{d\downarrow}^{}} 
.
\end{align} 

We use the imaginary-frequency  formulation 
for the impurity Green's function:  
\begin{align}
G_\sigma(i\omega)
 \, \equiv & \    - \int_0^{\beta} \!\!  d\tau \, 
e^{i \omega \tau} 
\left\langle T_\tau \, d_\sigma^{}(\tau)\,d_\sigma^{\dagger}
\right\rangle  
 \nonumber \\
 = & \  
\frac{1}{i\omega -\epsilon_{d\sigma}^{} +i\Delta \,\mathrm{sgn}\,\omega 
- \Sigma_\sigma(i\omega) } \;. 
 \label{eq:GM}
\end{align}
Here, 
$\langle \mathcal{O} \rangle \equiv 
\mathrm{Tr}\,[\mathcal{O}\,e^{-\beta\mathcal{H}} ]/\Xi$  
denotes the thermal average 
with 
$\Xi \equiv \mathrm{Tr}\,[e^{-\beta\mathcal{H}}]$ 
and  $\beta\equiv 1/T$, 
and $\Sigma_\sigma(i\omega)$ is the self-energy caused by the Coulomb interaction 
$U$. 
The retarded Green's function can be obtained carrying out 
the analytic continuation $i\omega \to \epsilon +i0^+$ for $\omega>0$,
and the density of states is  given by  
\begin{align}
  \rho_{d\sigma}^{}(\epsilon) \,\equiv\, -\,\frac{1}{\pi} \,
\mathrm{Im}\, G_{\sigma}(\epsilon+i0^+)  
\,. 
\label{eq:Aspec}
\end{align}

In the following, we mainly consider the zero-temperature limit $T \to 0$, 
where the Matsubara frequency $i\omega$ becomes continuous. 
We will suppress the frequency argument for 
the density of states at the Fermi energy $\omega=0$ for $T=0$: 
\begin{align}
\rho_{d\sigma}^{} \equiv   \rho_{d\sigma}^{}(0)=\frac{\sin^2 \delta_{\sigma}}{\pi \Delta},
\quad 
\cot \delta_{\sigma} 
= \frac{\epsilon_{d\sigma}^{} + \Sigma_\sigma(0)}{\Delta}.
\end{align}
The phase shift $\delta_{\sigma}$  is a primary parameter, 
which characterizes the Fermi-liquid ground state.
The Friedel sum rule relates  the phase shift to  the occupation number,  
which also corresponds to the first derivative of the free energy 
  $\Omega \equiv - T \log \Xi$,  
\begin{align}
&\langle n_{d\sigma}^{} \rangle \,=\,
\frac{\partial \Omega}{\partial \epsilon_{d\sigma}} 
 \,\xrightarrow{\,T\to 0\,} \,  
\frac{\delta_{\sigma}}{\pi} \;. 
\end{align}
Note that $\Omega$ is an even function of $h$.

\subsection{Linear-response susceptibilities}
\label{subsec:susceptibilities}

The  leading Fermi-liquid corrections 
can be described by  the static susceptibilities  
following Yamada-Yosida:\cite{YamadaYosida2}
\begin{align}
\chi_{\sigma\sigma'}^{} \equiv \  
- \frac{\partial^2 \Omega}
{\partial \epsilon_{d\sigma'}^{}\partial \epsilon_{d\sigma}^{}} 
 \,=\,
- \,\frac{\partial \langle n_{d\sigma}^{} \rangle }{\partial \epsilon_{d\sigma'}^{}} 
\,\xrightarrow{\,T\to 0\,} \,  
 \rho_{d\sigma}^{} \, \widetilde{\chi}_{\sigma\sigma'}^{}  .
\label{eq:chi_org}
\end{align}
Note that $\chi_{\uparrow\downarrow}^{}=\chi_{\downarrow\uparrow}^{}$, 
and  $\widetilde{\chi}_{\sigma\sigma'}^{} $ the enhancement factor defined by 
\begin{align}
\widetilde{\chi}_{\sigma\sigma'} 
\equiv 
\delta_{\sigma\sigma'} +
\frac{\partial  \Sigma_{\sigma}(0)}{\partial \epsilon_{d\sigma'}^{}} \;.
\label{eq:chi_tilde}
\end{align}
The susceptibility can be written as a static 2-body correlation function
\begin{align}
\chi_{\sigma\sigma'}^{}   =     
\int_0^{\beta} \! d \tau \, 
\left\langle \delta n_{d\sigma}^{}(\tau)\,\delta  n_{d\sigma'}^{}\right\rangle   , 
\quad  \ \ 
\delta n_{d\sigma}^{} \equiv n_{d\sigma}^{} - \langle n_{d\sigma}^{}  \rangle.
\end{align}
The  usual spin and charge susceptibilities  are  given by 
\begin{align}
\chi_{c}  \equiv &\,     - \frac{\partial^2 \Omega}{\partial \epsilon_{d}^2} \,=\, 
 \chi_{\uparrow\uparrow}^{} +\chi_{\downarrow\downarrow}^{} 
+  \chi_{\uparrow\downarrow}^{} +  \chi_{\downarrow\uparrow}^{}, \\
\chi_{s}  \,\equiv& \,   
- \frac{1}{4} \frac{\partial^2 \Omega}{\partial h^2} \, = \,  
\frac{1}{4} \left(\,  
 \chi_{\uparrow\uparrow}^{} +\chi_{\downarrow\downarrow}^{} 
-  \chi_{\uparrow\downarrow}^{} -  \chi_{\downarrow\uparrow}^{}
\,\right) .
\end{align}
We can also choose another set of parameters to describe the leading 
Fermi-liquid corrections defined at $T=0$,\cite{HewsonRPT2001}
\begin{align}
\frac{1}{z_{\sigma}^{}}
 \equiv & \    
1-\left.
\displaystyle \frac{\partial \Sigma_\sigma(i\omega)}{\partial i\omega}
\right|_{\omega\to 0} , 
\quad \ 
\widetilde{U} \equiv z_{\uparrow}^{}z_{\downarrow}^{}  
 \,\Gamma_{\uparrow\downarrow;\downarrow\uparrow}(0, 0; 0, 0) , 
\\ 
\widetilde{\rho}_{d\sigma}^{} \equiv& \ \frac{\rho_{d\sigma}^{}}{z_{\sigma}^{} } 
=  
\rho_{d\sigma}^{} \widetilde{\chi}_{\sigma\sigma} 
=  
\chi_{\sigma\sigma} 
\;, 
 \end{align}
where  $z_{\sigma}^{}$ is the  renormalization factor,
  $\widetilde{U}$ the  residual interaction  between quasi-particles, 
and  $\widetilde{\rho}_{d\sigma}^{}$ the  renormalized density of states.
Note that in the limit of $T\to 0$,  
the Matsubara frequency $i\omega$ can be treated a continuous variable 
along the imaginary axes, as mentioned above.   
 At finite magnetic fields, 
the Wilson ratio  $R_{W}^{}$  may be defined by 
 \begin{align}
R_W^{} \,\equiv \,  1+
\sqrt{\widetilde{\rho}_{d\uparrow}^{}\widetilde{\rho}_{d\downarrow}^{}} \, 
\widetilde{U} 
\ = \  1-
\frac{\chi_{\uparrow\downarrow}^{}}
{\sqrt{\chi_{\uparrow\uparrow}^{}\chi_{\downarrow\downarrow}^{}}} \;.
\label{eq:Wilson_ratio_general} 
\end{align}
Correspondingly,  the Kondo energy scale $T^*$  may also be defined such that 
$R_W^{}  =  1- 4T^* \chi_{\uparrow\downarrow}^{}$, i.e., 
\begin{align}
T^* \equiv 
\frac{1}{4\sqrt{\chi_{\uparrow\uparrow}^{}  \chi_{\downarrow\downarrow}^{}}} \;.
 \label{eq:Kondo_scale_general} 
 \end{align}

\begin{figure}[t]
 \leavevmode
\begin{minipage}{1\linewidth}
\includegraphics[width=0.5\linewidth]{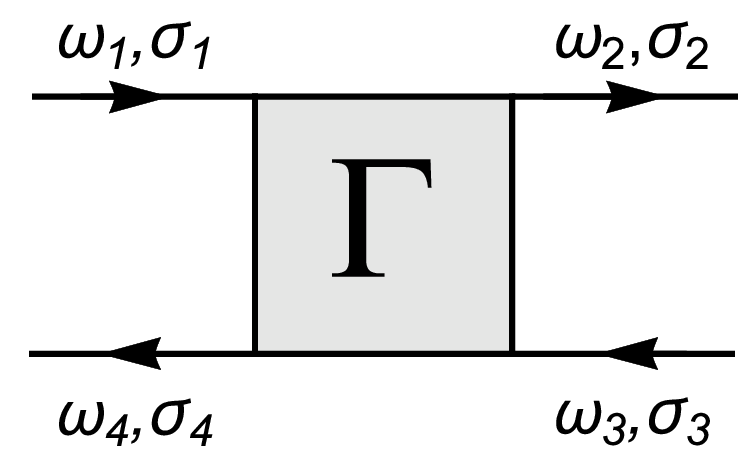}
\end{minipage}
 \caption{Vertex function 
$\Gamma_{\sigma_1\sigma_2;\sigma_3\sigma_4}^{}
(i\omega_1, i\omega_2; i\omega_3, i\omega_4)$ 
satisfies the anti-symmetry property:   
Eq.\ \eqref{eq:vertex_paralle_anti_sym}
with $\omega_1+\omega_3=\omega_2+\omega_4$. 
}
 \label{fig:vertex}
\end{figure}

\subsection{Non-linear susceptibilities: 3-body fluctuations}
\label{subsec:notes}

The next leading  Fermi-liquid corrections 
 are determined by  the static nonlinear  susceptibilities, as we will describe later,
\begin{align}
\chi_{\sigma_1\sigma_2\sigma_3}^{[3]} \equiv \  
- \,
\frac{\partial^3 \Omega }{\partial \epsilon_{d\sigma_1}\partial \epsilon_{d\sigma_2}\partial \epsilon_{d\sigma_3}} 
\,=\,  \frac{\partial \chi_{\sigma_2\sigma_3}}
{\partial \epsilon_{d\sigma_1}} .
\label{eq:canonical_correlation_3_dif}
\end{align}
It also corresponds to the thee-body correlations of the impurity occupation 
\begin{align}
\chi_{\sigma_1\sigma_2\sigma_3}^{[3]} 
\!  = 
- \!
\int_{0}^{\beta} \!\!\! d\tau_3 \!\! 
\int_{0}^{\beta} \!\!\! d\tau_2\, 
\langle T_\tau 
\delta n_{d\sigma_3} (\tau_3) \,
\delta n_{d\sigma_2} (\tau_2) \,
\delta n_{d\sigma_1}
\rangle .
\label{eq:canonical_correlation_3}
\end{align}
We have provided a derivation of 
the nonlinear response function  
in Appendix \ref{sec:nonlinear_static_susceptibility}.
More generally,  the  $n$-th derivative of $\Omega$ 
for $n=4,5,6\cdots$ corresponds to 
 the $n$-body correlation function 
$\chi_{\sigma_1\sigma_2\sigma_3\cdots}^{[n]}$. 
The Fermi-liquid corrections can be classified according to $n$,  
and  the derivative of the Ward identity reveals 
 a hierarchy of Fermi-liquid relations, as described below.
The $n$-body correlation function 
has permutation symmetry for the spin indexes  
$\chi_{\sigma_1\sigma_2\sigma_3 \cdots}^{[n]} 
 = 
\chi_{\sigma_2\sigma_1\sigma_3\cdots}^{[n]} 
 =  
\chi_{\sigma_3\sigma_2\sigma_1 \cdots}^{[n]} 
 =  \cdots$, and thus it has $n+1$ independent components 
at finite magnetic fields.  
Thus, for  the three-body functions 
$3$ components  among $2^3$ are independent  at finite magnetic fields:  
for instance, the following  for $\sigma=\uparrow, \downarrow$, 
\begin{align}
\frac{\partial \chi_{\sigma\sigma}}{\partial \epsilon_{d\sigma}^{}} 
   =  
\frac{\partial \chi_{\sigma\sigma}}{\partial \epsilon_{d}^{}} 
- \frac{\partial \chi_{\uparrow\downarrow}}{\partial \epsilon_{d\sigma}^{}} ,
\quad \ 
\frac{\partial \chi_{\uparrow\downarrow}}{\partial \epsilon_{d\sigma}^{}} 
 =    \frac{1}{2}
\left(
\frac{\partial \chi_{\uparrow\downarrow}}{\partial \epsilon_{d}^{}} 
-\sigma 
\frac{\partial \chi_{\uparrow\downarrow}}{\partial h} \right) . 
\end{align}
Note that  $\chi_{\uparrow\downarrow}$ is an even function of $h$, 
and thus $\partial \chi_{\uparrow\downarrow}/\partial h \xrightarrow{\,h\to 0\,} 0$. 
The derivative of  the renormalization factor 
$\widetilde{\chi}_{\sigma\sigma'}$ 
also has a similar permutation symmetry  but in a constrained way,   
\begin{align}
\frac{\partial^2 \Sigma_{\sigma}(0)}
{\partial \epsilon_{d\sigma_2}^{}\partial \epsilon_{d\sigma_1}^{}}
\,=& \  
\frac{\partial \widetilde{\chi}_{\sigma\sigma_1}}
{\partial \epsilon_{d\sigma_2}^{}} 
\,=\, 
\frac{\partial \widetilde{\chi}_{\sigma\sigma_2}}
{\partial \epsilon_{d\sigma_1}^{}} 
\label{eq:chi_tilde_diff} 
\end{align}
namely,  
the spin index $\sigma$ that corresponds 
to the index for the self-energy can  not generally be exchanged with other indexes.  
It can explicitly be expressed,  using  the derivative of the susceptibilities, as   
\begin{align}
& \frac{\partial \widetilde{\chi}_{\sigma_1\sigma_2}^{}}
{\partial \epsilon_{d\sigma_3}^{}} 
\,= \,
\frac{1}{\rho_{d\sigma_1}^{}}   
\left(
\frac{\partial \chi_{\sigma_1\sigma_2}^{}}{\partial \epsilon_{d\sigma_3}^{}} 
\,+\, 
2 \pi \,\cot \delta_{\sigma_1}\,
\chi_{\sigma_1\sigma_3}^{} \chi_{\sigma_1\sigma_2}^{} 
\right),  
\label{eq:chitilde_chi}
\\ 
& \ 
\frac{\partial }{\partial \epsilon_{d\sigma'}^{}} 
\left(\frac{1}{\rho_{d\sigma}^{}}\right)
\,= \,
2 \pi  \cot \delta_{\sigma}  \, \widetilde{\chi}_{\sigma\sigma'} 
\,.
\label{eq:Dren_to_Dsus_org2}
\end{align}
We also  note that the correspondence between the above  parameters 
and the coefficients used in FMvDM's  phenomenological description 
can be listed as
\begin{align}
\frac{\alpha_{1\sigma}}{\pi}\,= & \  
  \chi_{\sigma\sigma}^{} \,,
 \qquad \qquad 
\frac{\phi_{1}^{}}{\pi}\,=\, 
 - \chi_{\uparrow\downarrow}^{} \,,
\label{eq:phi_1_def_Nozi}
\\
\frac{\alpha_{2\sigma}}{\pi} 
= & \  
-\frac{1}{2}\, \frac{\partial  \chi_{\sigma\sigma}^{} }{\partial \epsilon_{d\sigma}} 
, 
\qquad
\frac{\phi_{2\sigma}^{}}{\pi}
= 
 2 \, \frac{\partial  \chi_{\uparrow\downarrow}^{}}{\partial \epsilon_{d\sigma}} 
\;. 
\label{eq:phi2_def_FMvDM}
\end{align}

\subsection{Example: $T^2$ correction of electric resistance}

 Before going into details, we would like to show 
  an example of how the non-linear susceptibilities 
$\chi_{\sigma_1\sigma_2\sigma_3}^{[3]}$ 
enter  the  Fermi-liquid corrections.
 Specifically, we consider the  electric resistance $R_\mathrm{MA}^{}$ 
 of a dilute magnetic alloy  (MA),   
\begin{align}
\frac{1}{R_\mathrm{MA}^{}} 
=   & \  
\frac{1}{2R_\mathrm{MA}^{0}} \sum_\sigma \mathcal{L}_{0,\sigma}^{} \,, 
\end{align}
\begin{align}
\mathcal{L}_{0,\sigma} = 
\int_{-\infty}^{\infty}  
d\omega\, 
\frac{1}{\pi \Delta \rho_{d\sigma}^{}(\omega,T)}
\left( -\frac{\partial f(\omega)}{\partial \omega}\right)  .  
\label{eq:L_thermal_0}
\end{align}
Here,   $R_\mathrm{MA}^{0}$  is 
the unitary-limit value of the electric resistance, and 
$f(\omega)= [e^{\beta \omega}+1]^{-1}$ is the Fermi function. 
Calculating the density of states $\rho_{d\sigma}^{}(\omega,T)$ 
 up to  terms of order $\omega^2$ and $T^2$ with the self-energy presented 
in  Eqs.\ \eqref{eq:self_T_imaginary} and \eqref{eq:self_real_T_mag},
$\mathcal{L}_{0,\sigma}$ can be deduced up to order $T^2$,    

\begin{align}
\mathcal{L}_{0,\sigma}  =  & \ 
\frac{1}{\pi \Delta \rho_{d\sigma}^{}}
\left[\,
1+
\frac{C_{0,\sigma}^\mathrm{MA}}{\pi \Delta \rho_{d\sigma}^{}}
  \left(\pi T\right)^2  
\,\right]  \, +  O(T^4) , 
\\
C_{0,\sigma}^\mathrm{MA}  =  & \ 
\frac{\pi^2}{3} 
\Biggl[\,
\bigl(2+\cos 2\delta_{\sigma}  \bigr)\,
\chi_{\sigma\sigma}^2
- 
2\cos 2 \delta_{\sigma}\,
\chi_{\uparrow\downarrow}^2
\nonumber \\
& 
\qquad + 
\frac{\sin 2\delta_{\sigma}}{2\pi}\,
\left(
 \frac{\partial \chi_{\sigma\sigma}}{\partial \epsilon_{d\sigma}^{}} 
 + \frac{\partial \chi_{\uparrow\downarrow}}{\partial \epsilon_{d,-\sigma}} 
\right)
\, \Biggr] \;.
\label{eq:C0_result}
\end{align}
We see that additional contributions of three-body fluctuations 
emerge in  the coefficient $C_{0,\sigma}^\mathrm{MA}$ 
away from half-filling through the derivative of the linear susceptibilities.  
They vanish in the particle-hole symmetric case  where 
$\epsilon_{d\sigma}^{}=-U/2$,  $h=0$,  and 
the phase shift takes the unitary-limit value  $\delta_{\sigma}=\pi/2$:   
then the coefficient is given 
by Yamada-Yosida's formula,\cite{YamadaYosida2,YamadaYosida4}
\begin{align}
C_{0,\sigma}^\mathrm{MA}  
\longrightarrow
\frac{\pi^2}{3} 
\Bigl(
\chi_{\uparrow\uparrow}^2
+ 2 \chi_{\uparrow\downarrow}^2
\Bigr) \,. 
\end{align}

\section{Hierarchy of Fermi-liquid relations}
\label{sec:hierarchy_of_FL_relations}

\begin{figure}[t]
 \leavevmode
\begin{minipage}{1\linewidth}
 \includegraphics[width=0.25\linewidth]{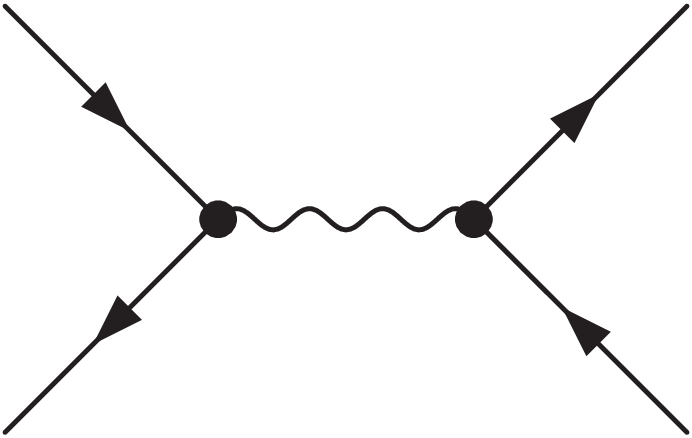}
\rule{0.04\linewidth}{0cm}
\includegraphics[width=0.3\linewidth]{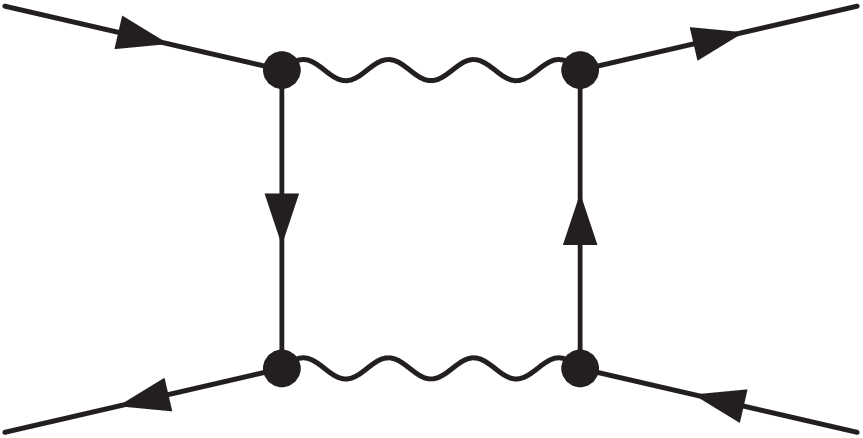}
\rule{0.04\linewidth}{0cm}
\includegraphics[width=0.3\linewidth]{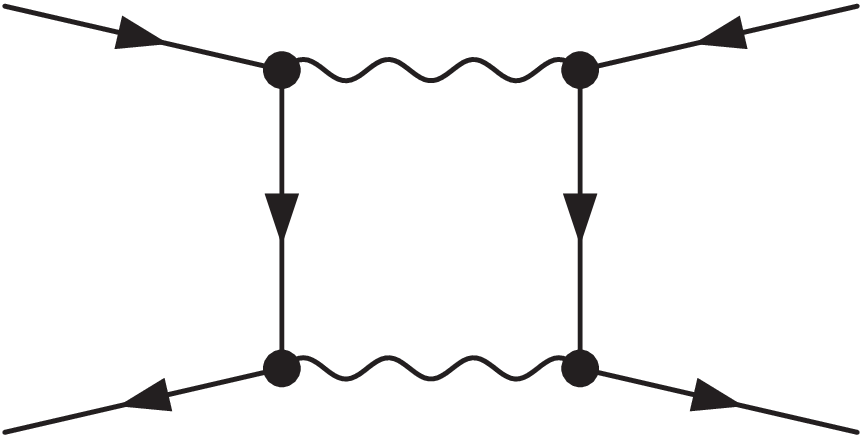}
\end{minipage}
 \caption{
Order $U$ and   $U^2$ vertex function 
 $\Gamma_{\sigma,-\sigma;-\sigma,\sigma}$:  
the anti-parallel spin component.
}
 \label{fig:vertex_u_up_dw}
\end{figure}

\begin{figure}[t]
 \leavevmode
\begin{minipage}{1\linewidth}
\includegraphics[width=0.34\linewidth]{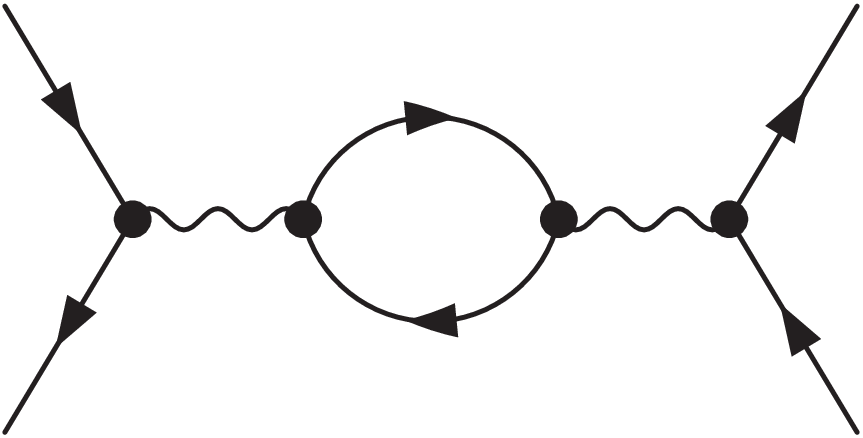}
\rule{0.12\linewidth}{0cm}
\includegraphics[width=0.34\linewidth]{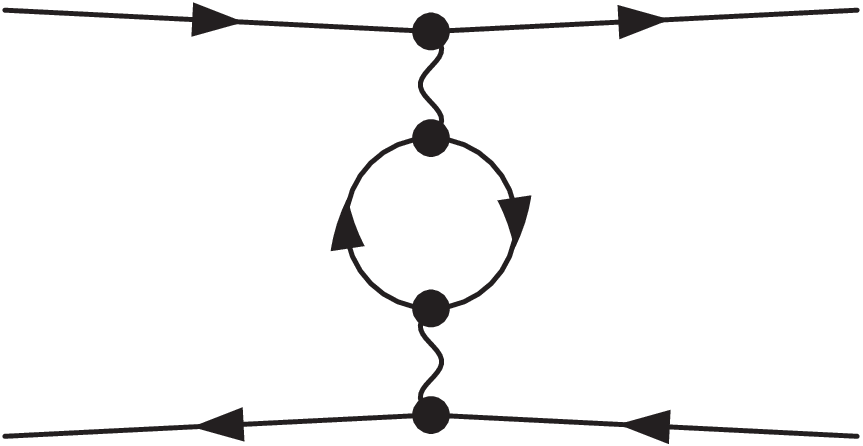}
\end{minipage}
 \caption{
Order $U^2$ vertex function  $\Gamma_{\sigma\sigma;\sigma\sigma}^{}$:  
  the parallel spin component.  
}
 \label{fig:vertex_u2_up_up}
\end{figure}

In this section, 
we describe how a series of the Fermi-liquid relations 
can be  derived   from  the  Ward identity 
which reflects  the local current conservation of each spin component 
$\sigma$,\cite{YamadaYosida2,Yoshimori} 
\begin{align}
-\delta_{\sigma\sigma'}  \frac{\partial \Sigma_{\sigma}(i \omega)}{\partial i \omega}
\, = \, 
\frac{\partial \Sigma_{\sigma}(i\omega)}{\partial \epsilon_{d\sigma'}}
+
\Gamma_{\sigma \sigma';\sigma' \sigma}(i\omega , 0; 0 , i\omega) 
\,\rho_{d\sigma'}.
\label{eq:YYY}
\end{align}
Here,  $\Gamma_{\sigma_2 \sigma_3;\sigma_4 \sigma_4}
(i\omega_1, i\omega_2; i\omega_3, i\omega_4)$ 
is the four-point vertex function, the frequencies  and suffixes of which  
are assigned in such a way shown in Fig. \ref{fig:vertex}: 
some examples of  the lowest-order diagrams are also 
shown in Figs.\  \ref{fig:vertex_u_up_dw} and \ref{fig:vertex_u2_up_up}.
The Ward identity describes a relation between 
the vertex function and the differential coefficients of the self-energy.

\subsection{
Leading Fermi-liquid corrections and the higher hierarchies 
}

The Ward identity for  $\omega=0$ is also called the Fermi-liquid relation. 
Specifically,  the anti-parallel $\sigma'=-\sigma$  
and parallel $\sigma'=\sigma$ spin components of  
 Eq.\ \eqref{eq:YYY} can be expressed 
in the following forms,\cite{YamadaYosida2,Yoshimori}  
respectively, 
\begin{align}
\Gamma_{\sigma,-\sigma;-\sigma\sigma}(0 , 0; 0 , 0) \,
\rho_{d,-\sigma}^{}
= - \widetilde{\chi}_{\sigma,-\sigma} ,
\qquad 
\frac{1}{z_{\sigma}} =
\widetilde{\chi}_{\sigma\sigma} .
\label{eq:YY2_results}
\end{align}
Note that 
$\Gamma_{\sigma\sigma;\sigma\sigma}(0 , 0; 0 , 0)=0$ 
due to the Pauli exclusion rule, and 
$\Gamma_{\uparrow\downarrow;\downarrow\uparrow}(0 , 0; 0 , 0)=
\Gamma_{\downarrow\uparrow;\uparrow\downarrow}(0 , 0; 0 , 0)$. 
Reflecting the property  $1/z_{\sigma} = \widetilde{\chi}_{\sigma\sigma}$, 
 the  frequency derivative and the $\epsilon_{d\sigma}$ derivative 
 of the density of states are identical except for the sign, 
\begin{align}
&\rho_{d\sigma}' \equiv \left.
\frac{\partial \rho_{d\sigma}^{}(\epsilon)^{}}
{\partial \epsilon} \right|_{\epsilon=0}^{} 
 = 
- 
\frac{\partial \rho_{d\sigma}^{}}{\partial \epsilon_{d\sigma}^{}} 
. 
 \label{eq:rho_d_ed}
\end{align}

The leading Fermi-liquid corrections are characterized 
by the parameters $\widetilde{\chi}_{\sigma\sigma}$ and 
$\widetilde{\chi}_{\sigma,-\sigma}$, i.e.,  the first derivatives 
of the self-energy. 
For instance,  Eq.\ \eqref{eq:chi_org} means that 
the susceptibilities are enhanced  from the one for the free quasi-particles 
$\rho_{d\sigma}^{}$  by the factor $\widetilde{\chi}_{\sigma\sigma'}$.
Furthermore,  as a first step,  the low-frequency expansion  of the 
self-energy  up to the $\omega$-linear terms  is given in terms 
of the phase shift and the renormalization factor, 
 \begin{align}
\epsilon_{d\sigma} +  \Sigma_{\sigma}(i\omega) 
   \,= \, 
 \Delta\, \cot \delta_{\sigma}
 + \bigl( 1-\widetilde{\chi}_{\sigma\sigma} \bigr)\, i\omega 
+ O(\omega^2) .
\label{eq:self_w_linear} 
\end{align}

\begin{widetext}
A series of the Fermi-liquid relations can be deduced, step by step,  
 from  the higher-order derivatives of  the Ward identity  for $n=1,2,3, \ldots$,  
\begin{align}
- \delta_{\sigma\sigma'}\, 
\frac{\partial^{n+1} \Sigma_{\sigma}(i \omega)}{\partial (i \omega)^{n+1}}
%
\ =  \ 
\frac{\partial}{\partial \epsilon_{d\sigma'}} 
\left(
\frac{\partial^n
\Sigma_{\sigma}(i\omega)}{\partial (i\omega)^n }
\right)
+
\frac{\partial^n}{\partial (i\omega)^n}
\Gamma_{\sigma \sigma';\sigma' \sigma}(i\omega , 0; 0 , i\omega) 
\,\rho_{d\sigma'}.
\label{eq:nth_B}
\end{align}
Here,  for the first term on the right-hand side,   
the derivative with respect to $i\omega$ and  
that with respect to $\epsilon_{d\sigma'}$ have been commuted.
Equation \eqref{eq:nth_B} means that  
the $n+1$-th derivative of  $\Sigma_{\sigma}(i\omega)$ in the left-hand side 
will be calculated from the $n$-th one   
if  the additional vertex term  for the parallel spins
\begin{align}
\frac{\partial^n}{\partial (i\omega)^n}
\Gamma_{\sigma \sigma;\sigma\sigma}(i\omega , 0; 0 , i\omega)  
\end{align}
is explicitly  known. Therefore, the  derivative of 
$\Gamma_{\sigma \sigma;\sigma\sigma}(i\omega , 0; 0 , i\omega)$ 
plays a central role to proceed iteratively to the next step.
The Fermi-liquid relations of the $n+1$-th step  
can be written in  the following form, 
for the anti-parallel $\sigma'=-\sigma$  
and the parallel $\sigma'=\sigma$ spin components of Eq.\ \eqref{eq:nth_B}, 
respectively,  
\begin{align}
&  \ \ 
\left.
\frac{\partial^n}{\partial (i\omega)^n}
\Gamma_{\sigma,- \sigma;-\sigma, \sigma}(i\omega , 0; 0 , i\omega) 
\,\rho_{d,-\sigma}
\right|_{\omega\to 0}
\ = \ \ 
\left. 
-\frac{\partial}{\partial \epsilon_{d,-\sigma}} 
\left(
\frac{\partial^n
\Sigma_{\sigma}(i\omega)}{\partial (i\omega)^n }
\right) 
\right|_{\omega\to 0}  
\label{eq:nth_C},
\\
\nonumber \\ 
& \left.
-\frac{\partial^{n+1} \Sigma_{\sigma}(i \omega)}{\partial (i \omega)^{n+1}}
\right|_{\omega\to 0} 
=   \   
(-1)^n
\frac{\partial^{n+1} \Sigma_{\sigma}(0)}{\partial \epsilon_{d\sigma}^{n+1}} 
\ \ +\ \  \sum_{k=1}^{n-1}(-1)^k
\!\!
\left.
\frac{\partial^{k}}{\partial \epsilon_{d\sigma}^{k}}
\left(
\frac{\partial^{n-k}}{\partial (i\omega)^{n-k} }
\Gamma_{\sigma \sigma;\sigma\sigma}(i\omega , 0; 0 , i\omega) \,\rho_{d\sigma}
\,\right) 
\right|_{\omega\to 0}
\nonumber \\ 
& 
\qquad\qquad \qquad \qquad \quad \ \ 
 +
\left.\frac{\partial^n}{\partial (i\omega)^n}
\Gamma_{\sigma \sigma;\sigma\sigma}(i\omega , 0; 0 , i\omega) 
\,\rho_{d\sigma}
\right|_{\omega\to 0}
.
\label{eq:nth_D}
\end{align}
\end{widetext}
The first term in the right-hand side of Eq.\ \eqref{eq:nth_D} 
corresponds to the $n$-th derivative of 
$\widetilde{\chi}_{\sigma\sigma}$ with respect to  $\epsilon_{d\sigma}$, 
and can be  related to the ($n+2$)-body correlation function 
$\chi_{\sigma\sigma\sigma\ldots}^{[n+2]}$, 
which correspond to a generalization of  
Eqs.\ 
\eqref{eq:canonical_correlation_3_dif} and \eqref{eq:canonical_correlation_3}. 
Therefore, the ($n+1$)-th step  Fermi-liquid relations  are 
determined by the static  nonlinear susceptibilities of  $\delta n_{d\sigma}$ 
up to the ($n+2$)-th order.

\begin{figure}[t]
\leavevmode 
\begin{minipage}{1\linewidth}
\includegraphics[width=0.3\linewidth]{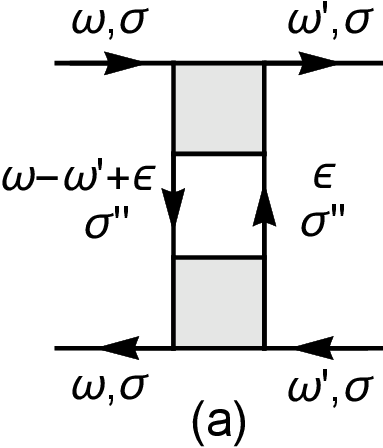}
\rule{0.02\linewidth}{0cm}
\includegraphics[width=0.3\linewidth]{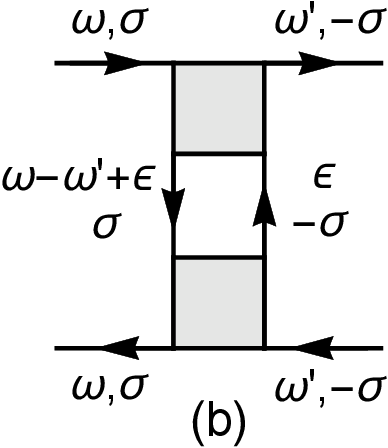}
\rule{0.02\linewidth}{0cm}
\includegraphics[width=0.3\linewidth]{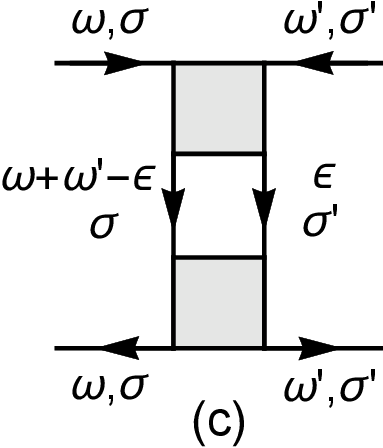}
\end{minipage}
 \caption{Feynman diagrams, which cause the singularities 
of the vertex function 
$\Gamma_{\sigma\sigma';\sigma'\sigma}^{}(i\omega, i\omega'; i\omega', i\omega)$ 
 for small $\omega$ and $\omega'$.
The  intermediate particle-hole excitation in  (a) and (b)  
yields the non-analytic  $\mathrm{sgn}\, (\omega-\omega')$  term. 
 The  particle-particle excitation in (c)  yields 
the  $\mathrm{sgn}\, (\omega+\omega')$ term.
As the vertex function, which is represented by the shaded square,  
vanishes   $\Gamma_{\sigma\sigma;\sigma\sigma}^{}(0, 0; 0, 0)=0$ 
for parallel spins  at zero frequencies, 
the non-analytic terms emerge for the spin configurations 
of  (a) $\sigma''= - \sigma$,  
 (b) $\sigma=\uparrow,\downarrow$, and  
(c) $\sigma'=-\sigma$.   
}
  \label{fig:vertex_singular_su2}
\end{figure}

\subsection{Next-leading Fermi-liquid corrections: the $n=1$ hierarchy }
\label{subsec:next_leading_n=1}

The next leading  (2nd step) Fermi-liquid relations are generated 
from the first derivative of the Ward identity,  namely 
Eqs.\ \eqref{eq:nth_C} and \eqref{eq:nth_D}  for $n=1$: 
\begin{align}
& 
\left.
\frac{\partial}{\partial i\omega}
\Gamma_{\sigma,-\sigma;-\sigma\sigma}
(i\omega , 0; 0 , i\omega) 
\,\rho_{d,-\sigma}^{} \right|_{\omega \to 0}^{} 
\nonumber \\ 
= & \ \  
\frac{\partial}{\partial \epsilon_{d,-\sigma}^{}} 
\left(
\frac{\partial
\Sigma_{\sigma}(0)}{\partial \epsilon_{d\sigma}^{}}
\right)
,
\label{eq:2ndB_up_down}
\\
\nonumber \\
&\left.
\frac{\partial^2 \Sigma_{\sigma}(i \omega)}{\partial (i \omega)^2}
\right|_{\omega\to 0}^{} 
\nonumber \\
= & \ 
\frac{\partial}{\partial \epsilon_{d\sigma}^{}} 
\left(
\frac{\partial
\Sigma_{\sigma}(0)}{\partial \epsilon_{d\sigma}^{}}
\right) 
  - 
\left.
\frac{\partial}{\partial i\omega}
\Gamma_{\sigma\sigma;\sigma\sigma}(i\omega , 0; 0 , i\omega) 
\, \rho_{d\sigma}^{}
\right|_{\omega\to 0}^{}.
\label{eq:2ndE}
\end{align}
For  the  second derivative of the self-energy 
on the  right-hand side of  these two equations, 
we have used  the relation $1/z_{\sigma} =
\widetilde{\chi}_{\sigma\sigma}$ given in Eq.\  \eqref{eq:YY2_results}.
These  two terms can also be written in terms of 
the first derivative of 
$\widetilde{\chi}_{\sigma\sigma}$ with 
respect to  $\epsilon_{d\sigma'}$, 
and are related to the three-body correlation function 
$\chi_{\sigma\sigma\sigma'}^{[3]}$.

From Eqs.\ \eqref{eq:YY2_results} and \eqref{eq:2ndB_up_down}, 
the vertex function for anti-parallel spins 
$\Gamma_{\sigma,-\sigma;-\sigma\sigma}
(i\omega , 0; 0 , i\omega)$ can be deduced up to the $\omega$-linear term, 
\begin{align}
&\Gamma_{\sigma,-\sigma;-\sigma\sigma}
(i\omega , 0; 0 , i\omega) 
\,\rho_{d,-\sigma}^{}
\nonumber  \\
& 
\qquad \qquad \quad 
=  \ 
-\widetilde{\chi}_{\sigma,-\sigma} + 
\frac{\partial \widetilde{\chi}_{\sigma,-\sigma}}
{\partial \epsilon_{d\sigma}^{}} \, i \omega   +  O(\omega^2) 
.
\label{eq:GammaUD_next}
\end{align}
It shows that the $\omega$-linear term does not accompany 
 a singular  $\omega \,\mathrm{sgn} (\omega)$ dependence 
which converts into the imaginary part  
by the analytic continuation 
 $i\omega \to \omega +i0^+$.\cite{YamadaYosida4,ShibaKorringa} 
This has been perturbatively understood as follows.
For the anti-parallel spins vertex,  
the non-analytic  $\omega \,\mathrm{sgn} (\omega)$ dependence 
 disappears  as a result of the cancellation between the contribution 
of the particle-hole pair and that  of the particle-particle pair. 
 These pairs first emerge  in the order $U^2$ processes  
described  in Fig.\ \ref{fig:vertex_u_up_dw}, 
where  the particle and hole carry different spins, $\sigma$ and $-\sigma$. 
The total contributions, which include all the higher-order processes 
described in Fig.\ \ref{fig:vertex_singular_su2} (b) and  (c) for $\sigma'=-\sigma$, 
 cancel each other out:  it can be  confirmed  explicitly  through  Eq.\ \eqref{eq:Im_(b)+(c)}.

We next consider the relation for the parallel spin component 
 given in Eq.\ \eqref{eq:2ndE}.
The first term in the right-hand side is real and is given by 
$\partial \widetilde{\chi}_{\sigma\sigma}/ \partial \epsilon_{d\sigma}$.   
Therefore,  the discontinuous  $\mathrm{sgn} (\omega)$  dependence 
emerges only from the second term,  
namely first derivative of 
$\Gamma_{\sigma\sigma;\sigma\sigma}(i\omega , 0; 0, i\omega)$ 
with respect to $i\omega$. 
It was also shown by Yamada-Yosida that 
 the discontinuous  $\mathrm{sgn} (\omega)$ dependence in the derivative 
 emerges from the  intermediate one particle-hole pair 
carrying spin $-\sigma$, which is opposite 
to the spin $\sigma$ of the external line:  
diagrams for the order $U^2$ processes 
and for the generalized ones    are shown in Figs.\  \ref{fig:vertex_u2_up_up},  
\ref{fig:vertex_singular_su2} (a),  and 
\ref{fig:vertex_singular_su2} (c) for $\sigma'=\sigma$, respectively.
The contribution 
has been obtained in the form\cite{YamadaYosida4,Yoshimori} 
\begin{align}
&
\mathrm{Im} 
\left. 
\frac{\partial}{\partial i\omega}
\Gamma_{\sigma\sigma;\sigma\sigma}(i\omega , 0; 0, i\omega) 
\,\rho_{d\sigma}^{}\right|_{\omega\to 0}^{} =\,
 \pi  \frac{\chi_{\uparrow\downarrow}^2}{\rho_{d\sigma}^{}} 
\, \mathrm{sgn}(\omega)  
 \;. 
\label{eq:vert_UU_im_w}
\end{align}
From this  result and the relation  Eq.\ \eqref{eq:2ndE}, 
 the $\omega^2$ imaginary part of the self-energy
also has been deduced:\cite{YamadaYosida4,Yoshimori}  
\begin{align}
\mathrm{Im}
\left.
\frac{\partial^2 \Sigma_{\sigma}(i\omega)}{\partial (i\omega)^2}\, 
\right|_{\omega \to 0}^{} 
=\,   
- \pi \, 
\frac{\chi_{\uparrow\downarrow}^2}{\rho_{d\sigma}^{}}  
\, \mathrm{sgn}(\omega)  
\;.
\label{eq:self_im_w2}
\end{align}


In contrast to the non-analytic component, 
the analytic component of the  $\omega$-linear part of 
$\Gamma_{\sigma\sigma;\sigma\sigma}(i\omega , 0; 0, i\omega)$ 
has not been studied in detail  so far.  
This part will contribute to low-energy transport away from half-filling if it is finite. 
In the present paper we calculate 
the {\it regular\/} part  using  a Green's-function product expansion, 
and show that it identically vanishes,    
\begin{align}
&
\mathrm{Re} 
\left. 
\frac{\partial}{\partial i\omega}
\Gamma_{\sigma\sigma;\sigma\sigma}(i\omega , 0; 0, i\omega) 
\,\rho_{d\sigma}^{}\right|_{\omega\to 0}^{} =\, 0 \;.
\label{eq:vert_UU_real_w}
\end{align}
This is one of the key features of the vertex function, 
and  is caused  by its anti-symmetry property. 
 We provide  a microscopic proof  later in the present paper.  
An  important identity, which relates 
the real parts of  two different second derivatives of the self-energy,  
 follows from  Eqs.\ \eqref{eq:2ndE} and \eqref{eq:vert_UU_real_w}, 
\begin{align}
\mathrm{Re}
\left.
\frac{\partial^2 \Sigma_{\sigma}(i\omega)
}{\partial( i\omega)^2}\, 
\right|_{\omega\to 0}^{}
=\, 
\frac{\partial^2 \Sigma_{\sigma}(0)}{\partial \epsilon_{d\sigma}^2} 
\;.
\label{eq:self_real_w2}
\end{align}
This relation agree with  FMvDM's  formula 
 given in Eq.\ (B8b) of  Ref.\ \onlinecite{FilipponeMocaVonDelftMora},   
which was obtained by extending Nozi\`{e}res'  phenomenological description.

From the knowledge 
of  Eqs.\  \eqref{eq:vert_UU_im_w} and \eqref{eq:vert_UU_real_w},   
the low-frequency behavior of the parallel-spin component of the vertex function 
can be explicitly written  up to the  $\omega$-linear part: 
\begin{align}
\Gamma_{\sigma\sigma;\sigma\sigma}(i\omega , 0; 0, i\omega) 
\,\rho_{d\sigma}^{2}
=  \,  
 i \pi \,
\chi_{\uparrow\downarrow}^2
\, i\omega \, \mathrm{sgn} (\omega) 
+ O(\omega^2) . 
\label{eq:GammaUU_general_causal}
\end{align}
Note that the non-analytic  $\omega$-linear term 
corresponds to the absolute value $|\omega|= \omega\, \mathrm{sgn} (\omega)$, 
which has a cusp at $\omega=0$. 
Then,  using Eqs.\ \eqref{eq:self_im_w2} and \eqref{eq:self_real_w2}
 with Eq.\ \eqref{eq:chi_tilde_diff},
the self-energy can be determined up to the $(i\omega)^2$ term  
which extends Eq.\ \eqref{eq:self_w_linear}, as 
 \begin{align}
\epsilon_{d\sigma} +  \Sigma_{\sigma}(i\omega) 
  = &  \  
 \Delta\, \cot \delta_{\sigma}
 + \bigl( 1-\widetilde{\chi}_{\sigma\sigma}   \bigr)\, i\omega 
\nonumber \\ 
& 
\!\!\!\! \!\!\!\! \!\!\!\! \!\!
 + \frac{1}{2} 
\left(\frac{\partial \widetilde{\chi}_{\sigma\sigma}}
 {\partial \epsilon_{d\sigma}^{}} 
 - i \pi   
\frac{\chi_{\uparrow\downarrow}^2}{\rho_{d\sigma}^{}}
\, \mathrm{sgn} (\omega) \right)
 (i\omega)^2 
 + O(\omega^3) . 
\end{align}
The next-leading Fermi-liquid correction  that enters  through 
 ${\partial \widetilde{\chi}_{\sigma\sigma}}/{\partial \epsilon_{d\sigma'}}$    
 vanishes  in the particle-hole  symmetric case at zero magnetic field.
This is because  the spin and charge susceptibilities 
 take extreme values:  
${\partial \chi_s}/{\partial \epsilon_d^{}} =0$ and
${\partial \chi_c}/{\partial \epsilon_d^{}} =0$,  
at  $\xi_d \equiv  \epsilon_d +U/2=0$ and $h=0$.

\subsection{Higher-order Fermi-liquid corrections for $n=2$}

We can also deduce the second derivative of the vertex function for 
the anti-parallel spins 
from Eqs.\ \eqref{eq:vert_UU_im_w} and \eqref{eq:self_real_w2} 
through Eq.\ \eqref{eq:nth_C} for $n=2$,
\begin{align}
& 
\!\!\!\!\!\!
\left.
\frac{\partial^2}{\partial (i\omega)^2}
\Gamma_{\sigma, -\sigma;-\sigma, \sigma}(i\omega , 0; 0 , i\omega) 
\,\rho_{d,-\sigma} 
\right|_{\omega \to 0}^{}
\nonumber \\
& =  \ 
- \, 
\frac{\partial}{\partial \epsilon_{d,-\sigma}^{}} 
\left.
\left(
\frac{\partial^2
\Sigma_{\sigma}(i\omega)}{\partial (i\omega)^2 } 
\right)
\right|_{\omega\to 0}^{}
\nonumber \\
 & = \ 
\frac{\partial}{\partial \epsilon_{d,-\sigma}^{}} 
\left(
- 
\frac{\partial^2 \Sigma_{\sigma}(0)}{\partial \epsilon_{d\sigma}^2} 
 + i \pi \, \frac{\chi_{\uparrow\downarrow}^2}{\rho_{d\sigma}^{}} 
 \, \mathrm{sgn} (\omega) 
\right) .
\end{align}
Therefore,  adding  the  $\omega^2$ term to  Eq.\ \eqref{eq:GammaUD_next},
we obtain  the low-energy expansion  in an extended form:  
\begin{widetext} 
\begin{align}
 &
\!\!
\Gamma_{\sigma, -\sigma;-\sigma, \sigma}(i\omega, 0; 0 ,i\omega) 
\,\rho_{d\sigma}^{}\rho_{d,-\sigma}^{} 
= 
\,-\chi_{\uparrow\downarrow} + 
\rho_{d\sigma}^{}
\frac{\partial \widetilde{\chi}_{\sigma,-\sigma}}
{\partial \epsilon_{d\sigma}^{}} \ i\omega  
 +\frac{\rho_{d\sigma}^{}}{2} \,
\frac{\partial}{\partial \epsilon_{d,-\sigma}^{}} 
\left[
- 
\frac{\partial \widetilde{\chi}_{\sigma\sigma}}{\partial \epsilon_{d\sigma}^{}} 
 + i \pi\, \frac{\chi_{\uparrow\downarrow}^2}{\rho_{d\sigma}^{}} 
 \, \mathrm{sgn}(\omega) 
\right]  (i\omega)^2  + \cdots
\label{eq:GammaUD_general_matsubara}
\end{align}
\end{widetext} 
 This is also one of the most important  results of the present work.
The  $\omega^2$ term 
involves  higher-order  corrections which  correspond to the static 
four-body susceptibilities   $\chi_{\sigma\sigma\sigma,-\sigma}^{[4]}$.
At zero field $h=0$,  
the coefficients for the imaginary and real part of the $\omega^2$ term 
can be expressed in the form 
of  Eqs.\  \eqref{eq:vert_u_d_w4_im} and \eqref{eq:vert_u_d_w4_re2},  
given  in Appendix \ref{sec:w2_vertex_ud_coefficient}.
Specifically  in the particle-hole symmetric case at zero magnetic field,  
\begin{align}
&
\frac{\Gamma_{\sigma,-\sigma;-\sigma,\sigma}
(i\omega, 0; 0 , i\omega) 
}{\pi\Delta}
\nonumber \\
& 
 \qquad \quad 
= \ 
-\widetilde{\chi}_{\uparrow\downarrow} 
- \frac{1}{2} 
 \left.
\frac{\partial^2 \widetilde{\chi}_{\sigma,-\sigma} }{\partial \epsilon_{d\sigma}^2} 
\right|_{h \to 0 \atop \xi_d \to 0}^{}   (i\omega)^2 
   +  \cdots   .
\label{eq:GammaUD_general_causal_symmetric_case}
\end{align}
The real part of the  $\omega^2$ term remains finite with the coefficient 
\begin{align}
&\left.
\frac{1}{\pi\Delta}
\frac{\partial^2 \widetilde{\chi}_{\sigma,-\sigma} }{\partial \epsilon_{d\sigma}^2} 
\right|_{h \to 0 \atop \xi_d \to 0}^{}
 \nonumber\\
& \qquad 
=   
\left.
\frac{1}{4}
\left( 
 \frac{\partial^2 \chi_{\uparrow\downarrow}^{} }{\partial \epsilon_{d}^2} 
 + \frac{\partial^2 \chi_{\uparrow\downarrow}^{}}{\partial h^2}  
  \right)
\right|_{h \to 0 \atop \xi_d \to 0}^{}
\!\! +2\pi^2 
\chi_{\uparrow\downarrow}^{} \chi_{\uparrow\uparrow}^{2} .
\end{align}
 Fermi-liquid corrections of this order, $n=2$,  also emerge 
in the order $\omega^2$ contributions of the  parallel spins component. 
Therefore,  to take into account  all corrections of this order, 
one needs to calculate 
\begin{align}
\left.
\frac{\partial^2}{\partial (i\omega)^2}
\Gamma_{\sigma \sigma;\sigma \sigma}(i\omega , 0; 0 , i\omega) 
\,\rho_{d\sigma}^{} 
\right|_{\omega \to 0}^{} \;.
\end{align}
Then,  $\frac{\partial^{3}
 \Sigma_{\sigma}(i\omega)}{\partial (i\omega)^{3} }$  follows 
through Eq.\ \eqref{eq:nth_D} for $n=2$.

\section{
Double-frequency expansion of 
$\Gamma_{\sigma\sigma';\sigma'\sigma}(i\omega, i\omega'; i\omega',i\omega )$ 
}
\label{sec:vertex_omega_omega_dash}

\begin{widetext}
In this section, 
we show that the double-frequency expansion of 
$\Gamma_{\sigma\sigma';\sigma'\sigma}
(i\omega, i\omega'; i\omega',i\omega )$ 
 up to the linear terms in  $\omega$ and $\omega'$ 
 can also be expressed in terms of the Fermi-liquid parameters.
The results are shown in 
Eqs.\ \eqref{eq:GammaUU_general_omega_dash_abs} and 
\eqref{eq:GammaUD_general_matsubara_omega_dash_abs}:
for the parallel component $\sigma'=\sigma$,
\begin{align}
& 
\Gamma_{\sigma\sigma;\sigma\sigma}(i\omega , i\omega'; i\omega', i\omega) 
\,\rho_{d\sigma}^{2}
\,= \,
 - \pi \,
\chi_{\uparrow\downarrow}^2
\left| \omega - \omega' \right|
+ \cdots , 
\label{eq:GammaUU_general_omega_dash_abs}
\end{align}
and  for  $\sigma'=-\sigma$ it is   
\begin{align}
\Gamma_{\sigma, -\sigma;-\sigma, \sigma}(i\omega, i\omega'; i\omega' ,i\omega) 
\,\rho_{d\sigma}^{}\rho_{d,-\sigma}^{}
\, = \,   
\,-\chi_{\uparrow\downarrow} + 
\rho_{d\sigma}^{}
\frac{\partial \widetilde{\chi}_{\sigma,-\sigma}}
{\partial \epsilon_{d\sigma}^{}} \, i\omega  
+ 
\rho_{d,-\sigma}^{}
\frac{\partial \widetilde{\chi}_{-\sigma,\sigma}}
{\partial \epsilon_{d,-\sigma}^{}} \, i\omega'   
- \pi \,\chi_{\uparrow\downarrow}^2
\Bigl[\,
\left|\omega-\omega'\right| - \left|\omega+\omega'\right|
\,\Bigr]
+ \cdots .
\label{eq:GammaUD_general_matsubara_omega_dash_abs}
\end{align}
Note that  
$\left| \omega \pm \omega' \right| 
= -i( i\omega  \pm  i\omega') \  \mathrm{sgn} (\omega \pm \omega')$. 
\end{widetext}

These asymptotically exact results 
capture  the essential features of the Fermi liquid, 
and are analogous to Landau's quasi-particle 
interaction $f(\bm{p}\, \sigma,\bm{p}' \sigma')$ 
and Nozi\`{e}res' function $\phi_{\sigma\sigma'}(\varepsilon,\varepsilon')$. 
\cite{AGD,NozieresFermiLiquid}  One important difference is that 
the vertex function also has  a non-analytic part which 
directly determines  the damping of the quasi-particles. 
We give the derivations of  
Eqs.\ \eqref{eq:GammaUU_general_omega_dash_abs} and 
\eqref{eq:GammaUD_general_matsubara_omega_dash_abs} 
in the following.

\begin{figure}[t]
 \leavevmode
\begin{minipage}{1\linewidth}
\includegraphics[width=0.45\linewidth]{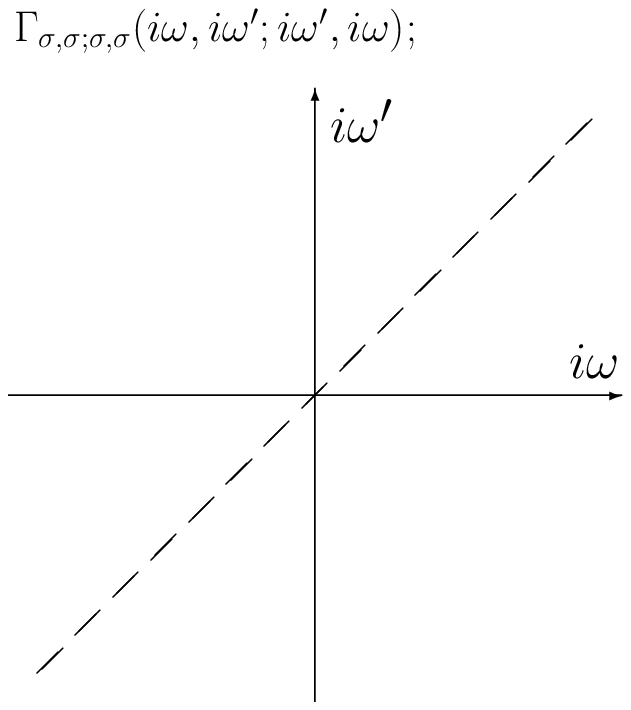}
\rule{0.05\linewidth}{0cm}
\includegraphics[width=0.45\linewidth]{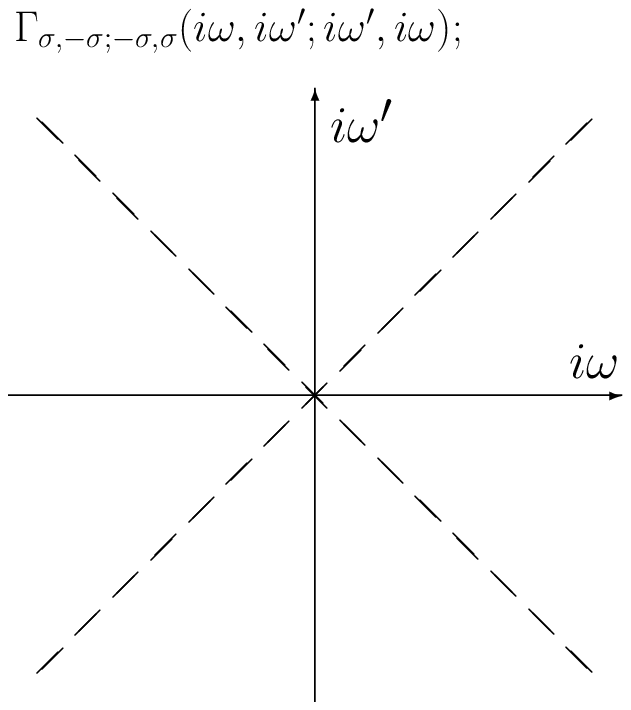}
\end{minipage}
 \caption{
Non-analytic $|\omega-\omega'|$ and $|\omega+\omega'|$ 
contributions of    
$\Gamma_{\sigma\sigma;\sigma\sigma}^{}
(i\omega, i\omega'; i\omega', i\omega)$  
and  
$\Gamma_{\sigma,-\sigma;-\sigma,\sigma}^{}
(i\omega, i\omega'; i\omega', i\omega)$.  
}
 \label{fig:w_w'_plane_discontinuities}
\end{figure}

\subsection{Anti-symmetry properties of  
 $\Gamma_{\sigma\sigma';\sigma'\sigma}(i\omega,i\omega';i\omega',i\omega)$}
\label{subsec:singular_vertex_w_wdash_linear_1}

Two important features of the vertex function, 
the anti-symmetric property and analytic property,  
play an essential role in the proof,  which we provide in this section.  
The fermionic antisymmetric commutation relation 
imposes a strong restriction on the vertex function,  
\begin{align}
&
\!\!\!\!\!\!\!\!\!\!\!\!\!\!\!\!\!\!\!\!\!\!\!\!\!\!\!\!\!\!\!
\Gamma_{\sigma_1\sigma_2;\sigma_3\sigma_4}
(i\omega_1, i\omega_2; i\omega_3, i\omega_4) 
\nonumber \\
=&   \   \  
\Gamma_{\sigma_3\sigma_4;\sigma_1\sigma_2}
(i\omega_3, i\omega_4; i\omega_1, i\omega_2) 
\nonumber \\
=&   \, 
-\Gamma_{\sigma_3\sigma_2;\sigma_1\sigma_4}
(i\omega_3, i\omega_2; i\omega_1, i\omega_4) 
\nonumber \\
=& \,  
-\Gamma_{\sigma_1\sigma_4;\sigma_3\sigma_2}
(i\omega_1, i\omega_4; i\omega_3, i\omega_2) . 
\label{eq:vertex_paralle_anti_sym}
\end{align}
Obviously,  $\Gamma_{\sigma\sigma;\sigma\sigma}(0,0; 0, 0)=0$ 
follows from Eq.\ \eqref{eq:vertex_paralle_anti_sym} at zero frequencies. 
The anti-symmetry property also  imposes strong constrains in the linear 
$\omega$ and $\omega'$ dependences.   
The analytic properties of the vertex function is another key to 
determine the explicit form of the linear terms. 
The vertex function 
  $\Gamma_{\sigma\sigma';\sigma'\sigma}(i\omega,i\omega';i\omega',i\omega)$
has some singularities in the $\omega$-$\omega'$ plane.\cite{Eliashberg,EliashbergJETP15} 
Specifically, the non-analytic  terms emerge through 
the three diagrams given in Fig.\ \ref{fig:vertex_singular_su2} 
for small $\omega$  and  $\omega'$.
The intermediate particle-hole and 
particle-particle pair excitations in the Anderson impurity yield 
the non-analytic terms of the form  $|\omega - \omega'|$ and  $|\omega + \omega'|$, 
respectively,  which divide the $\omega$-$\omega'$ plane as shown 
  in Fig.\ \ref{fig:w_w'_plane_discontinuities}.
Thus,  the linear terms of the double-frequency expansion 
can be expressed as a  linear combination 
of $\omega$, $\omega'$,   $|\omega - \omega'|$, and  $|\omega + \omega'|$; 
\begin{align}
&\Gamma_{\sigma\sigma;\sigma\sigma}(i\omega , i\omega'; i\omega', i\omega) 
\,\rho_{d\sigma}^{2}
\nonumber \\
& \quad = \, 
 a_{\sigma\sigma}^{}  \bigl( i\omega +i\omega'  \bigr)
+ b_{\sigma\sigma}^{-}\,  |\omega-\omega'| 
+  b_{\sigma\sigma}^{+}\,  |\omega+\omega'|  +  \cdots 
\label{eq:omega_omega_dash_expand_UU}
\\
& \Gamma_{\sigma, -\sigma;-\sigma, \sigma}(i\omega, i\omega'; i\omega' ,i\omega) 
\,\rho_{d\sigma}^{}\rho_{d,-\sigma}^{}
\nonumber \\
& \quad
= \, 
-\chi_{\uparrow\downarrow} 
+  a_{\sigma,-\sigma}^{}  \,i\omega 
+  a_{-\sigma,\sigma}^{}  \,i\omega' 
+ b_{\uparrow\downarrow}^{-}\,  |\omega-\omega'| 
\nonumber \\
&\quad \quad \   
+  b_{\uparrow\downarrow}^{+}\,  |\omega+\omega'|  +  \cdots  
 .  
\label{eq:omega_omega_dash_expand_UD}
\end{align}
Here,  $a_{\sigma\sigma'}^{}$  and $b_{\sigma\sigma'}^{\pm}$ 
with  $b_{\uparrow\downarrow}^{\pm} 
= b_{\downarrow\uparrow}^{\pm}$ 
are the  expansion coefficients for the  analytic and non-analytic terms, 
respectively.
 Equations \eqref{eq:omega_omega_dash_expand_UU} and 
 \eqref{eq:omega_omega_dash_expand_UD}
  are constructed in such way that each satisfies  one of the requirements   
$
\Gamma_{\sigma\sigma';\sigma'\sigma}(i\omega , i\omega'; i\omega', i\omega) 
= 
\Gamma_{\sigma'\sigma;\sigma\sigma'}(i\omega' , i\omega; i\omega, i\omega') 
$.
In order to satisfy the remaining requirements of the anti-symmetry 
given in Eq.\ \eqref{eq:vertex_paralle_anti_sym}, 
the coefficient  $a_{\sigma\sigma}^{}$ for the parallel spin components 
must vanish  as shown in Appendix \ref{sec:homogeneous polynomial},
\begin{align}
 a_{\sigma\sigma}^{} \, \equiv \,0 \;.
\end{align}
Thus, the parallel-spin vertex does not have the analytic term.
Equation \eqref{eq:vert_UU_real_w} follows from this result. 
Therefore,  as  Eq.\ \eqref{eq:GammaUD_general_matsubara} 
has been deduced from  Eq.\ \eqref{eq:vert_UU_real_w} 
and the Ward identities,
we can use  Eq.\ \eqref{eq:GammaUD_general_matsubara} 
to determine  $a_{\sigma,-\sigma}^{}$ taking $\omega'=0$, 
\begin{align}
 a_{\sigma,-\sigma}^{} \, \equiv \,\rho_{d\sigma}^{}
\frac{\partial \widetilde{\chi}_{\sigma,-\sigma}}
{\partial \epsilon_{d\sigma}^{}} \;.
\end{align}

\subsection{Non-analytic part  of  
 $\Gamma_{\sigma\sigma';\sigma'\sigma}(i\omega,i\omega';i\omega',i\omega)$
for small $\omega$ and $\omega'$}
\label{subsec:singular_vertex_w_wdash_linear_2}

In the following, 
we calculate the non-analytic part of 
$\Gamma_{\sigma\sigma';\sigma'\sigma}(i\omega, i\omega'; i\omega', i\omega) $, 
and  directly derive the $|\omega-\omega'|$   and $|\omega+\omega'|$ 
contributions including the coefficients.
The analytic  and non-analytic parts 
of  the first derivative of the vertex function 
with respect to $i\omega$, or $i\omega'$,  
take real and pure-imaginary values, respectively.
Specifically, we consider  the imaginary part of the first derivative 
with respect to $i\omega'$ in detail.

For the parallel spin vertex, the non-analytic term emerges from  
the diagram of Fig.\ \ref{fig:vertex_singular_su2} (a). 
For small $\omega$ and $\omega'$,  it is given by
\begin{align}
&   \!\!
 \mathrm{Im}\, 
\rho_{d\sigma}^{}\,
\frac{\partial}{\partial i\omega'}
\,
\Gamma_{\sigma\sigma;\sigma\sigma}(i\omega,i\omega';i\omega',i\omega) 
\nonumber \\
& =    
-\,
\left|\Gamma_{\uparrow\downarrow;\downarrow\uparrow}(0,0;0,0)
\right|^2 
\nonumber \\
& \quad 
\times 
\rho_{d\sigma}^{}
\, \mathrm{Im} 
\int \frac{d \varepsilon}{2\pi}\,
G_{-\sigma}^{}(i\varepsilon) \,
\frac{\partial}{\partial i\omega'} \,
G_{-\sigma}^{}(i\omega -i\omega'+ i\varepsilon)
\nonumber
\\
& = \,   
-\,
\pi 
 \left| 
  \Gamma_{\uparrow\downarrow;\downarrow\uparrow}(0,0;0,0)\right|^2 \,
\rho_{d\sigma}^{}\, \rho_{d,-\sigma}^2
\ \mbox{sgn}(\omega-\omega') \,
 .
\label{eq:Ward_2_Im_diagram_1}
\end{align}
Here, we have used the differential formula for  the  Green's function, 
given in  Eq.\ \eqref{eq:G_full_omega}.
The non-analytic terms of the anti-parallel spin vertex function  
emerge from the other two diagrams 
shown in Figs.\ \ref{fig:vertex_singular_su2} (b) and (c): 
\begin{align}
&    
\!\!
 \mathrm{Im}\, 
\rho_{d,-\sigma}^{}\,
\frac{\partial}{\partial i\omega'}
\,
\Gamma_{\sigma,-\sigma;-\sigma,\sigma}(i\omega,i\omega';i\omega',i\omega) 
\nonumber \\
& =    
-\,
\left|\Gamma_{\uparrow\downarrow;\downarrow\uparrow}(0,0;0,0)
\right|^2 
\nonumber \\
& \quad 
\times 
\rho_{d,-\sigma}^{}
\, \mathrm{Im} \Biggl[ 
\int \frac{d \varepsilon}{2\pi}\,
G_{-\sigma}^{}(i\varepsilon) \,
\frac{\partial}{\partial i\omega'} \,
G_{\sigma}^{}(i\omega -i\omega'+ i\varepsilon)
\nonumber \\
& \qquad \qquad \quad 
+ 
\int \frac{d \varepsilon}{2\pi}\,
G_{-\sigma}^{}(i\varepsilon) \,
\frac{\partial}{\partial i\omega'} \,G_{\sigma}^{}(i\omega +i\omega'-i\varepsilon)\,
\,\Biggr]
\nonumber
\\
& = \,   
-\,
\pi 
 \left| 
  \Gamma_{\uparrow\downarrow;\downarrow\uparrow}(0,0;0,0)\right|^2 \,
\nonumber\\
& \qquad \times 
\rho_{d\sigma}^{}\, \rho_{d,-\sigma}^2
\Bigl[\, 
\,\mbox{sgn}(\omega-\omega') \,
+\, 
\,\mbox{sgn}(\omega+\omega') \,
\Bigr]
 .
\label{eq:Ward_2_Im_diagram_2}
\end{align}
Equations \eqref{eq:Ward_2_Im_diagram_1} and \eqref{eq:Ward_2_Im_diagram_2} 
determine the non-analytic 
  $|\omega-\omega'|$ and  $|\omega+\omega'|$ terms 
of  $\Gamma_{\sigma\sigma';\sigma'\sigma}(i\omega,i\omega';i\omega',i\omega)$: 
 \begin{align}
 b_{\sigma\sigma}^{-} = 
 b_{\uparrow\downarrow}^{-} =
  -b_{\uparrow\downarrow}^{+}  = -\pi \chi_{\uparrow\downarrow}^2, 
 \qquad 
 b_{\sigma\sigma}^{+} \,= \,0 , 
 \end{align}
and we obtain Eqs.\  \eqref{eq:omega_omega_dash_expand_UU} and 
 \eqref{eq:omega_omega_dash_expand_UD}. 
These non-analytic contributions divide the $\omega$-$\omega'$ plane of 
 $\Gamma_{\sigma\sigma';\sigma'\sigma}(i\omega,i\omega';i\omega',i\omega)$  
into the separate analytic regions as  in Fig.\ \ref{fig:w_w'_plane_discontinuities}.
Furthermore, from Eqs.\ \eqref{eq:Ward_2_Im_diagram_1} 
and \eqref{eq:Ward_2_Im_diagram_2},  
the single-frequency results described in 
Eqs.\ \eqref{eq:GammaUD_next} and \eqref{eq:vert_UU_im_w},   
which correspond to the result of Yamada-Yosida,\cite{YamadaYosida4}    
can be deduced,  using 
$\Gamma_{\sigma\sigma';\sigma'\sigma}(i\omega,i\omega';i\omega',i\omega)  
= \Gamma_{\sigma'\sigma;\sigma\sigma'}(i\omega',i\omega;i\omega,i\omega')$:  
\begin{align}
& \left.
 \mathrm{Im}\, 
\frac{\partial}{\partial i\omega}
\Gamma_{\sigma\sigma;\sigma\sigma}(i\omega,0;0,i\omega) \,
\rho_{d\sigma}^{}
\right|_{\omega\to 0}
= \, \pi \, \frac{\chi_{\uparrow\downarrow}^2}{\rho_{d\sigma}^{}} 
\, \mbox{sgn}(\omega) ,
\label{eq:Im_(a)}
\\
& \left.
 \mathrm{Im}\, 
\frac{\partial}{\partial i\omega}
\Gamma_{\sigma,-\sigma;-\sigma,\sigma}(i\omega,0;0,i\omega) 
\, \rho_{d,-\sigma}^{}\,
\right|_{\omega\to 0}
 = \, 0   .
\label{eq:Im_(b)+(c)}
\end{align}

\newpage

\section{The $T^2$  real part of self-energy} 
\label{sec:T2_self_energy}

We next consider the  $T^2$ correction of  
 the retarded self-energy $\Sigma^r(\omega)$, especially the real part. 
Before describing the derivation,  we  show the result first.  
Including  the $T^2$ correction,  
the low-energy asymptotic form of  the self-energy can be expressed 
in the following form. The imaginary part  is given by 
\begin{align}
&\mathrm{Im}\, \Sigma_{\sigma}^r(\omega) 
\,  = \,  -\,   \frac{\pi}{2}\,   
\frac{\chi_{\uparrow\downarrow}^2}{\rho_{d\sigma}^{}}
\,
   \Bigl[\,
    \omega^2 
 +(\pi T)^2  
    \,\Bigr] + \cdots   ,
\label{eq:self_T_imaginary}
\end{align}
and the real part is
\footnote{
We note that our result for the $T^2$ real part 
disagrees with  FMvDM's result:\cite{FilipponeMocaVonDelftMora}
the coefficient for the spin  $\sigma$ component is determined 
by  ${\partial \chi_{\uparrow\downarrow}}/{\partial \epsilon_{d,-\sigma}}$  
in Eq.\ \eqref{eq:self_real_T_mag} 
whereas it is  ${\partial \chi_{\uparrow\downarrow}}/{\partial \epsilon_{d\sigma}}$ 
that appears in  FMvDM's formula given 
in  Eqs.\ (B2a) and (B8a) of  Ref.\ \onlinecite{FilipponeMocaVonDelftMora} 
[See  also {\it paper III\/}\cite{ao2017_3_PRB} for details]}  
\begin{align}
& \epsilon_{d\sigma}+ 
\mathrm{Re}\, \Sigma_{\sigma}^r(\omega) 
\,  = \,   
\Delta\, \cot \delta_{\sigma}
\,+ \bigl( 1-\widetilde{\chi}_{\sigma\sigma} \bigr)\, \omega 
\nonumber \\ 
& \qquad \ \ 
 + \frac{1}{2}\,\frac{\partial \widetilde{\chi}_{\sigma\sigma}}
{\partial \epsilon_{d\sigma}}\, \omega^2 
 +  \frac{1}{6}\,
\frac{1}{\rho_{d\sigma}^{}} 
\frac{\partial \chi_{\uparrow\downarrow}}{\partial \epsilon_{d,-\sigma}} 
\, \left( \pi T\right)^2 
 +  \cdots  .
\label{eq:self_real_T_mag}
\end{align}
At zero magnetic field $h=0$, 
the 3-body correlations can be rewritten in 
terms of the derivative with respect 
to the spin-independent impurity level  $\epsilon_{d}^{}$, 
\begin{align}
& 
\epsilon_{d\sigma}+ 
\mathrm{Re}\, \Sigma_{\sigma}^r(\omega) 
\ \xrightarrow{\,h\to 0\,} 
\ \ 
\Delta\, \cot \delta_{\sigma}
\,+ \bigl( 1-\widetilde{\chi}_{\uparrow\uparrow} \bigr)\, \omega 
\nonumber \\
& \qquad \qquad 
 + 
\frac{1}{2\rho_{d}^{}}
\left(
\frac{\partial  \chi_{\uparrow\uparrow}^{}}{\partial \epsilon_{d}} 
- 
\frac{1}{2}
\frac{\partial \chi_{\uparrow\downarrow}^{}}{\partial \epsilon_{d}} 
+ 
2 \pi \cot \delta\   \chi_{\uparrow\uparrow}^{2} 
\right) 
\, \omega^2 
\nonumber \\
& 
 \qquad \qquad +  \frac{1}{12}\,
\frac{1}{\rho_{d}^{}} 
\frac{\partial \chi_{\uparrow\downarrow}}{\partial \epsilon_{d}} 
\, \left( \pi T\right)^2 
 +  \cdots  \;.
\end{align}

\subsection{Ward identity for the  $T^2$ corrections}
\label{subsec:T2_coefficient}

We  calculate  the finite-temperature $T^2$ correction of the self-energy 
using  the  Euler-Maclaurin formula 
\begin{align}
\sum_{n=0}^{\infty} q \Big(n+\frac{1}{2}\Bigr)
\,\simeq
\int_{0}^{\infty} \! dx \,q(x)  + \frac{1}{24} \left. \frac{dq(x)}{dx}\right|_{x \to 0}
.
\label{eq:euler_mac_sum}
\end{align}
Specifically,  summation over the Matsubara frequency $\omega_n = (2n +1) \pi T$ 
 of a function $\mathcal{Q}(i\omega_n)$, which has a discontinuity at  $\omega \to 0$, 
 can be calculated by using the formula  separately 
for $\omega_n>0$ and $\omega_n<0$, following Yamada-Yosida,\cite{YamadaYosida4}  
%
%
\begin{align}
& T \!  \sum_{n=-\infty}^{\infty} \mathcal{Q}(i\omega_n)
- \int_{-\infty}^{\infty} \! \frac{d \omega}{2\pi}   \ \mathcal{Q}(i\omega) \, 
\nonumber \\
& =\     
 \frac{(\pi  T)^2}{6}
\left(
 \lim_{\omega \to 0^+}
- \lim_{\omega \to 0^-}
 \right)
\left( \frac{-1}{2\pi i}
\frac{\partial \mathcal{Q}(i\omega) }{\partial i \omega }
\right) + O(T^4).
\label{eq:euler_mac_Matsubara}
\end{align}

The leading  correction  
for the self-energy  $\Sigma_{\sigma}(i\omega,T)$ of order $T^2$ 
is obtained by taking a functional derivative of the self-energy 
with respect to the full interacting Green's function 
as shown in Sec.\ 19.5 of the book of  Abrikosov, Gorkov and Dzyashinski (AGD),  
specifically the formula for order $T^2$ correction is given 
in  Eq.\ (19.22) of AGD.\cite{AGD} 
It can be derived by using  the Luttinger-Ward functional,\cite{LuttingerWard} 
and  taking into account   the  corrections emerging  through 
the summation over the Matsubara frequency and that through 
the other $T$-dependent part of  the interacting  $G_\sigma(i\omega_n,T)$, 
where the second argument represents the temperature dependence 
emerging through the summation over the internal Matsubara frequencies. 
Alternatively, the $T^2$ correction can also be calculated using 
 the expansion with respect to the non-interacting propagator,  
which does not have an extra $T$ dependence 
other than the one included in the discrete frequency.
In the bare-expansion formulation,  
all the temperature-dependent term of the self-energy  $\Sigma_{\sigma}(i\omega,T)$  
emerge through the summations over the Matsubara frequency. 
Therefore, the leading-correction can be calculated by taking 
the variational derivative of the self-energy  with respect 
to bare internal  $G^0_\sigma(i\omega_n)$ and then evaluating 
  the difference between the summation and the integration 
over the imaginary frequency. 
As the  variational calculation picks up a single  internal propagator 
from the self-energy diagrams  in all the possible ways,  
 the $T^2$ correction is determined by  
\begin{widetext}
\begin{align}
 \Sigma_{\sigma}(i\omega, T) 
-
\Sigma_{\sigma}(i\omega, 0) 
=& \         
\left[\,
 T \sum_{i\omega'} 
- 
\int_{-\infty}^{\infty} \! \frac{d \omega'}{2\pi}
\,\right]   
 \sum_{\sigma'}\,
\Gamma_{\sigma \sigma';\sigma' \sigma}(i\omega , i\omega'; i\omega' , i\omega) 
\,G_{\sigma}(i\omega')   
  + O(T^4) 
\nonumber \\
 =  & \ 
 \frac{(\pi  T)^2}{6}\, \Psi_{\sigma}^{}(i\omega)
 \ + \ O(T^4) 
\;,
\label{eq:T2_matsubara} 
\end{align} 
where 
\begin{align}
\Psi_{\sigma}^{}(i\omega) 
\,\equiv & \ 
\left(
 \lim_{\omega' \to 0^+}
- \lim_{\omega' \to 0^-}
 \right)
\frac{-1}{2\pi i} \, 
\frac{\partial}{\partial i\omega'} \, 
 \sum_{\sigma'}\,
\Gamma_{\sigma \sigma';\sigma' \sigma}(i\omega , i\omega'; i\omega' , i\omega) 
\,G_{\sigma'}(i\omega') 
\nonumber \\
= & \ 
 \lim_{\omega' \to 0}
\frac{\partial}{\partial i\omega'} \, 
 \sum_{\sigma'}\,
\Gamma_{\sigma \sigma';\sigma' \sigma}(i\omega , i\omega'; i\omega' , i\omega) 
\,\frac{G_{\sigma'}(i\omega'+i0^+) -G_{\sigma'}(i\omega'-i0^+)}{-2\pi i} 
 \nonumber \\
 = & \ 
 \sum_{\sigma'}\,
 \lim_{\omega' \to 0}
\frac{\partial}{\partial i\omega'} \, 
\Gamma_{\sigma \sigma';\sigma' \sigma}(i\omega , i\omega'; i\omega' , i\omega) 
\,\rho_{d\sigma'}^{}(0) 
+
 \sum_{\sigma'}\,
\Gamma_{\sigma \sigma';\sigma' \sigma}(i\omega , 0; 0, i\omega) 
\,\rho_{d\sigma'}'(0)
. 
%
\label{eq:T2_eV=0}
\end{align}
\end{widetext}
Note that at finite external frequencies  $\omega \neq 0$,  
 the limit of the internal frequency $\omega' \to 0$ 
does not depend on the directions of the approach,   $0^+$ or $0^-$,  
for both 
$\Gamma_{\sigma \sigma';\sigma' \sigma}(i\omega , i\omega'; i\omega' , i\omega) $
and 
$\partial/\partial  i \omega' \Gamma_{\sigma \sigma';\sigma' \sigma}
(i\omega , i\omega'; i\omega' , i\omega) $.  
Therefore, the discontinuity along  $\omega'=0$ emerges 
only through  $G_{\sigma'}(i\omega')$.

The $T^2$ correction can also be calculated using the corresponding causal function 
 $\Psi_{\sigma}^{--}(\omega)$, which   
can be obtained at $T=0$ via an analytic continuation of  $\Psi_{\sigma}^{}(w)$  
to the real axis   
 $w = \omega +i  0^+ \mathrm{sgn} (\omega)$, 
\begin{align}
\Psi_{\sigma}^{--}(\omega) 
 =   
 \lim_{\omega' \to 0}
 \frac{\partial}{\partial \omega'}\,  
 \sum_{\sigma'}  
 \Gamma_{\sigma\sigma';\sigma'\sigma}(\omega,\omega';\omega',\omega)\,
 \rho_{d\sigma'}^{}(\omega') 
.
\label{eq:d2_T0}
\end{align}
In the zero-frequency limit, the causal and  Matsubara  
take the same values 
 $\lim_{\omega \to 0} \Psi_{\sigma}^{--}(\omega)  
=  \lim_{\omega \to 0} \Psi_{\sigma}^{}(i\omega)$. 
This function $\Psi_{\sigma}^{--}(\omega)$  also plays an important role 
in  a non-equilibrium steady state driven by 
a bias voltage $eV$.
\footnote{
This function is shown to be identical to the correlation function  
that determines  the  $(eV)^2$  correction of the self-energy 
defined in Ref.\ \onlinecite{ao2001PRB}: 
 $
\protect \widehat{D}^2 
 \Sigma_{\mathrm{eq},\sigma}^{--}(\omega) 
\equiv \Psi_{\sigma}^{--}(\omega)
$ [see Ref.\ \onlinecite{ao2017_3_PRB} for details]. 
}

\subsection{Calculation of 
  $\left.\Psi_{\sigma}^{}(i \omega)\right|_{\omega \to 0}^{}$ }

We show in the following that  
the coefficient for  the $T^2$ correction of the 
self-energy can be expressed in terms of  $\chi_{\uparrow\downarrow}$ and 
its derivative with respect to $ \epsilon_{d,-\sigma}^{}$,  as  
\begin{align}
 \lim_{\omega \to 0} 
\Psi_{\sigma}^{}(i\omega)  
\, = \, 
 \frac{1}{\rho_{d\sigma}^{}} 
\frac{\partial \chi_{\uparrow\downarrow}}{\partial \epsilon_{d,-\sigma}^{}} 
 - i 3\pi \, \frac{\chi_{\uparrow\downarrow}^2}{\rho_{d\sigma}^{}}
\, \mbox{sgn}(\omega) 
.
\label{eq:Psi_result}
\end{align}

Taking the limit of Eq.\ \eqref{eq:T2_eV=0}, 
\begin{align}
 & \lim_{\omega \to 0} \Psi_{\sigma}^{}(i\omega)  
\nonumber \\
&=  \ 
 \lim_{\omega \to 0}
 \lim_{\omega' \to 0}
 \sum_{\sigma'}\,
\frac{\partial}{\partial i\omega'} \, 
\Gamma_{\sigma \sigma';\sigma' \sigma}(i\omega , i\omega'; i\omega' , i\omega) 
\,\rho_{d\sigma'}^{}
\nonumber \\
&\quad  +  \,
\Gamma_{\sigma,-\sigma;-\sigma,\sigma}(0,0;0,0)
\,
\rho_{d,-\sigma}'.
\label{eq:T2_correction_new_calc}
\end{align}
The derivative of 
$\Gamma_{\sigma \sigma';\sigma' \sigma}(i\omega , i\omega'; i\omega' , i\omega) $
 in the right-hand side can be calculated using the asymptotic form  
given in Eqs.\ \eqref{eq:GammaUU_general_omega_dash_abs} and 
\eqref{eq:GammaUD_general_matsubara_omega_dash_abs}. 
We can also use the result of  the derivative of the non-analytic parts 
with respect to $i\omega'$ given in 
Eqs.\ \eqref{eq:Ward_2_Im_diagram_1} and \eqref{eq:Ward_2_Im_diagram_2}. 
Separating  the analytic and non-analytic parts of the derivative of the 
vertex function,  we obtain
\begin{align}
& \lim_{\omega \to 0} \Psi_{\sigma}^{}(i\omega)  
\nonumber \\
& =   
\frac{\rho_{d,-\sigma}^{}}{\rho_{d\sigma}^{}}
 \frac{\partial \widetilde{\chi}_{-\sigma,\sigma}}{\partial \epsilon_{d,-\sigma}^{}}
 \,-\, 
 \frac{\rho_{d,-\sigma}'}{\rho_{d,-\sigma}^{}}
 \frac{\chi_{\uparrow\downarrow}
 }{\rho_{d\sigma}^{}}\, 
\nonumber \\
& \qquad   
  - \, 
 i \pi \,
\frac{\chi_{\uparrow\downarrow}^2}{\rho_{d\sigma}^{}}
\lim_{\omega \to 0}
\lim_{\omega' \to 0}
\Bigl[\, 
2\,\mathrm{sgn} (\omega-\omega')  
\,+\,\mathrm{sgn} (\omega+\omega')  
\,\Bigr]
\nonumber \\
&  =  \ 
 \frac{1}{\rho_{d\sigma}^{}} 
\frac{\partial \chi_{\uparrow\downarrow}}{\partial \epsilon_{d,-\sigma}^{}} 
 - i 3\pi \, \frac{\chi_{\uparrow\downarrow}^2}{\rho_{d\sigma}^{}}
\, \mbox{sgn}(\omega) 
.
\label{eq:T2_correction_new_calc_2}
\end{align}
In order to rewrite  the real part in the above form,
we have used Eqs.\ \eqref{eq:chitilde_chi}, \eqref{eq:Dren_to_Dsus_org2} and
 \eqref{eq:rho_d_ed}.


\section{Low-frequency expansion for a  particle-hole pair excitation}
\label{sec:green_function_product_expansion}

In this section we describe a perturbative approach to directly calculate 
the $\omega$-linear contribution of the vertex function 
$\Gamma_{\sigma \sigma;\sigma \sigma}(i\omega , 0; 0 , i\omega)$ 
for the parallel spins. 
We  calculate  the {\it regular\/} part which does not 
accompany the non-analytic  $\mathrm{sgn}\, \omega$ dependence, 
 mentioned in the above. 
To this end, we use a Green's function's product expansion 
for one intermediate particle-hole pair excitation 
shown in Fig.\  \ref{fig:vertex_singular_su2} (a), 
and then introduce a differential operator $\widehat{\partial}_{i\omega}^{+}$,  
which can extract the  {\it regular\/} component of the  $\omega$-linear part.

\subsection{Green's function product expansion}
\label{subsec:greens_function_product}

 The Green's function has a discontinuity  at $\omega=0$, and thus 
one needs an extra care for taking a derivative.  
The following are some differential formulas for 
the full Green's function which 
we will use later,
\begin{align}
&\frac{\partial G_\sigma(i\omega)}{\partial \epsilon_{d\sigma'}}  
= 
 \left\{G_\sigma(i\omega)\right\}^2 
\left[\delta_{\sigma\sigma'}  +
\frac{\partial  \Sigma_\sigma(i\omega) }{\partial \epsilon_{d\sigma'}} 
\right] , 
\label{eq:G_full_ed}
\\
&\frac{\partial G_\sigma(i\omega)}{\partial i \omega}  
=  
 -\left\{G_\sigma(i\omega)\right\}^2 
\left[1-  \frac{\partial  \Sigma_\sigma(i\omega) }{\partial i \omega} 
\right]
- 2 \pi \rho_{d\sigma}^{}  \, \delta(\omega) 
, 
\label{eq:G_full_omega}
\\
&\frac{\partial \left\{ G_\sigma(i\omega) \right\}^2}{\partial i \omega}  
= 
 -2\left\{G_\sigma(i\omega)\right\}^3 
\left[1-  \frac{\partial  \Sigma_\sigma(i\omega) }{\partial i \omega} 
\right]
\nonumber \\
& \qquad \qquad \qquad  
- 2 \pi \rho_{d\sigma}^{}  \,\Bigl[
G_\sigma(i0^+)+G_\sigma(-i0^+)
\Bigr] \delta(\omega) . 
\label{eq:G2_full_deff}
\end{align}
Furthermore, a product of full Green's functions, 
which correspond  to one particle-hole-pair carrying the parallel spin $\sigma$, 
can be expanded for a small relative frequency  $\omega \to 0$,  
\begin{align}
& \lim_{\omega \to \pm 0}
 G_\sigma(i\varepsilon)\, 
\frac{\partial G_\sigma(i\varepsilon+i\omega) }{\partial i \omega}  
\nonumber \\
& =  
 -\left\{G_\sigma(i\varepsilon)\right\}^3 
\left[1-  \frac{\partial  \Sigma_\sigma(i\varepsilon) }
{\partial i \varepsilon} 
\right]
-2 \pi \rho_{d\sigma}^{} \,  G_\sigma(\mp i0^+)\,\delta(\varepsilon) 
\nonumber \\
&  =   
\frac{1}{2}
\,\frac{\partial \left\{ G_\sigma(i\varepsilon) \right\}^2}{\partial i \varepsilon}  
- i \,2
 \, \left(\pi \rho_{d\sigma}^{}\right)^2 \delta(\varepsilon) 
\, \mathrm{sgn}(\omega) 
\;. 
\label{eq:discont_full}
\end{align}
We have used  Eq.\ \eqref{eq:G2_full_deff} to obtain the last line. 
The second term in the last line gives an  imaginary part which corresponds 
to  the $\omega$-linear part of the particle-hole hole 
pair propagator.\cite{YamadaYosida4,ShibaKorringa}
To our knowledge, however, 
the first term in the right-hand has not been paid much attention so far 
while it plays an central role for the main result of the present paper.
We refer to this first term as {\it regular\/} part.  
An interesting observation of Eq.\ \eqref{eq:discont_full} 
is  that this {\it regular\/}  part in the right-hand  
looks as if a function that is obtained from the left-hand side 
using  a naive  chain rule. 
The simplification occurs only for 
this type the  particle-hole pair carrying the parallel spins,  
corresponding to the intermediate state shown   
in Fig.\  \ref{fig:vertex_singular_su2} (a).  
We note that, for the anti-parallel spin component of the vertex function 
 $\Gamma_{\sigma,-\sigma;-\sigma,\sigma}(i\omega,0;0,i\omega)$, 
 different types of  intermediate pairs, as the ones described 
in Figs.\ \ref{fig:vertex_singular_su2} (b) and (c) for $\sigma'=-\sigma$ 
emerge. Their contributions at low-frequencies are 
determined by the  Green's-function  products of the form,  
\begin{align}
&\lim_{\omega \to \pm 0}
 G_{-\sigma}(i\varepsilon)\, 
\frac{\partial G_{\sigma}(i\varepsilon+i\omega) }{\partial i \omega}  
\;,  \\
&\lim_{\omega \to \pm 0}
 G_{\sigma''}(i\varepsilon)\, 
\frac{\partial G_{\sigma}(i\omega-i\varepsilon) }{\partial i \omega}  \;.
\label{eq:discont_full_other}
\end{align}
However, our main purpose here  is to calculate the $\omega$-linear part 
of the vertex correction $\Gamma_{\sigma\sigma;\sigma\sigma}(i\omega,0;0,i\omega)$, and  these pairs do not contribute to  this part.

\subsection{
Operator formulation for the next-leading correction 
}
\label{subsec:chain_rule_partial+}

In the perturbation expansion for the vertex function, 
one needs to treat such a function as $F(i\omega)$  
that is continuous at $\omega=0$  but its first derivative jumps;  
namely, $F(i0^+)=F(-i0^+)$ and 
\begin{align}
F'(i 0^+) \neq  F'(-i0^+), 
\qquad 
F'(i \omega) \equiv \frac{\partial F(i\omega) }{\partial i \omega}. 
\end{align}
We introduce the following two operators for the derivative with respect to $i\omega$:  
\begin{align}
\widehat{\partial}_{i\omega}^{\pm}\,  F(i\omega) 
\,\equiv & \ 
\frac{1}{2}
\left(\lim_{\omega \to  0^+} \pm \lim_{\omega \to  0^-} \right)
\frac{\partial F(i\omega) }{\partial i \omega}  
\nonumber \\ 
= & \  
\frac{1}{2} \Bigl[\,  F'(i0^+) \pm  F'(-i0^+) \,\Bigr] .
\end{align}
The operator $\widehat{\partial}_{i\omega}^{+}$ extracts 
the {\it regular\/} part of $F'(i\omega)$  while 
 $\widehat{\partial}_{i\omega}^{-}$ gives 
the discontinuous  $\mathrm{sgn}\,(\omega)$ part.
For a  function that is continuous at $\omega =0$, 
the operator   $\widehat{\partial}_{i\omega}^{+}$  gives 
 the usual differential coefficient,
\begin{align}
\widehat{\partial}_{i\omega}^{+}  \Bigl[G_\sigma(i\varepsilon+i\omega) 
\Bigr]\, 
= & \ 
\frac{1}{2}
\left(\lim_{\omega \to  0^+} + \lim_{\omega \to  0^-} \right) 
\frac{\partial G_\sigma(i\varepsilon+i\omega) }{\partial i \omega} 
\nonumber \\
= & \ 
\,\frac{\partial G_\sigma(i\varepsilon)}{\partial i \varepsilon}  ,
\\
\nonumber \\
\widehat{\partial}_{i\omega}^{-}  \Bigl[G_\sigma(i\varepsilon+i\omega) 
\Bigr]\, 
= & \ 
\frac{1}{2}
\left(\lim_{\omega \to  0^+} - \lim_{\omega \to  0^-} \right) 
\frac{\partial G_\sigma(i\varepsilon+i\omega) }{\partial i \omega} 
\nonumber \\
= & \ 0 ,
\end{align}
for $\varepsilon \neq 0$.
The Green's-function product  defined  in  Eq.\ \eqref{eq:discont_full} 
includes the discontinuous  $\mathrm{sgn}(\omega)$ term, and thus  
\begin{align}
\widehat{\partial}_{i\omega}^{+} 
\Bigl[\,  G_\sigma(i\varepsilon)\, 
G_\sigma(i\varepsilon+i\omega) \,\Bigr]
\,=& \  
\frac{1}{2}
\,\frac{\partial \left\{ G_\sigma(i\varepsilon) \right\}^2}
{\partial i \varepsilon}  ,
\label{eq:d+_for_ph_pair}
\\
\widehat{\partial}_{i\omega}^{-} 
\Bigl[\,  G_\sigma(i\varepsilon)\, 
G_\sigma(i\varepsilon+i\omega) \,\Bigr] \,=& \  
- i  2  \left(\pi \rho_{d\sigma}\right)^2 \delta(\varepsilon) \;.
\label{eq:d-_for_ph_pair}
\end{align}
Note that Eq.\ \eqref{eq:d+_for_ph_pair} 
can be rewritten as a differential rule for 
  $\widehat{\partial}_{i\omega}^{+}$,  
\begin{align}
\widehat{\partial}_{i\omega}^{+}
\Bigl
[\,
 G_{\sigma}^{}(i\varepsilon)\,G_{\sigma}^{}(i\varepsilon+i\omega)
\,\Bigr] \,
=  \,  
\frac{1}{2}\, 
\widehat{\partial}_{i\omega}^{+} \Bigl[\,
\left\{G_{\sigma}^{}(i\varepsilon+i\omega)\right\}^2 
\Bigr]
\;. 
\label{eq:d+_for_ph_pair2}
\end{align}
We refer to this as a  {\it generalized\/}  chain rule  
for $\widehat{\partial}_{i\omega}^{+}$ 
because for a continuous function $P(i\varepsilon)$   
 it becomes equivalent to the usual chain rule 
$P(i\varepsilon) \frac{\partial P (i\varepsilon)}{\partial i\varepsilon} 
= \frac{1}{2} \frac{\partial}{\partial i\varepsilon} \left\{P (i\varepsilon)\right\}^2$.

The frequency $\varepsilon$ in these examples  
appears in Feynman diagrams  as  an internal  frequency 
which will be  integrated out. 
 For example,  
the  discontinuous part of  one particle-hole bubble  can be extracted using  
 $\widehat{\partial}_{i\omega}^{-}$,  
\begin{align}
& \widehat{\partial}_{i\omega}^{-} 
\left[\, 
\int_{-\infty}^{\infty} \!\! 
d\varepsilon \  
 G_\sigma(i\varepsilon)\, 
G_\sigma(i\varepsilon+i\omega) 
\,\right]
\nonumber \\
& \quad 
=   
\int_{-\infty}^{\infty} \!\! 
d\varepsilon   
\left[ - i 2  
\left(\pi \rho_{d\sigma}\right)^2 \delta(\varepsilon)  
\right]
\ =  
- i 2 \left(\pi \rho_{d\sigma}\right)^2 .  
\end{align}
The  corresponding regular part  can  be 
calculated using $\widehat{\partial}_{i\omega}^{+}$ 
with the {\it generalized\/} chain rule,   
\begin{align}
& 
\widehat{\partial}_{i\omega}^{+} 
\left[\, 
\int_{-\infty}^{\infty} \!\! 
d\varepsilon \  
 G_\sigma(i\varepsilon)\, 
G_\sigma(i\varepsilon+i\omega) 
\,\right]
\nonumber \\
& \qquad =  \ 
\frac{1}{2} \,
\widehat{\partial}_{i\omega}^{+} \left[\,
\int_{-\infty}^{\infty} \!\! 
d\varepsilon \ 
\left\{G_{\sigma}^{}(i\varepsilon+i\omega)\right\}^2
\right] 
\nonumber \\
& \qquad =   \ 
\frac{1}{2} \,
\widehat{\partial}_{i\omega}^{+} \left[\,
\int_{-\infty}^{\infty} \!\! 
d\varepsilon' \ 
\left \{G_{\sigma}^{}(i\varepsilon')\right\}^2
\right] 
\ = \ 0 \;.
\end{align}
In the last line, 
 the internal frequency was replaced by  $\varepsilon'=\varepsilon+\omega$.

We can perturbatively calculate differentiation  $\widehat{\partial}_{i\omega}^{+}$ 
 operating upon  $\Gamma_{\sigma\sigma;\sigma\sigma}(i\omega,0;0,i\omega)$,  
taking into account the {\it generalized\/} chain rule defined 
in Eq.\ \eqref{eq:d+_for_ph_pair2}. 
For instance, an integration which includes  a function  $P(i\varepsilon+i\omega)$ 
that is continuous at $\omega=0$,  can be carried out  such that  
\begin{align}
& 
\widehat{\partial}_{i\omega}^{+} 
\left[\, 
\int_{-\infty}^{\infty} \!\! 
d\varepsilon \  
 G_\sigma(i\varepsilon)\, 
G_\sigma(i\varepsilon+i\omega) \, 
P(i\varepsilon+i\omega)\, 
\,\right]
\nonumber \\ 
& =  
\int_{-\infty}^{\infty} \!\! 
d\varepsilon \, 
\Biggl( \,
\widehat{\partial}_{i\omega}^{+} 
\Bigl[\, 
 G_\sigma(i\varepsilon)\, 
G_\sigma(i\varepsilon+i\omega) \, 
\Bigr] P(i\varepsilon)
\nonumber \\
& \qquad \qquad \qquad 
 + 
 \left\{G_\sigma(i\varepsilon)\right\}^2 
\widehat{\partial}_{i\omega}^{+} 
\Bigl[ P(i\varepsilon+i\omega) \Bigr] 
\, \Biggr) 
\nonumber \\ 
& = 
\int_{-\infty}^{\infty} \!\! 
d\varepsilon 
\left(
\frac{1}{2}
\,\frac{\partial \left\{ G_\sigma(i\varepsilon) \right\}^2}
{\partial i \varepsilon}  
P(i\varepsilon)
+ 
\left\{ G_\sigma(i\varepsilon) \right\}^2
\,\frac{\partial P(i\varepsilon )\, 
}{\partial i \varepsilon }
\right) . 
\end{align}

 We note  that the particle-particle pair excitation, 
illustrated in Fig.\ \ref{fig:vertex_singular_su2} (c) for $\sigma'=\sigma$,  
does not give an  $\omega$-linear term  
because the scattering amplitude vanishes  
 $\Gamma_{\sigma\sigma;\sigma\sigma}(0,0;0,0)=0$ 
at zero frequencies, as  mentioned.
It appears through 
 a particle-particle Green's-function product that is  associated 
with a function $Q(i\varepsilon,i\omega)$,  which vanishes  at  $\omega=0$, 
\begin{align}
& 
\!\!\!\!\!\!\!\!\!
\widehat{\partial}_{i\omega}^{+}
\Bigl[\,
 G_{\sigma}^{}(i\varepsilon)\,G_{\sigma}^{}(i\omega-i\varepsilon)\, 
Q(i\varepsilon,i\omega)
\,\Bigr] 
\nonumber \\
&= \ 
\widehat{\partial}_{i\omega}^{+}
\Bigl
[\,
 G_{\sigma}^{}(i\varepsilon)\,G_{\sigma}^{}(i\omega-i\varepsilon)\, 
\,\Bigr] \, 
Q(i\varepsilon,0) 
\nonumber \\
& \qquad 
+
 G_{\sigma}^{}(i\varepsilon)\,G_{\sigma}^{}(-i\varepsilon)\, 
\widehat{\partial}_{i\omega}^{+}
\Bigl[\,
Q(i\varepsilon,i\omega)
\,\Bigr] 
\nonumber \\
& =  \ 
 G_{\sigma}^{}(i\varepsilon)\,G_{\sigma}^{}(-i\varepsilon)\  
\widehat{\partial}_{i\omega}^{+}
\Bigl[\,
Q(i\varepsilon,i\omega)
\,\Bigr] 
 ,
\label{eq:d+_for_pp_pair_example} 
\end{align}
where  $Q(i\varepsilon,0)=0$.

\subsection{The $\omega$-linear part of 
$\Gamma_{\sigma\sigma;\sigma\sigma}(i\omega,0;0,i\omega)$}
\label{subsec:zeta_R-zeta_I}

Our strategy to calculate the $\omega$-linear part is as follows.
For small  $\omega$, 
 the vertex correction for the parallel spins  can be expanded  in the form  
\begin{align}
\Gamma_{\sigma\sigma;\sigma\sigma}(i\omega , 0; 0 , i\omega) 
 \, =  
\Bigl[\,
\zeta_{\sigma}^{R}\, - \,i\, \zeta_{\sigma}^{I} \, \mathrm{sgn} (\omega) 
 \Bigr]\, i\omega
+  O(\omega^2) .
\label{eq:zeta_coeff}
\end{align}
The coefficients $\zeta_{\sigma}^{R} $ and $\zeta_{\sigma}^{I}$ 
correspond to the real and imaginary parts, respectively, 
of the function which is obtained through the analytic continuation  
 $i\omega \to \omega + i0^+$ in the upper-half complex plain.
These coefficients can be extracted   
using the operators $\widehat{\partial}_{i\omega}^{\pm}$, defined in the above
\begin{align}
 \zeta_{\sigma}^{R} = & \  
\widehat{\partial}_{i\omega}^{+}\, 
\Gamma_{\sigma\sigma;\sigma\sigma}(i\omega , 0; 0 , i\omega) ,
\\
\zeta_{\sigma}^{I} =& \  
i\,  \widehat{\partial}_{i\omega}^{-}\, 
\Gamma_{\sigma\sigma;\sigma\sigma}(i\omega , 0; 0 , i\omega) .
\end{align}

The $\omega$-linear imaginary part 
arises from the Feynman diagram shown in  
Fig.\ \ref{fig:vertex_singular_su2} (a) as mentioned. 
It can be calculated immediately by using  the Green's-function product expansion 
with Eq.\ \eqref{eq:d-_for_ph_pair},  
\begin{align}
& \widehat{\partial}_{i\omega}^{-}\, 
\Gamma_{\sigma\sigma;\sigma\sigma}(i\omega , 0; 0 , i\omega) 
\nonumber \\
& =  
- \sum_{\sigma'}
\int_{-\infty}^{\infty} \frac{d \varepsilon}{2\pi}\, 
\Bigl\{\Gamma_{\sigma\sigma';\sigma'\sigma}
(0 , i\varepsilon; i \varepsilon, 0)\Bigr\}^2 
\nonumber \\  
& \qquad \qquad \qquad 
\times \, 
\widehat{\partial}_{i\omega}^{-} \Bigl[
 G_{\sigma'}^{}(i\varepsilon)
\,G_{\sigma'}(i\varepsilon+i\omega) 
\Bigr]
\nonumber \\ 
& = \  i \pi\, 
\bigl|\Gamma_{\sigma,-\sigma;-\sigma\sigma}(0 , 0; 0 , 0)\bigr|^2
\rho_{d,-\sigma}^{2}\, 
 \  = \ 
i \pi \,\widetilde{\chi}_{\sigma,-\sigma}^2 
.
\end{align}
Therefore,  $\zeta_{\sigma}^{I} =  -\pi \widetilde{\chi}_{\sigma,-\sigma}^2$, 
which reproduces the result of Yamada-Yosida, as mentioned for  
Eqs.\ \eqref{eq:vert_UU_im_w} 
and \eqref{eq:Im_(a)}.\cite{YamadaYosida4,Yoshimori,ao2001PRB}
In the rest of the present paper, 
we calculate the real part perturbatively 
in  a skeleton-diagrammatic expansion which is a resummation scheme 
using the full Green's function $G_{\sigma}(i\omega)$. 
Through the direct perturbative calculations, we show  in the following  sections 
  that  the real part identically vanishes  $\zeta_{\sigma}^{R}=0$.


\section{
Skeleton diagram expansion for 
$\Gamma_{\sigma\sigma;\sigma\sigma}(i\omega, 0; 0, i\omega)$ 
}
\label{sec:skeleton_diagram_expansion_for_vertex}

In this section, 
we  perturbatively calculate 
the $\omega$-linear analytic part of 
$\Gamma_{\sigma\sigma;\sigma\sigma}(i\omega, 0; 0, i\omega)$ 
in order to clarify how the cancellation, 
which causes  $\zeta_{\sigma}^{R}=0$, occurs. 
As mentioned in Sec.\ \ref{sec:vertex_omega_omega_dash},
the antisymmetry property of the vertex function plays 
an important role in the Fermi-liquid properties, 
especially  for the parallel-spin component
\begin{align}
&
\Gamma_{\sigma\sigma;\sigma\sigma}
(i\omega_1, i\omega_2; i\omega_3, i\omega_4) 
\,=\,  
-\Gamma_{\sigma\sigma;\sigma\sigma}
(i\omega_3, i\omega_2; i\omega_1, i\omega_4)  
\nonumber \\
& 
=   
\Gamma_{\sigma\sigma;\sigma\sigma}
(i\omega_3, i\omega_4; i\omega_1, i\omega_2) 
=
-\Gamma_{\sigma\sigma;\sigma\sigma}
(i\omega_1, i\omega_4; i\omega_3, i\omega_2) . 
\label{eq:vertex_crossing_symmetry}
\end{align}
We diagrammatically demonstrate in the following 
that it is essential for such cancellations to take into account together 
the diagrams which are related to each other through 
the antisymmetry property.

\subsection{Symmetry operation}
In order to clearly describe the antisymmetry property,  
we introduce the operator $\widehat{C}_\mathrm{in}^{}$ 
that  exchanges the two frequencies, which enter into the vertex part, 
and   $\widehat{C}_\mathrm{out}^{}$ 
that exchanges the two frequencies  getting out: 
\begin{align}
&
\!\!\!\!
\widehat{C}_\mathrm{in}^{}\, 
\Gamma_{\sigma\sigma;\sigma\sigma}
(i\omega_1, i\omega_2; i\omega_3, i\omega_4) 
=  
\Gamma_{\sigma\sigma;\sigma\sigma}
(i\omega_3, i\omega_2; i\omega_1, i\omega_4) 
,
\\
&
\!\!\!\!
\widehat{C}_\mathrm{out}^{}
\Gamma_{\sigma\sigma;\sigma\sigma}
(i\omega_1, i\omega_2; i\omega_3, i\omega_4) 
=    
\Gamma_{\sigma\sigma;\sigma\sigma}
(i\omega_1, i\omega_4; i\omega_3, i\omega_2) 
,
\\
&
\!\!\!\!
\widehat{C}_\mathrm{in}^{}\widehat{C}_\mathrm{out}^{}
\Gamma_{\sigma\sigma;\sigma\sigma}
(i\omega_1, i\omega_2; i\omega_3, i\omega_4)
\nonumber \\
& \qquad \qquad \qquad \qquad \   
= \ 
\Gamma_{\sigma\sigma;\sigma\sigma}
(i\omega_3, i\omega_4; i\omega_1, i\omega_2) 
. 
\label{eq:Cin_Cout}
\end{align}
These  operators have the properties 
$\widehat{C}_\mathrm{in}^{2}
= \widehat{C}_\mathrm{out}^{2}=1$, 
and 
$\widehat{C}_\mathrm{in}^{}\widehat{C}_\mathrm{out}^{}
= \widehat{C}_\mathrm{out}^{}\widehat{C}_\mathrm{in}^{}$.  
The vertex function for 
the parallels spins can also be written in a form
that explicitly shows  the asymmetry property:  
\begin{align}
&
\Gamma_{\sigma\sigma;\sigma\sigma}
(i\omega_1, i\omega_2; i\omega_3, i\omega_4) 
\nonumber \\
&=  
\frac{1}{4} 
\biggl[  
\Gamma_{\sigma\sigma;\sigma\sigma}
(i\omega_1, i\omega_2; i\omega_3, i\omega_4) 
-\Gamma_{\sigma\sigma;\sigma\sigma}
(i\omega_3, i\omega_2; i\omega_1, i\omega_4) 
\nonumber \\
& \ \ + 
\Gamma_{\sigma\sigma;\sigma\sigma}
(i\omega_3, i\omega_4; i\omega_1, i\omega_2) 
- \Gamma_{\sigma\sigma;\sigma\sigma}
(i\omega_1, i\omega_4; i\omega_3, i\omega_2) 
 \biggr] . 
\label{eq:vertex_antisymetrize}
\end{align}

\begin{figure}[t]
 \leavevmode
\begin{minipage}{1\linewidth}
\includegraphics[width=0.36\linewidth]{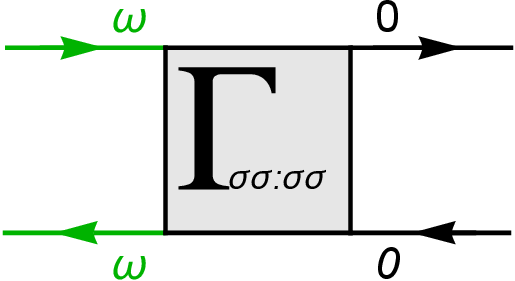}
\rule{0.06\linewidth}{0cm}
\includegraphics[width=0.36\linewidth]{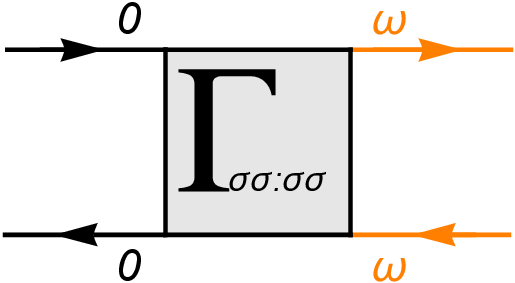}
\\
\includegraphics[width=0.36\linewidth]{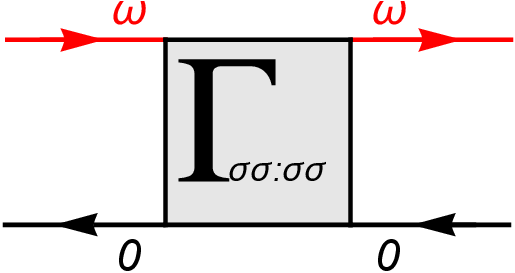}
\rule{0.06\linewidth}{0cm}
\includegraphics[width=0.36\linewidth]{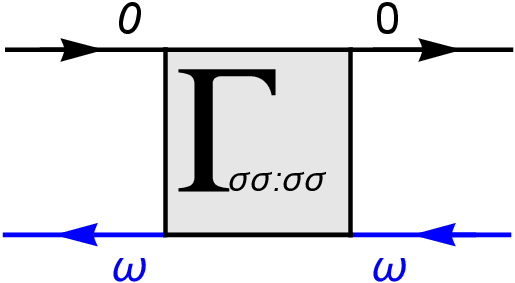}
\end{minipage}
 \caption{
(Color online) 
Four different ways that the external frequency $\omega$ 
enters and gets out of the vertex part. 
These diagrams corresponding to 
 $\Gamma_{\sigma\sigma;\sigma\sigma}(i\omega,0;0,i\omega)$,  
 $\Gamma_{\sigma\sigma;\sigma\sigma}(0,i\omega;i\omega,0)$,  
 $\Gamma_{\sigma\sigma;\sigma\sigma}(i\omega,i\omega;0,0)$, and   
 $\Gamma_{\sigma\sigma;\sigma\sigma}(0,0;i\omega,i\omega)$.  
}
 \label{fig:vertex_w00w}
\end{figure}

As the total frequencies  are conserved, 
 $ \omega_1 + \omega_3 = \omega_2 + \omega_4$, 
and three frequencies among the four are independent, 
we can choose the following  $\varepsilon$, $\varepsilon'$, and $\nu$ 
as three independent variables:  
\begin{align}
\omega_1 = \varepsilon + \nu, \quad
\omega_2 = \varepsilon' + \nu, \quad
\omega_3 = \varepsilon', \quad
\omega_4 = \varepsilon .
\end{align}
Using these three frequencies, 
we  write the vertex function in an abbreviated form:  
\begin{align}
& 
\Gamma_{\sigma\sigma}(i\varepsilon, i\varepsilon';i\nu) 
\, \equiv \, 
\Gamma_{\sigma\sigma;\sigma\sigma}
(i\varepsilon +i\nu, i\varepsilon' + i\nu;i\varepsilon', i\varepsilon ) 
\;.
\label{eq:3_frequencies_1}
\end{align}
As a function of 
$\varepsilon$, $\varepsilon'$ and  $\omega$,  
the vertex correction for interchanged frequencies, 
$\omega_2 \Leftrightarrow \omega_4$ and/or $\omega_1 \Leftrightarrow \omega_3$, can be expressed in the form, 
\begin{align}
&\widehat{C}_\mathrm{out}^{}
\Gamma_{\sigma\sigma;\sigma\sigma}
(i\varepsilon +i\nu, i\varepsilon' + i\nu;i\varepsilon', i\varepsilon ) 
\nonumber \\
& 
\qquad \qquad \qquad  \qquad 
= \ 
\Gamma_{\sigma\sigma}
(i\varepsilon'+i\nu,i\varepsilon';i\varepsilon -i\varepsilon' ),
\label{eq:3_frequencies_2}
\\
& \widehat{C}_\mathrm{in}^{}\,
\Gamma_{\sigma\sigma;\sigma\sigma}
(i\varepsilon +i\nu, i\varepsilon' + i\nu;i\varepsilon', i\varepsilon ) 
\nonumber \\
&
\qquad \qquad \qquad  \qquad 
=    \ 
\Gamma_{\sigma\sigma}
(i\varepsilon,i\varepsilon+i\nu ;i\varepsilon' -i\varepsilon),
\label{eq:3_frequencies_3}
\\
&
\widehat{C}_\mathrm{in}^{}
\widehat{C}_\mathrm{out}^{}\, 
\Gamma_{\sigma\sigma;\sigma\sigma}
(i\varepsilon +i\nu, i\varepsilon' + i\nu;i\varepsilon', i\varepsilon ) 
\nonumber \\
& 
\qquad \qquad \qquad  \qquad
=   \ 
\Gamma_{\sigma\sigma}
(i\varepsilon'+i\nu,i\varepsilon+i\nu; -i\nu).
\label{eq:3_frequencies_4}
\end{align}
Choosing the 
  frequencies such that  $\varepsilon'=\nu=0$ and   $\varepsilon=\omega$, 
Eq.\ \eqref{eq:vertex_antisymetrize} can be  expressed in the form 
\begin{align}
& \!\!\!\!\!
\Gamma_{\sigma\sigma;\sigma\sigma}
(i\omega, 0;0, i\omega ) 
\nonumber \\
& =  
\frac{1}{4} \, 
\biggl[ \,
\Gamma_{\sigma\sigma}(i\omega, 0;0) 
 + 
\Gamma_{\sigma\sigma}(0,i\omega; 0) 
-  
\Gamma_{\sigma\sigma}(0,0;i\omega) 
\nonumber \\
 & \qquad \ \ 
- 
\Gamma_{\sigma\sigma}
(i\omega,i\omega ; -i\omega) 
\, \biggr]
 . 
\label{eq:key1}
\end{align}
The assignment of the frequency $\omega$ for each term is 
indicated in  Fig.\ \ref{fig:vertex_w00w}.

\subsubsection{Total derivative with respect to  $\, \widehat{\partial}_{i\omega}^{+}$
}
\label{subsec:total_derivative}

The derivative  
$\widehat{\partial}_{i\omega}^{+}\Gamma_{\sigma\sigma;\sigma\sigma}(i\omega, 0;0, i\omega )$ can be carried out using the {\it generalized\/}  chain rule  
which is quite similar to the usual chain rule for differentiation 
as described in Eq.\ \eqref{eq:d+_for_ph_pair2}. 
If  the total derivative can be defined for 
$\widehat{\partial}_{i\omega}^{+}$ such that
\begin{align}
& \!\!\!\!
\widehat{\partial}_{i\omega}^{+}\,  
\Gamma_{\sigma\sigma} (i\omega,i\omega ; -i\omega) 
\nonumber \\
=& \ 
\widehat{\partial}_{i\omega}^{+}\,  
\Gamma_{\sigma\sigma} (i\omega, 0; 0) 
+
\widehat{\partial}_{i\omega}^{+}\,  
\Gamma_{\sigma\sigma} (0,i\omega; 0) 
-
\widehat{\partial}_{i\omega}^{+}\,  
\Gamma_{\sigma\sigma} (0, 0; i\omega) ,
\label{eq:key1_dif_org}
\end{align}
then the result  $\zeta_{\sigma}^{R}=0$ will follow 
from Eq.\ \eqref{eq:key1} as   
\begin{align}
&
\widehat{\partial}_{i\omega}^{+}\,  
\Gamma_{\sigma\sigma;\sigma\sigma}
(i\omega, 0;0, i\omega ) 
\nonumber \\
&=\, \frac{1}{4} \,
\widehat{\partial}_{i\omega}^{+}  
\biggl[ \, 
\Gamma_{\sigma\sigma}(i\omega, 0;0) 
 + 
\Gamma_{\sigma\sigma}(0,i\omega; 0) 
 -  
\Gamma_{\sigma\sigma}(0,0;i\omega) 
\nonumber \\
& \qquad \qquad 
 -  
\Gamma_{\sigma\sigma}
(i\omega,i\omega ; -i\omega) 
\biggr] \ = \,0
 . 
\label{eq:key1_dif}
\end{align}
This observation can be regarded  as another interpretation of 
 the property of the analytic part 
of  $\Gamma_{\sigma\sigma;\sigma\sigma}(i\omega, 0;0, i\omega )$ 
 discussed in Sec.\ \ref{sec:vertex_omega_omega_dash}, i.e.,   
 $a_{\sigma\sigma}^{}=0$,  which follows from the fact that 
an anti-symmetric function, which satisfies Eq.\ \eqref{eq:vertex_crossing_symmetry}, 
cannot be constructed by a homogeneous polynomial of a linear form 
as shown in Appendix \ref{sec:homogeneous polynomial}. 
We  carry out perturbative calculations up to order $U^4$ 
below  to show that  $\zeta_{\sigma}^{R}=0$.

\subsection{Anti-symmetrized skeleton diagram expansion}

We perturbatively calculate  the  regular part 
of  $\omega$-linear contribution, 
$\zeta_{\sigma}^{R}$ defined in Eq.\ \eqref{eq:zeta_coeff},
operating  $\widehat{\partial}_{i\omega}^{+}$  
upon $\Gamma_{\sigma\sigma;\sigma\sigma}^{}(i\omega,0;0,i\omega)$, 
and show diagrammatically how  the cancellations 
that results in  $\zeta_{\sigma}^{R}=0$ occur. 
In order to carry out the calculations in a fully anti-symmetrized way,   
a standard  Bethe-Salpeter type resummation, 
in which the full vertex function is decomposed into 
the irreducible part and the iterative ladders of particle-hole-pair propagators,  is not useful.   
We calculate together the contributions of each set that consists of four related diagrams,   
 generated from one of them carrying out the symmetry operations  
 $\widehat{C}_\mathrm{in}^{}$, $\widehat{C}_\mathrm{out}^{}$, and 
 $\widehat{C}_\mathrm{in}^{}\widehat{C}_\mathrm{out}^{}$.  
We explicitly show how the cancellation occurs  
in the skeleton-diagram expansion,  
for which the solid lines represent the exact interacting 
Green's functions  $G_{\sigma}^{}$,  up to order  $U^4$.\cite{Note1}

In the following, we consider the parallel-spin 
vertex function  for $\sigma=\uparrow$ to make the equations simpler  
as the corresponding result obviously holds  for the opposite spin  $\sigma=\downarrow$. 
We choose one arbitrary diagram from the four related diagrams mentioned in the above,  
and refer to it  as {\it representative\/} of the set.
The contribution of the  {\it representative\/}   diagram alone 
 $\Gamma_{\uparrow\uparrow;\uparrow\uparrow}^{(\mathrm{rep})}  
(i\omega_1, i\omega_2;i\omega_3,i\omega_4)$ 
 is not an anti-symmetric function  but  the total contribution of the set 
$\Gamma_{\uparrow\uparrow;\uparrow\uparrow}^{(\mathrm{set})}$
acquires  the  anti-symmetry property,   
\begin{align}
&
\!\!\!\!\!\!\!\!\!\!
\Gamma_{\uparrow\uparrow;\uparrow\uparrow}^{(\mathrm{set})}
(i\omega, 0;0, i\omega ) 
\nonumber \\
=&  \ \ 
\Gamma_{\uparrow\uparrow;\uparrow\uparrow}^{(\mathrm{rep})}
(i\omega, 0;0, i\omega ) 
+\Gamma_{\uparrow\uparrow;\uparrow\uparrow}^{(\mathrm{rep})}
(0,i\omega;i\omega,0 ) 
 \nonumber \\
 & 
-\Gamma_{\uparrow\uparrow;\uparrow\uparrow}^{(\mathrm{rep})}
(i\omega,  i\omega;0,0 ) 
-\Gamma_{\uparrow\uparrow;\uparrow\uparrow}^{(\mathrm{rep})}
(0,0;i\omega, i\omega ) \;.
\label{eq:key1_set}
\end{align}

There is a class of vertex diagrams that graphically have two axes 
of the reflection symmetry: 
one in the horizontal direction and the other in the vertical direction. 
The simplest example  is the  order $U^2$ 
diagrams shown in Fig.\ \ref{fig:vertex_u2_up_up}.
Such sets  with an additional symmetry consist of  two independent diagrams. 
Thus, for such sets, we multiply an extra factor $1/2$ 
to the right-hand side of Eq.\ \eqref{eq:key1_set} 
for compensating the double counting  of the two identical diagrams. 
We examine the contributions of  
the first few diagrams in the skeleton-diagram expansion below.


\begin{figure}[t]
 \leavevmode
\begin{minipage}{1\linewidth}
\includegraphics[width=0.47\linewidth]{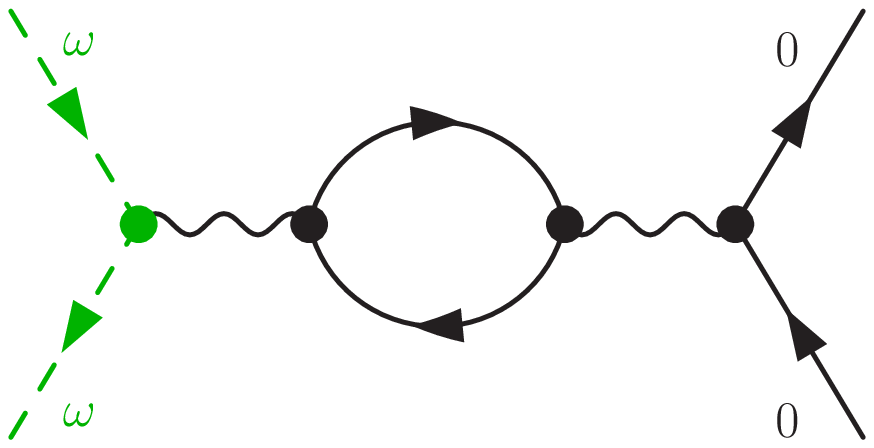}
\rule{0.02\linewidth}{0cm}
\includegraphics[width=0.47\linewidth]{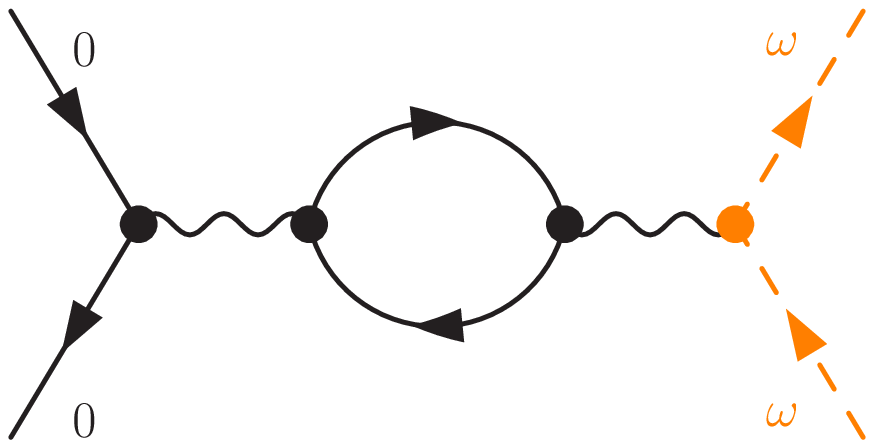}
\end{minipage}

 \rule{0cm}{0.5cm}

\begin{minipage}{1\linewidth}
\includegraphics[width=0.47\linewidth]{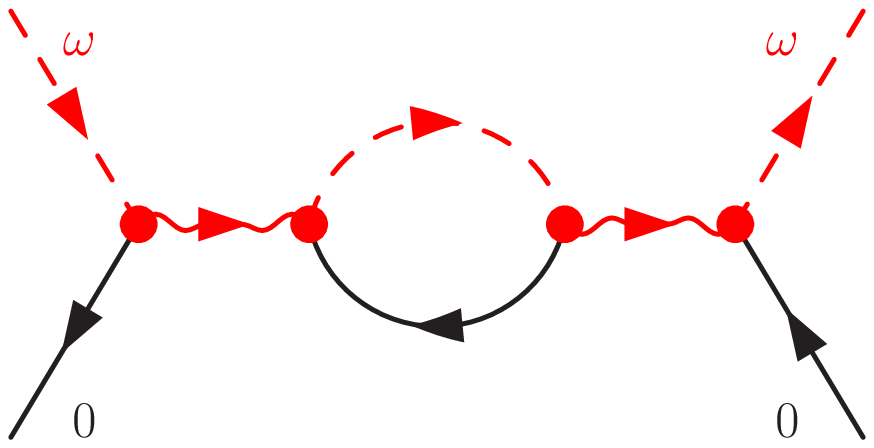}
\rule{0.02\linewidth}{0cm}
\includegraphics[width=0.47\linewidth]{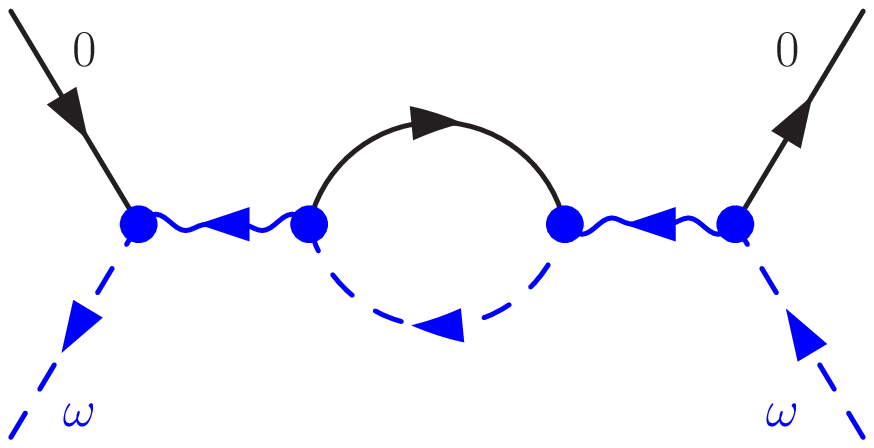}
\end{minipage}

 \caption{
(Color online) 
A set of four diagrams for $\Gamma_{\uparrow\uparrow;\uparrow\uparrow}^{(2)}$ 
generated from 
the first one in the upper left panel by
 operating 
$\widehat{C}_\mathrm{in}^{}\widehat{C}_\mathrm{out}^{}$, 
$\widehat{C}_\mathrm{out}^{}$, and 
$\widehat{C}_\mathrm{in}^{}$. 
The dashed line represents the propagator that carries the external frequency $\omega$. 
The wavy lines, which carry $\omega$, are  
shown with the arrow,  indicating the direction $\omega$ flows. 
In this case,  the four diagrams are not  independent 
because the diagram has two different axes  of the reflection symmetry. }
 \label{fig:vertex_w00w_order_u2_color_each}
\end{figure}

\begin{figure}[t]
 \leavevmode
\begin{minipage}{1\linewidth}
\includegraphics[width=0.55\linewidth]{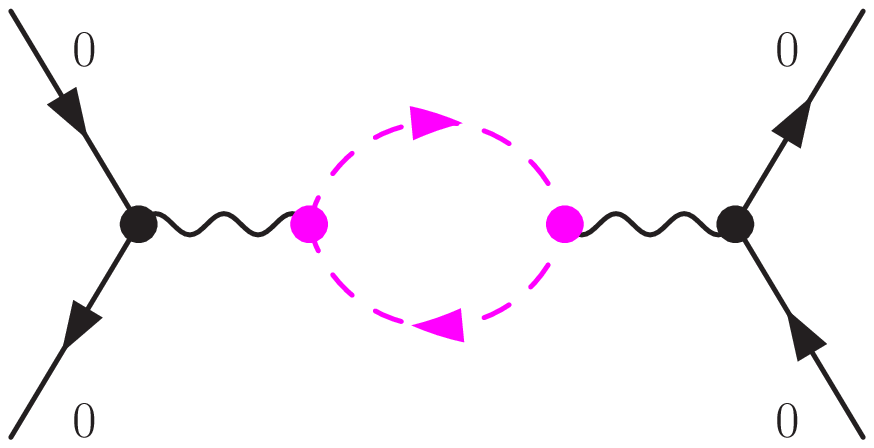}
\end{minipage}
 \caption{
(Color online) 
Schematic picture expressing 
the total  contribution   
$ \Gamma_{\uparrow\uparrow;\uparrow\uparrow}^{(2)}$ 
of the diagrams 
shown in Fig.\ \ref{fig:vertex_w00w_order_u2_color_each}. 
The dashed propagators 
carrying the external frequency  $\omega$ 
form a closed loop. 
}
 \label{fig:vertex_w00w_order_u2_color_sum}
\end{figure}

\subsubsection{Order $U^2$ contributions}



The diagrams for order $U^2$ skeleton diagrams for the  parallel spins vertex function 
are described in Fig.\ \ref{fig:vertex_u2_up_up}, 
and their contribution can be expressed in the form 
\begin{align}
&\Gamma_{\uparrow\uparrow;\uparrow\uparrow}^{(2)}
  (i\omega, 0; 0,i\omega) 
\,=\,  U^2 \Bigl[\, 
\chi_{\downarrow\downarrow}^{qp}(i\omega) - 
\chi_{\downarrow\downarrow}^{qp}(0) \,\Bigr]  
\;, 
\\
&\chi_{\sigma\sigma}^{qp}(i\omega) 
\,\equiv\,   
- \int_{-\infty}^{\infty} 
\! \frac{d\varepsilon}{2\pi} \, 
G_{\sigma}^{}(i\varepsilon+i\omega)\, 
G_{\sigma}^{}(i\varepsilon)\,.
%
\end{align}
We can rewrite this  equation  
in the form of  Eq.\ \eqref{eq:key1_set}  
to make  the anti-symmetry property explicit,
\begin{align}
&
\!\!\!\!\!\!\!\!
\widehat{\partial}_{i\omega}^{+}\, 
\Gamma_{\uparrow\uparrow;\uparrow\uparrow}^{(2)}
  (i\omega, 0; 0,i\omega) 
\nonumber \\
= & \ 
\frac{U^2}{2}
 \!
\int_{-\infty}^{\infty} \frac{d \varepsilon}{2\pi}\ 
\widehat{\partial}_{i\omega}^{+}
\Bigl
[\,
\left\{ G_{\downarrow}^{}(i\varepsilon)\right\}^2
+  
\,\left\{ G_{\downarrow}^{}(i\varepsilon)\right\}^2
\nonumber \\ 
&  \qquad \quad 
-  G_{\downarrow}^{}(i\varepsilon)\,
{\color{red}
G_{\downarrow}^{}(i\varepsilon+i\omega)
}
-  
{\color{blue}
G_{\downarrow}^{}(i\varepsilon+i\omega)\,
}
G_{\downarrow}^{}(i\varepsilon)
\,\Bigr] 
 \nonumber 
 \\
=& \ 
-\frac{U^2}{2}\,\
{\color[rgb]{1,0,1}
\widehat{\partial}_{i\omega}^{+} 
\left[\,
\int_{-\infty}^{\infty} \frac{d \varepsilon}{2\pi}\, 
\left\{G_{\downarrow}^{}(i\varepsilon+i\omega)\right\}^2
\right] 
}
\, = \, 0 . 
\label{eq:cancellation_order_u2}
\end{align}
The first and second terms of the integrand 
represent  the contributions of the diagrams shown  
in the upper panel of  Fig.\  \ref{fig:vertex_w00w_order_u2_color_each};
 the {\it representative} and the one generated by 
 $\widehat{C}_\mathrm{in}^{} \widehat{C}_\mathrm{out}^{}$.  
Similarly, the third and fourth terms represents the contributions of  the diagrams
 shown in the lower panel; the ones  generated by 
$\widehat{C}_\mathrm{in}^{}$ and $ \widehat{C}_\mathrm{out}^{}$.  
The second line of Eq.\ \eqref{eq:cancellation_order_u2} is obtained 
by using  the {\it generalized\/} chain rule given in Eq.\ \eqref{eq:d+_for_ph_pair2}. 
It shows that the contribution can be written as  a total 
 $\widehat{\partial}_{i\omega}^{+}$ derivative of 
a definite integral  over the circular frequency $\varepsilon$ along the loop 
which are symbolically illustrated in Fig.\  \ref{fig:vertex_w00w_order_u2_color_sum}. 
The $\omega$ dependence disappears  after 
 carrying out  the integration over   $\varepsilon$ as 
 it can be absorbed into the loop frequency  $\varepsilon$.  
This lowest order example already captures a general feature 
of   the cancellation of the regular part of 
 $\omega$-linear term occurring in the set of four anti-symmetrized diagrams.

\subsubsection{Order $U^3$ contributions}

\begin{figure}[t]
 \leavevmode
\begin{minipage}{1\linewidth}
\includegraphics[width=0.47\linewidth]{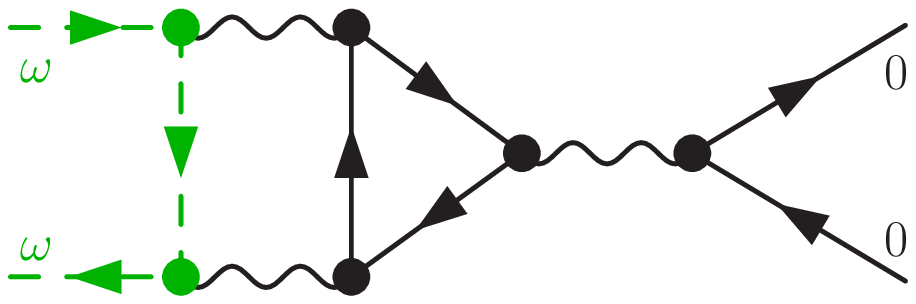}
\rule{0.02\linewidth}{0cm}
\includegraphics[width=0.47\linewidth]{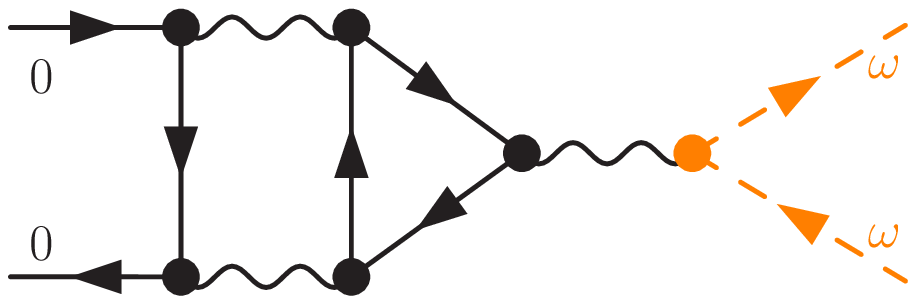}
\end{minipage}

 \rule{0cm}{0.5cm}

\begin{minipage}{1\linewidth}
\includegraphics[width=0.47\linewidth]{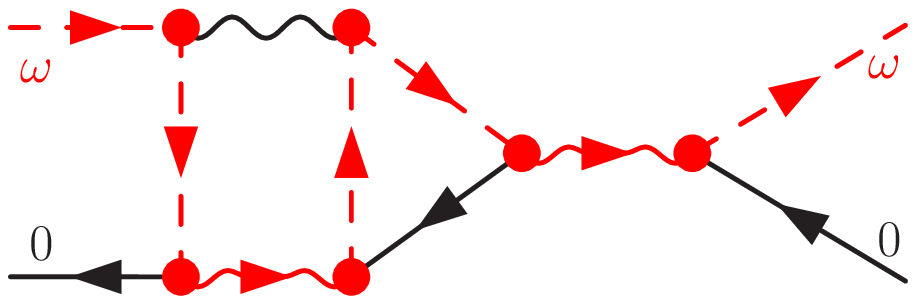}
\rule{0.02\linewidth}{0cm}
\includegraphics[width=0.47\linewidth]{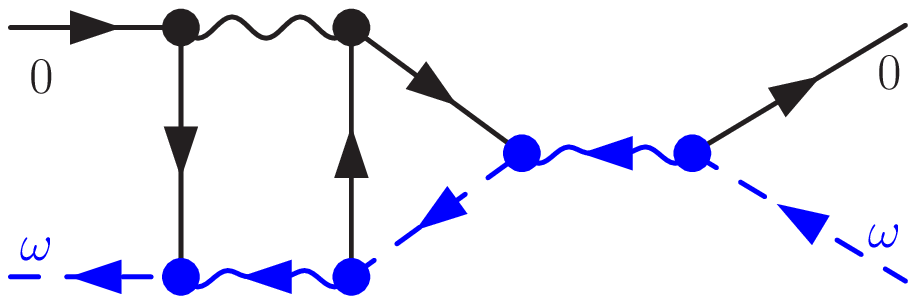}
\end{minipage}
 \caption{
(Color online) 
A set of four diagrams for 
$\Gamma_{\uparrow\uparrow;\uparrow\uparrow}^{(3A)}$ generated from 
the first one in the upper left panel by
 operating 
$\widehat{C}_\mathrm{in}^{}\widehat{C}_\mathrm{out}^{}$, 
$\widehat{C}_\mathrm{out}^{}$, and 
$\widehat{C}_\mathrm{in}^{}$. 
The dashed line represents the propagator that carries the external frequency $\omega$. 
The wavy lines, which carry $\omega$, are  
shown with the arrow,  indicating the direction $\omega$ flows. 
}
 \label{fig:vertex_w00w_order_u3ph_color_each}
\end{figure}

\begin{figure}[t]
 \leavevmode
\begin{minipage}{1\linewidth}
\includegraphics[width=0.55\linewidth]{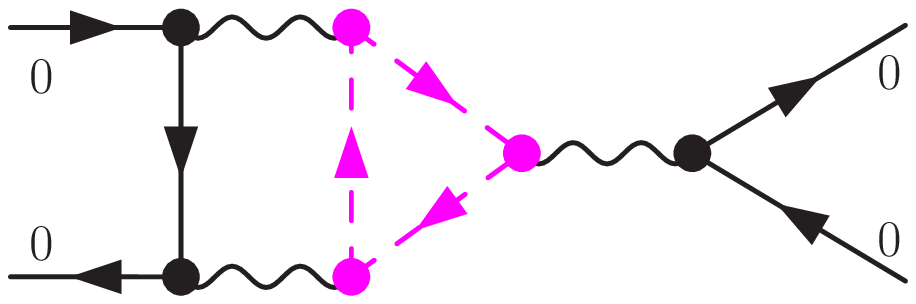}
\end{minipage}

 \caption{
(Color online) 
Schematic picture expressing 
the total  contribution  
 $\widehat{\partial}_{i\omega}^{+} 
\Gamma_{\uparrow\uparrow;\uparrow\uparrow}^{(3A)}$  
of the diagrams 
shown in Fig.\ \ref{fig:vertex_w00w_order_u3ph_color_each}. 
The dashed propagators 
carrying the external frequency  $\omega$ 
form a closed loop. 
}
\label{fig:vertex_w00w_order_u3ph_color_sum}
\end{figure}

\begin{figure}[t]
 \leavevmode
\begin{minipage}{1\linewidth}
\includegraphics[width=0.47\linewidth]{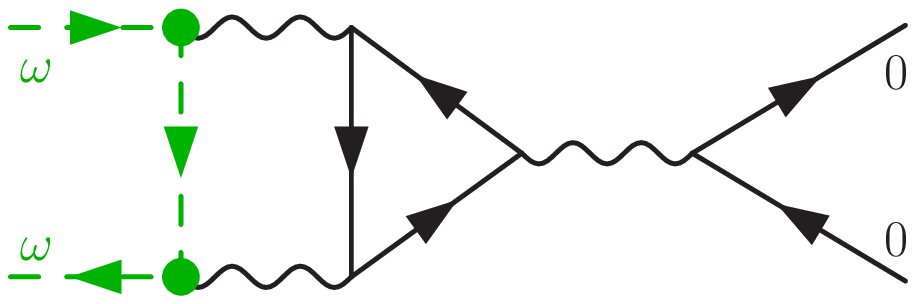}
\rule{0.02\linewidth}{0cm}
\includegraphics[width=0.47\linewidth]{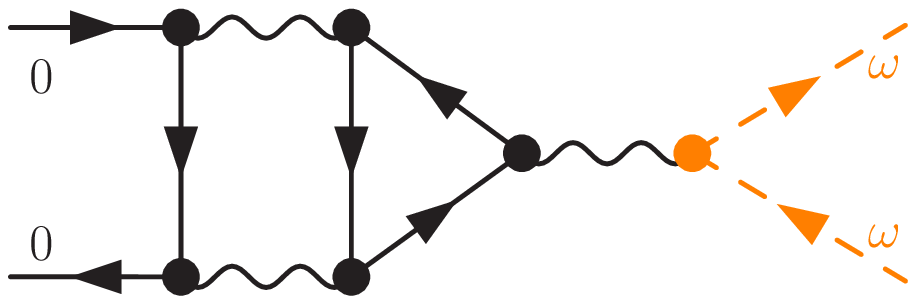}
\end{minipage}

 \rule{0cm}{0.5cm}

\begin{minipage}{1\linewidth}
\includegraphics[width=0.47\linewidth]{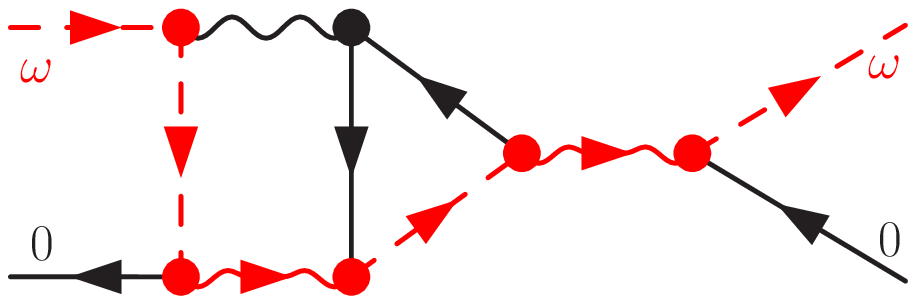}
\rule{0.02\linewidth}{0cm}
\includegraphics[width=0.47\linewidth]{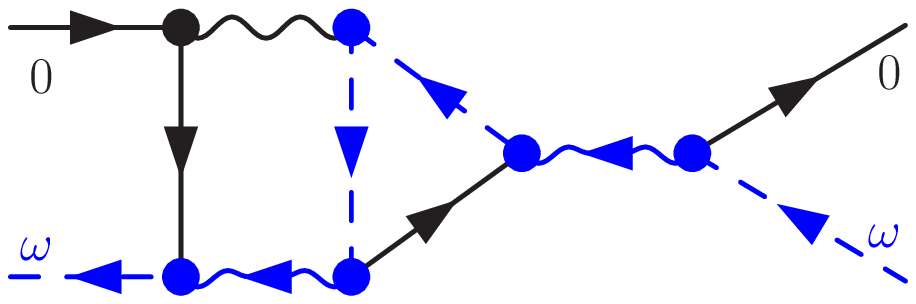}
\end{minipage}
 \caption{
(Color online) 
A set of four diagrams for 
$\Gamma_{\uparrow\uparrow;\uparrow\uparrow}^{(3B)}$ 
generated from 
the first one in the upper left panel by
 operating 
$\widehat{C}_\mathrm{in}^{}\widehat{C}_\mathrm{out}^{}$, 
$\widehat{C}_\mathrm{out}^{}$, and 
$\widehat{C}_\mathrm{in}^{}$. 
The dashed line represents the propagator 
which is assigned to carry the external frequency $\omega$. 
The wavy line, through which $\omega$ passes, 
is associated with the arrow showing 
the direction $\omega$ flows. 
}
 \label{fig:vertex_w00w_order_u3pp_color_each}
\end{figure}

There are two different sets in  order $U^3$ 
skeleton diagrams for the parallel-spin vertex function, 
as shown in 
Figs.\ \ref{fig:vertex_w00w_order_u3ph_color_each}
 and  \ref{fig:vertex_w00w_order_u3pp_color_each}.
The contribution of the set of four diagrams in 
 Fig.\ \ref{fig:vertex_w00w_order_u3ph_color_each}  
can be calculated as 
\begin{align}
 & 
\widehat{\partial}_{i\omega}^{+}
\Gamma_{\uparrow\uparrow;\uparrow\uparrow}^{(3A)}
  (i\omega, 0; 0,i\omega)  
\nonumber \\
& =  \ U^3 
\int_{-\infty}^{\infty}
\int_{-\infty}^{\infty}
\frac{d\varepsilon' d\varepsilon}{(2\pi)^2}\ 
\nonumber \\
& 
\ \times 
\widehat{\partial}_{i\omega}^{+}
 \Biggl[
{\color{red}
 G_{\uparrow}^{}(i\varepsilon' +i\omega)\, 
G_{\downarrow}^{}(i\varepsilon'+i\varepsilon+i\omega)\,
G_{\downarrow}^{}(i\varepsilon+i\omega)\, 
}
G_{\downarrow}^{}(i\varepsilon)
\nonumber \\ 
& 
\qquad 
\quad 
+
 G_{\uparrow}^{}(i\varepsilon' )\, 
G_{\downarrow}^{}(i\varepsilon'+i\varepsilon)\,
G_{\downarrow}^{}(i\varepsilon)\, 
{\color{blue}
G_{\downarrow}^{}(i\varepsilon+i\omega)\, 
}
\nonumber \\
& 
\qquad 
\quad 
- 
{\color[rgb]{0,0.6,0}
 G_{\uparrow}^{}(i\varepsilon' +i\omega)\, 
}
G_{\downarrow}^{}(i\varepsilon'+i\varepsilon)
\left\{G_{\downarrow}^{}(i\varepsilon)\right\}^2
\nonumber \\
& \qquad \quad 
-
 G_{\uparrow}^{}(i\varepsilon' )\, 
G_{\downarrow}^{}(i\varepsilon'+i\varepsilon)
\left\{G_{\downarrow}^{}(i\varepsilon)\right\}^2
\Biggr]
\nonumber \\
&= \  
U^3
\int_{-\infty}^{\infty}
\int_{-\infty}^{\infty}
\frac{d\varepsilon' d\varepsilon}{(2\pi)^2}\ 
 G_{\uparrow}^{}(i\varepsilon' )\ 
\nonumber \\
&
\ \times 
\widehat{\partial}_{i\omega}^{+}
\Biggl[\,
{\color{red}
G_{\downarrow}^{}(i\varepsilon'+i\varepsilon+i\omega)\,
G_{\downarrow}^{}(i\varepsilon+i\omega)\, 
}
G_{\downarrow}^{}(i\varepsilon)
\nonumber \\
& 
\quad  
 +
G_{\downarrow}^{}(i\varepsilon'+i\varepsilon)\,
G_{\downarrow}^{}(i\varepsilon)\, 
{\color{blue}
G_{\downarrow}^{}(i\varepsilon+i\omega)\, 
}
\nonumber \\
& 
\quad 
- 
G_{\downarrow}^{}(i\varepsilon'+i\varepsilon)
\left\{G_{\downarrow}^{}(i\varepsilon)\right\}^2
- 
G_{\downarrow}^{}(i\varepsilon'+i\varepsilon)
\left\{G_{\downarrow}^{}(i\varepsilon)\right\}^2
\, \Biggr]
\nonumber \\
& =  \ U^3
\int_{-\infty}^{\infty}
\frac{d\varepsilon'}{2\pi}\,
 G_{\uparrow}^{}(i\varepsilon' )
\nonumber \\
& \quad 
\times 
{\color[rgb]{1,0,1}
\widehat{\partial}_{i\omega}^{+}
\Biggl[\,
\int_{-\infty}^{\infty}
\frac{d\varepsilon}{2\pi}\ 
G_{\downarrow}^{}(i\varepsilon'+i\varepsilon+i\omega)\,
\left\{G_{\downarrow}^{}(i\varepsilon+i\omega)\right\}^2\, 
\Biggr] 
}
\nonumber \\
&
 = \  0 . 
\label{eq:gamU3a_skeleton}
\end{align}
To obtain the second line, 
we have carried out  the derivative with respect to $\omega$ 
assigned for the propagators along the direct line 
which  links  the two external propagators on the left side.   
It can  easily be seen that the derivative of the $\uparrow$ spin propagator 
 in the first term  and that of  the  third term cancel each other out.
Then, to obtain the next line,  
the operator $\widehat{\partial}_{i\omega}^{+}$  is applied to the 
 $\downarrow$ spin propagators 
taking into account  the {\it generalized\/} chain rule 
 given in Eq.\ \eqref{eq:d+_for_ph_pair2} 
for the particle-hole product 
$G_{\downarrow}^{}(i\varepsilon+i\omega)G_{\downarrow}^{}(i\varepsilon) $. 
In the last line, 
all the $\downarrow$-spin propagators along the closed loop, 
which is drawn with  dashed lines  
in Fig.\ \ref{fig:vertex_w00w_order_u3ph_color_sum},
capture the external frequency $\omega$ as their argument,  
and thus the $\omega$ dependence 
vanishes after carrying out the integration over 
the circular frequency $\varepsilon$  along the $\downarrow$-spin loop.
Therefore, this example also shows that 
the contribution of  the set of four diagrams  
on  the  regular part of the $\omega$-linear dependence   
can be absorbed into some internal loop frequencies and then it vanishes.

\begin{figure}[t]
 \leavevmode
\begin{minipage}{1\linewidth}
\includegraphics[width=0.55\linewidth]{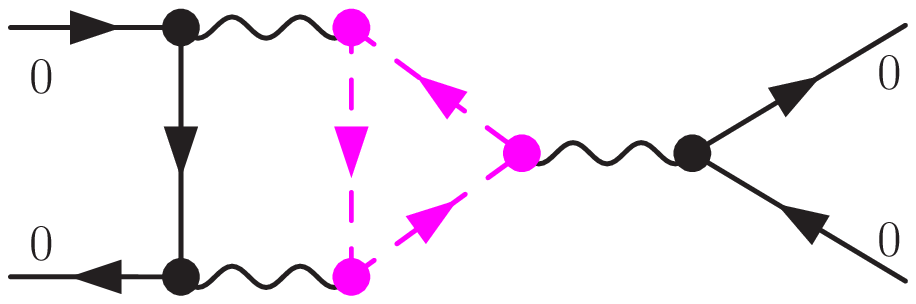}
\end{minipage}

 \caption{
(Color online) 
Schematic picture expressing 
the total  contribution  
  $\widehat{\partial}_{i\omega}^{+} 
\Gamma_{\uparrow\uparrow;\uparrow\uparrow}^{(3B)}$ 
of the diagrams 
shown in Fig.\ \ref{fig:vertex_w00w_order_u3pp_color_each}. 
The dashed propagators 
carrying the external frequency  $\omega$ 
form a closed loop.
}
 \label{fig:vertex_w00w_order_u3pp_color_sum}
\end{figure}

The other order $U^3$ set of four skeleton diagrams is  
 in Fig.\ \ref{fig:vertex_w00w_order_u3pp_color_each},
which has an  intermediate particle-particle pair in the vertical direction. 
The contribution of this set  can be calculated in a similar way 
to the case of  the particle-hole pair described in the above,    
\begin{align}
 & 
\widehat{\partial}_{i\omega}^{+}
\Gamma_{\uparrow\uparrow;\uparrow\uparrow}^{(3B)}
  (i\omega, 0; 0,i\omega)  
\nonumber \\ 
& = \ U^3 \int_{-\infty}^{\infty}
\int_{-\infty}^{\infty}
\frac{d\varepsilon' d\varepsilon}{(2\pi)^2}\ 
\nonumber \\
& 
\ \times \widehat{\partial}_{i\omega}^{+}
\Biggl[
{\color{red}
 G_{\uparrow}^{}(i\varepsilon' +i\omega)\, 
G_{\downarrow}^{}(i\varepsilon+i\omega)\,
} 
G_{\downarrow}^{}(i\varepsilon-i\varepsilon')\,
G_{\downarrow}^{}(i\varepsilon)\, 
\nonumber \\ 
& 
\qquad 
\quad 
+
 G_{\uparrow}^{}(i\varepsilon' )\, 
{\color{blue}
G_{\downarrow}^{}(i\varepsilon-i\varepsilon'+i\omega)\,
G_{\downarrow}^{}(i\varepsilon+i\omega)\, 
}
G_{\downarrow}^{}(i\varepsilon)\, 
\nonumber \\
& 
\qquad 
\quad 
- 
{\color[rgb]{0,0.6,0}
G_{\uparrow}^{}(i\varepsilon'+i\omega)\,
}
G_{\downarrow}^{}(i\varepsilon-i\varepsilon')
\left\{G_{\downarrow}^{}(i\varepsilon)\right\}^2
\nonumber \\
& \qquad \quad 
-
 G_{\uparrow}^{}(i\varepsilon' )\, 
G_{\downarrow}^{}(i\varepsilon-i\varepsilon')
\left\{G_{\downarrow}^{}(i\varepsilon)\right\}^2
\Biggr]
\nonumber \\
& = \ U^3 
\int_{-\infty}^{\infty}
\int_{-\infty}^{\infty}
\frac{d\varepsilon' d\varepsilon}{(2\pi)^2}\ 
 G_{\uparrow}^{}(i\varepsilon' )\ 
\nonumber \\
&
\ \times 
\widehat{\partial}_{i\omega}^{+}
\Biggl[\,
G_{\downarrow}^{}(i\varepsilon-i\varepsilon')\,
G_{\downarrow}^{}(i\varepsilon)\, 
{\color{red}
G_{\downarrow}^{}(i\varepsilon+i\omega)\, 
}
\nonumber \\
& 
\quad  
+
{\color{blue}
G_{\downarrow}^{}(i\varepsilon-i\varepsilon'+i\omega)\,
G_{\downarrow}^{}(i\varepsilon+i\omega)\, 
}
G_{\downarrow}^{}(i\varepsilon)\, 
\nonumber \\
& 
\quad 
- 
G_{\downarrow}^{}(i\varepsilon-i\varepsilon')
\left\{G_{\downarrow}^{}(i\varepsilon)\right\}^2
- 
G_{\downarrow}^{}(i\varepsilon-i\varepsilon')
\left\{G_{\downarrow}^{}(i\varepsilon)\right\}^2
\,\Biggr]
\nonumber \\
& = \ U^3 
\int_{-\infty}^{\infty}
\frac{d\varepsilon'}{2\pi}\ 
 G_{\uparrow}^{}(i\varepsilon' ) 
\nonumber \\
& \quad 
\times 
{\color[rgb]{1,0,1}
\widehat{\partial}_{i\omega}^{+}
\Biggl[\,
\int_{-\infty}^{\infty}
\frac{d\varepsilon}{2\pi}\ 
G_{\downarrow}^{}(i\varepsilon-i\varepsilon'+i\omega)\,
\left\{G_{\downarrow}^{}(i\varepsilon+i\omega) \right\}^2 \, 
\Biggr] 
}
\nonumber \\
& = \ 0 .
\label{eq:gamU3b_skeleton}
\end{align}
To obtain the second line, 
 the derivative with respect to $\omega$ 
assigned for the propagators along the direct line has been carried out. 
Then,   the derivative of the  $\uparrow$ spin propagators 
in the red first term  and that of  the green  third term cancel out.
The next line has been obtained 
 operating $\widehat{\partial}_{i\omega}^{+}$ upon  
the  $\downarrow$ spin propagators along the loop, 
using  the {\it generalized\/} chain rule for the particle-hole product 
$G_{\downarrow}^{}(i\varepsilon+i\omega)G_{\downarrow}^{}(i\varepsilon) $. 
The last line again shows that the contribution can be expressed 
in a total derivative with respect to the loop frequency 
as illustrated in Fig.\ \ref{fig:vertex_w00w_order_u3pp_color_sum}, and it vanishes.


\subsection{Cancellations in general cases}
\label{subsec:general_aspect}

We summarize how the cancellation which generally occurs 
for every  such set  of  four anti-symmetrized  skeleton diagrams in this subsection. 
To make the discussion clear,  
we assign the internal frequencies 
in such a way as described in the following items $i$)--$iii$), 
which has already been used in the above:

\begin{itemize}


\item[$i$)]
We choose the {\it representative\/}  
$\Gamma_{\uparrow\uparrow;\uparrow\uparrow}^{(\mathrm{rep})}
(i\omega, 0;0, i\omega )$ 
 to be  the contribution of such a diagram in which  
the external frequency $\omega$ flows  
along  a {\it direct} line  of   $\uparrow$-spin internal propagators 
towards the exit, i.e.,  we assign the frequencies along the {\it direct} line such that 
the external  $\omega$ does {\it not} flow 
into the wavy interaction lines which link  to  closed loops. 
One  example  is the diagram shown in  the upper left panel of 
 Fig.\ \ref{fig:vertex_w00w_order_u3ph_color_each},  
in which  the 
 dashed vertical  line on the left  is in  the  {\it direct} path on the left.

\item[$ii$)]
We choose the second diagram to be the one corresponding to 
$\Gamma_{\uparrow\uparrow;\uparrow\uparrow}^{(\mathrm{rep})}
(0,i\omega;i\omega,0 )$ which can be generated  by the symmetry operation 
 $\widehat{C}_\mathrm{in}^{}\widehat{C}_\mathrm{out}^{}\Gamma_{\uparrow\uparrow;\uparrow\uparrow}^{(\mathrm{rep})}
(i\omega, 0;0, i\omega )$.
In the diagram of this category,  
the external $\omega$ flows through the other  {\it direct} $\uparrow$  
spin path.  
An example of this type is shown in 
the upper right panel of Fig.\ \ref{fig:vertex_w00w_order_u3ph_color_each}, 
in which  the 
dashed  lines  on the right correspond to the direct path of this category. 

\item[$iii$)]
The remaining two terms   
$\Gamma_{\uparrow\uparrow;\uparrow\uparrow}^{(\mathrm{rep})}
(i\omega, i\omega;0,0)$ and 
$\Gamma_{\uparrow\uparrow;\uparrow\uparrow}^{(\mathrm{rep})}
(0,0;i\omega,i\omega)$ 
are generated 
 by operating  $\widehat{C}_\mathrm{out}^{}$ 
and $\widehat{C}_\mathrm{in}^{}$ upon the {\it representative\/}. 
In the diagrams of this two categories, 
the external frequency $\omega$ 
  enters  into  the vertex part on the one side and gets out from the other side,   
passing through interaction lines and closed loops  in the central region.  
An example that is derived from $\widehat{C}_\mathrm{out}^{}$ 
and  that from $\widehat{C}_\mathrm{in}^{}$ 
are shown in  the lower-left  and lower-right panels  
of Fig.\ \ref{fig:vertex_w00w_order_u3ph_color_each}, for which 
the frequencies are chosen such that 
  the external $\omega$ flows along  the dashed lines.
For simplicity, we assign the internal frequencies for 
$\Gamma_{\uparrow\uparrow;\uparrow\uparrow}^{(\mathrm{rep})}
(i\omega, i\omega;0,0)$ and that for 
$\Gamma_{\uparrow\uparrow;\uparrow\uparrow}^{(\mathrm{rep})}
(0,0;i\omega,i\omega)$ 
 in a synchronized way  such that 
the external $\omega$ passes through the {\it same interaction line\/}. 
Such interaction lines are denoted by dashed wavy lines 
with an arrow  that indicates the direction the $\omega$ flows. 

\end{itemize}


\begin{figure}[t]
\leavevmode

\begin{minipage}{1\linewidth}
\includegraphics[width=0.28\linewidth]{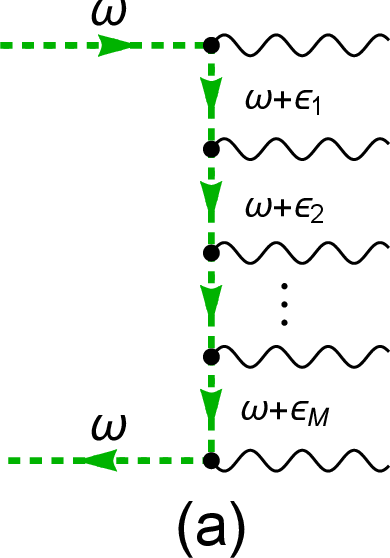}
\rule{0.1\linewidth}{0cm}
\includegraphics[width=0.28\linewidth]{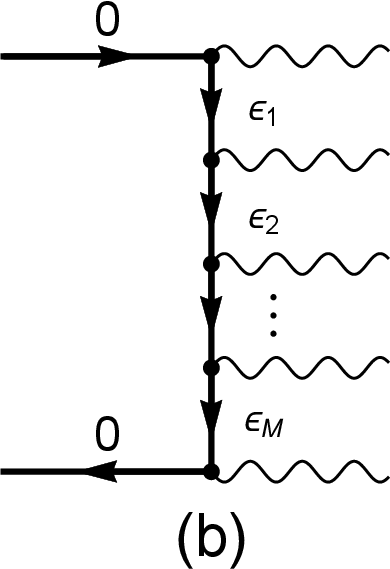}
\\
\includegraphics[width=0.28\linewidth]{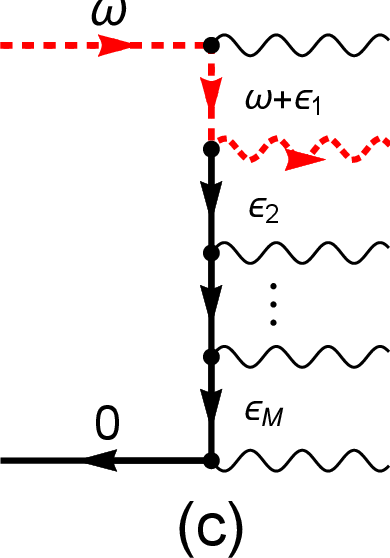}
\rule{0.1\linewidth}{0cm}
\includegraphics[width=0.28\linewidth]{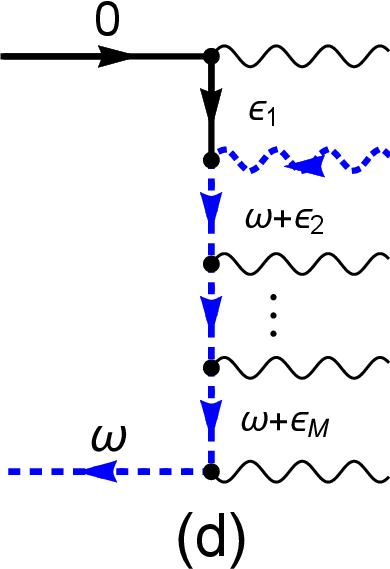}
\end{minipage}

 \caption{
(Color online)  
The {\it direct\/} line on the left side 
of 
 (a): 
$\Gamma_{\uparrow\uparrow;\uparrow\uparrow}^{(\mathrm{rep})}
(i\omega, 0;0, i\omega )$, 
(b):
$\Gamma_{\uparrow\uparrow;\uparrow\uparrow}^{(\mathrm{rep})}
(0,i\omega;i\omega,0 )$,
(c):
$\Gamma_{\uparrow\uparrow;\uparrow\uparrow}^{(\mathrm{rep})}
(i\omega, i\omega;0,0)$, 
and 
(d): 
$\Gamma_{\uparrow\uparrow;\uparrow\uparrow}^{(\mathrm{rep})}
(0,0;i\omega,i\omega)$.
The dashed lines denote the propagators and interaction lines 
 that  carry the external frequency $\omega$. 
}
 \label{fig:cancellation_in}
\end{figure}

\begin{figure}[t]
\leavevmode

\begin{minipage}{1\linewidth}
\includegraphics[width=0.28\linewidth]{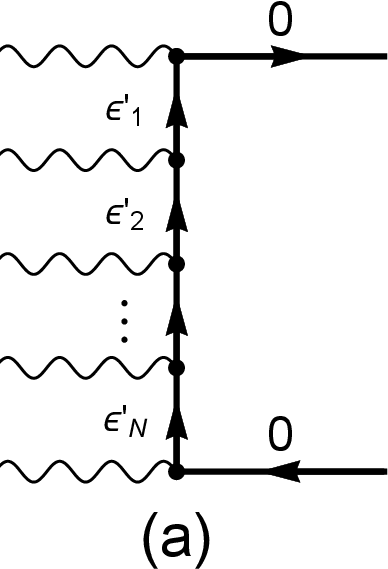}
\rule{0.1\linewidth}{0cm}
\includegraphics[width=0.28\linewidth]{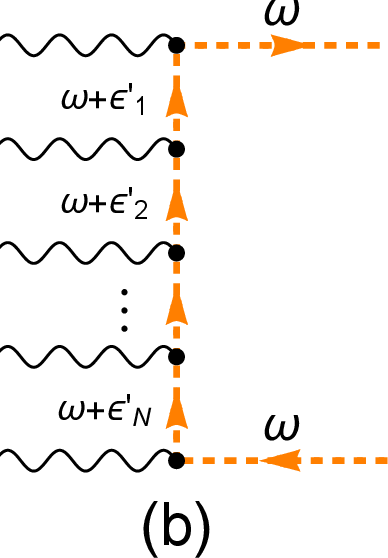}
\\
\includegraphics[width=0.28\linewidth]{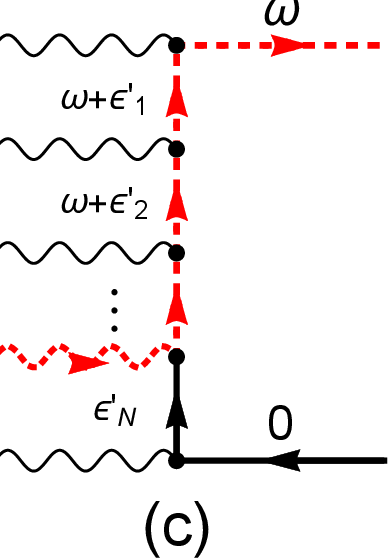}
\rule{0.1\linewidth}{0cm}
\includegraphics[width=0.28\linewidth]{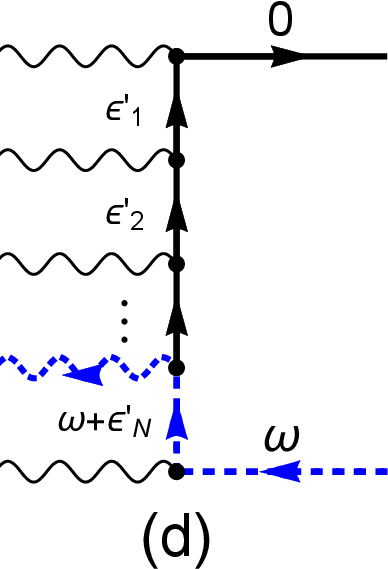}
 \end{minipage}
 \caption{
(Color online) 
The {\it direct\/} line on the right  side 
of 
  (a): 
$\Gamma_{\uparrow\uparrow;\uparrow\uparrow}^{(\mathrm{rep})}
(i\omega, 0;0, i\omega )$, 
(b):
$\Gamma_{\uparrow\uparrow;\uparrow\uparrow}^{(\mathrm{rep})}
(0,i\omega;i\omega,0 )$,
(c):
$\Gamma_{\uparrow\uparrow;\uparrow\uparrow}^{(\mathrm{rep})}
(i\omega, i\omega;0,0)$, 
and 
(d): 
$\Gamma_{\uparrow\uparrow;\uparrow\uparrow}^{(\mathrm{rep})}
(0,0;i\omega,i\omega)$.
The dashed lines denote the propagators and interaction lines 
 that  carry the external frequency $\omega$. 
}
\label{fig:cancellation_out}
\end{figure}


We calculate together the contributions of the four diagrams which constitutes the set, 
 $\widehat{\partial}_{i\omega}^{+}\Gamma_{\uparrow\uparrow;\uparrow\uparrow}^{(\mathrm{set})}  (i\omega, 0;0, i\omega )$   
in order to keep the anti-symmetry of the vertex function.  
It can be shown that  the contribution of the propagators which  
belong to one of the two {\it direct} $\uparrow$ lines vanishes, 
as seen in the middle part 
of  Eqs.\ \eqref{eq:gamU3a_skeleton} and \eqref{eq:gamU3b_skeleton}
for the order $U^3$ contributions.
This is owing to the anti-symmetry,  and can be confirmed by  
operating  $\widehat{\partial}_{i\omega}^{+}$ upon the {\it direct} lines 
as shown in Figs.\  \ref{fig:cancellation_in} and  \ref{fig:cancellation_out}.

\begin{figure}[t]
 \leavevmode
\begin{minipage}{1\linewidth}
\includegraphics[width=0.4\linewidth]{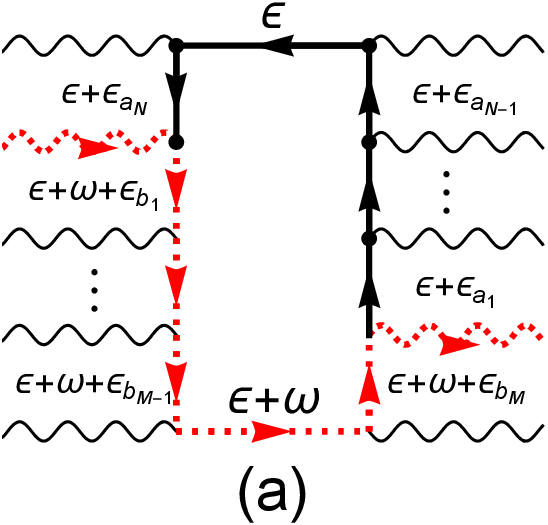}
\rule{0.1\linewidth}{0cm}
\includegraphics[width=0.4\linewidth]{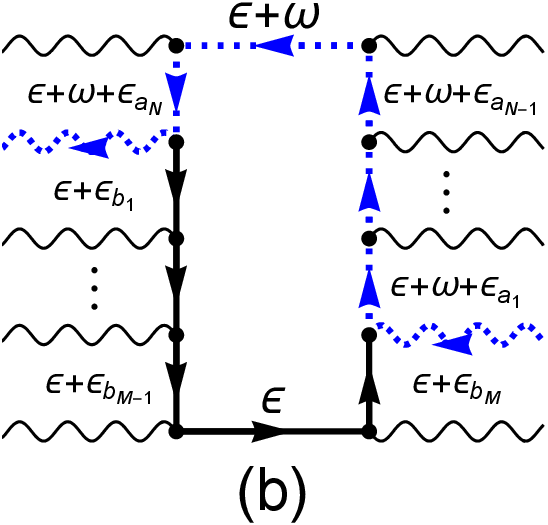}
\\
\includegraphics[width=0.4\linewidth]{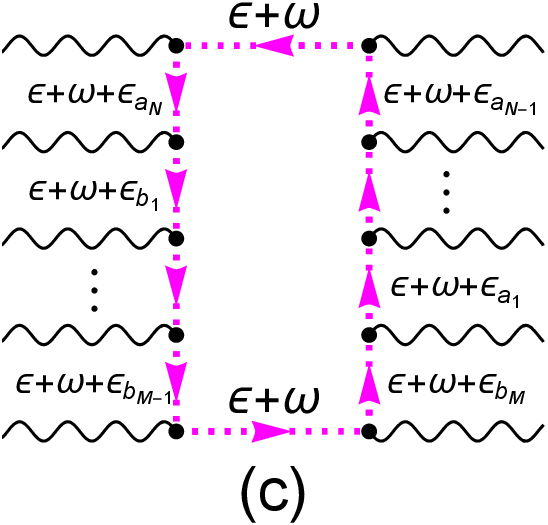}
\end{minipage}
 \caption{
(Color online) 
Example of a closed loop that includes one singular particle-hole product      
$G_{\sigma''}(i\varepsilon)\,G_{\sigma''}(i\varepsilon+i\omega)$  for 
(a):
$\Gamma_{\uparrow\uparrow;\uparrow\uparrow}^{(\mathrm{rep})}
(i\omega, i\omega;0,0)$, 
and 
(b): 
$\Gamma_{\uparrow\uparrow;\uparrow\uparrow}^{(\mathrm{rep})}
(0,0;i\omega,i\omega)$. 
Total contribution of  (a) and  (b)  coincides 
with  the contribution of (c), or symbolically 
$
\widehat{\partial}_{i\omega}^{+} 
\Bigl[  (a) \,+\, (b)  \Bigr] = 
\widehat{\partial}_{i\omega}^{+} 
\Bigl[  (c) \Bigr] 
$. 
In (c),  the external frequency $\omega$  flows along the loop, 
and thus
it can be integrated out with the circular frequency $\varepsilon$, i.e.,  
 $
\widehat{\partial}_{i\omega}^{+} 
\int_{-\infty}^\infty d\varepsilon\,  \Bigl[ (c) \Bigr]=0 $. 
}
 \label{fig:cancellation_loop}
\end{figure}

The remaining contributions can be generated,    
operating  $\widehat{\partial}_{i\omega}^{+}$ 
 upon the closed loops, which partially carry $\omega$.
In our construction of the diagrams, such contributions arise from 
$\Gamma_{\uparrow\uparrow;\uparrow\uparrow}^{(\mathrm{rep})}
(i\omega, i\omega;0,0)$ and  
$\Gamma_{\uparrow\uparrow;\uparrow\uparrow}^{(\mathrm{rep})}
 (0,0;i\omega,i\omega)$. The corresponding order $U^3$ diagrams are described 
in  the lower panel of    
Figs.\ \ref{fig:vertex_w00w_order_u3ph_color_each} and 
 \ref{fig:vertex_w00w_order_u3pp_color_each}.
The contributions of the two diagrams cancel out 
as  the external $\omega$ is absorbed into the circular frequency  along  the closed loop 
as shown in 
Figs.\  \ref{fig:vertex_w00w_order_u3ph_color_sum} 
and \ref{fig:vertex_w00w_order_u3pp_color_sum}.
Figure  \ref{fig:cancellation_loop} describes the same cancellation 
occurring  in a single but a more complicated loop.
As we construct 
Figs.\  \ref{fig:cancellation_loop} (a)  and (b) following rule $iii$) 
such that the external frequency $\omega$ passes through the same interaction lines,  
their contributions of the loop part can be written in the form
\begin{align}
I^{(a)}_{\sigma''}
=& \  \ 
\widehat{\partial}_{i\omega}^{+}
 \Biggl[
\int_{-\infty}^{\infty} \frac{d \varepsilon}{2\pi}\, 
{\color[rgb]{1,0,0} 
G_{\sigma''}^{}(i\varepsilon +i\omega) 
}
\,G_{\sigma''}(i\varepsilon)  
\nonumber \\
& \qquad 
 \times 
\prod_{j}^N 
 G_{\sigma''}^{}(i\varepsilon+i\varepsilon_{a_j})\,
{\color[rgb]{1,0,0}
\prod_{k}^M 
 G_{\sigma''}^{}(i\varepsilon+i\omega+i\varepsilon_{b_k})
}
\Biggr]
, 
\\
I^{(b)}_{\sigma''}
=& \  \  
\widehat{\partial}_{i\omega}^{+}
 \Biggl[
\int_{-\infty}^{\infty} \frac{d \varepsilon}{2\pi}\, 
G_{\sigma''}(i\varepsilon) \,
{\color[rgb]{0,0,1}
 G_{\sigma''}^{}(i\varepsilon +i\omega) }
\nonumber \\
& \qquad
\times 
{\color[rgb]{0,0,1}
\prod_{j}^N 
 G_{\sigma''}^{}(i\varepsilon+\omega+i\varepsilon_{a_j})
}
\,\prod_{k}^M 
 G_{\sigma''}^{}(i\varepsilon+i\varepsilon_{b_k})
\Biggr]
.
\end{align}
The sum of these two contributions can be described by 
a single diagram Fig.\  \ref{fig:cancellation_loop} (c),  
\begin{align}
 & I^{(a)}_{\sigma''} + I^{(b)}_{\sigma''}
\ = \   
\widehat{\partial}_{i\omega}^{+}\, 
 \Biggl[
\int_{-\infty}^{\infty} \frac{d \varepsilon}{2\pi}\, 
{\color[rgb]{1,0,1}
\left\{ G_{\sigma''}^{}(i\varepsilon +i\omega) \right\}^2 
}
\nonumber \\
&  
\times
{\color[rgb]{1,0,1}
\prod_{j}^N 
 G_{\sigma''}^{}(i\varepsilon+i\omega+i\varepsilon_{a_j})
\,\prod_{k}^M 
 G_{\sigma''}^{}(i\varepsilon+i\omega+i\varepsilon_{b_k})
}
\Biggr] =0 .
\label{eq:cancellation_loop}
\end{align}
Here, the integration over $\varepsilon$ gives a constant 
that is independent of $\omega$. 

For more complicated diagrams,  the external frequency $\omega$ 
 passes through a number of  different closed loops, 
but the contribution from each of such closed loops vanishes  
in a similar way to Eq.\ \eqref{eq:cancellation_loop}.
We have also calculated all  the skeleton-diagrams 
for  $\Gamma_{\sigma\sigma;\sigma\sigma} (i\omega, 0; 0, i\omega)$ 
up to order $U^4$.\cite{Note1} 
It explicitly shows that the cancellation of the analytic $\omega$-linear part 
occurs for each and every set of four anti-symmetrized diagrams.

\section{Summary}
\label{sec:summary}

In summary,  
we have provided a precise derivation of 
 the higher-order Fermi-liquid relations of the Anderson impurity model. 
One of the most important results is the double-frequency expansion of the vertex function 
 $\Gamma_{\sigma\sigma';\sigma'\sigma}^{}(i\omega,i\omega';i\omega',i\omega)$,  
given in Eqs.\  \eqref{eq:GammaUU_general_omega_dash_abs} 
and \eqref{eq:GammaUD_general_matsubara_omega_dash_abs}.
These two equations have been deduced from  
the analytic and anti-symmetric properties of the vertex function: 
the linear terms with respect to these frequencies must be described by a 
linear combination of the analytic  $\omega$  and $\omega'$ contributions 
and the non-analytic $|\omega-\omega'|$ and $|\omega+\omega'|$ contributions.  
In addition, the anti-symmetry imposes the restriction: 
the vertex function for parallel spins $\sigma'=\sigma$ 
does not have the analytic $\omega$ and $\omega'$ contributions. 
The coefficients for these terms have been determined using the Ward identities.
The explicit form of 
$\Gamma_{\sigma\sigma';\sigma'\sigma}^{}(i\omega,i\omega';i\omega',i\omega)$ 
captures the essential features of the Fermi liquid away from half-filling, 
and is analogous to Landau's quasi-particle 
interaction $f(\bm{p}\, \sigma,\bm{p}' \sigma')$ 
and Nozi\`{e}res' function $\phi_{\sigma\sigma'}(\varepsilon,\varepsilon')$. 
\cite{AGD,NozieresFermiLiquid}  One important difference is that 
the vertex function also has  a non-analytic part, which 
directly determines  the damping of the quasi-particles.

In the second half of the present paper, 
we have also provided a complementary perturbative proof 
for the low-frequency behavior of the vertex function.
For this purpose,  we have introduced an operator  
that can extract the next-leading contribution from 
a singular Green's-function product expansion, Eq.\ \eqref{eq:discont_full}, 
for the intermediate particle-hole pair.
Specifically, we  have calculated all  the skeleton-diagrams 
for  $\Gamma_{\sigma\sigma;\sigma\sigma} (i\omega, 0; 0, i\omega)$ 
up to order $U^4$,\cite{Note1} 
and have directly confirmed  that a cancellation of the analytic $\omega$-linear part 
occurs in a set of four related  Feynman diagrams 
which anti-symmetrize the vertex corrections.

The higher-order Fermi-liquid corrections away from half-filling
are determined not only  by the linear susceptibilities  $\chi_{\sigma\sigma'}$ but  also 
the non-linear susceptibilities  $\chi_{\sigma_1,\sigma_2, \ldots, \sigma_n}^{[n]}$  
  for  $n=3$ and $n=4$.
We have also revisited the $T^2$-correction 
of the self-energy $\Sigma_{\sigma}^{}(i\omega)$
for calculating the real part which becomes finite away from half-filling,  
and have  shown that the coefficient is given by  
  $(\pi^2/6\rho_{d\sigma}^{})\,
\partial \chi_{\uparrow\downarrow}/\partial \epsilon_{d,-\sigma}$.  
Our result for the $\omega^2$ real part, 
$\mathrm{Re} 
\left. \partial^2 \Sigma_{\sigma}(i\omega)/
\partial (i\omega)^2\right|_{\omega=0} = 
\partial^2 \Sigma_{\sigma}(0)/\partial \epsilon_{d\sigma}^2$,
reproduces exactly the FMvDM's formula.\cite{FilipponeMocaVonDelftMora} 
We will give a more detailed comparison in a separate paper,  
i.e., {\it paper III}.\cite{ao2017_3_PRB}

In {\it paper III},  
we  will also  present an extension of the microscopic description  
to  the non-equilibrium steady state  driven 
by the bias voltage $eV$ using the Keldysh formalism. 
Furthermore,  we calculate the Fermi-liquid parameters 
using the NRG, and will demonstrate applications 
to various systems such as  the  non-linear 
magneto-conductance  through a quantum dot, 
thermo-electric transport of dilute magnetic alloys, 
and the Anderson impurity with a number orbitals.

\begin{acknowledgments}
We wish to  thank J.\ Bauer and R.\ Sakano 
for valuable discussions, and C.\ Mora and J.\ von Delft for 
sending us Ref.\ \onlinecite{FilipponeMocaVonDelftMora} prior to publication.
This work was supported by JSPS KAKENHI (No.\ 26400319) and 
  a Grant-in-Aid for Scientific Research (S) (No.\ 26220711).
\end{acknowledgments}


\appendix

\section{Static non-linear response functions}
\label{sec:nonlinear_static_susceptibility}

We  show that  $\chi_{\sigma_1\sigma_2\sigma_3}^{[3]} $ can be 
expressed in terms of  three-body correlation functions of the electron configuration 
of the impurity site,  defined with respect  to thermal equilibrium.
We consider the  Hamiltonian, 
 $\mathcal{H}_\mathrm{tot}^{}  = \mathcal{H} + \mathcal{H}_\mathrm{ex}$,  
which includes a static external part  $\mathcal{H}_\mathrm{ex}$. 
Following the standard perturbation theory,  
the imaginary-time evolution operator ${\cal U}\equiv 
 e^{\beta \mathcal{H}}e^{-\beta \mathcal{H}_\mathrm{tot}^{}}$ 
 can be expanded in  a power series of  $\mathcal{H}_\mathrm{ex}$:  
\begin{align}
& 
\!\!\!
{\cal U}\, =\, 
%
1 - \int_{0}^{\beta} \! d\tau_1 \, \mathcal{H}_\mathrm{ex} (\tau_1) 
\nonumber \\
&  \quad + \,
\frac{1}{2!}\int_{0}^{\beta} \! d\tau_1 \!
\int_{0}^{\beta} \! d\tau_2  \,
T_{\tau} \bigl[\mathcal{H}_\mathrm{ex} (\tau_1) \,
\mathcal{H}_\mathrm{ex} (\tau_2) \bigr] + \cdots,
\label{eq:U_beta}
\end{align}
where  $\beta=1/T$. 
The average of an operator $\mathcal{O}$ is defined by  
\begin{align}
& \langle \mathcal{O} \rangle_\mathrm{tot}^{} \equiv  
\frac{ \mathrm{Tr} \left[e^{-\beta\,\mathcal{H}_\mathrm{tot}^{}}\, \mathcal{O}\right]}{\mathrm{Tr}\, e^{-\beta\,\mathcal{H}_\mathrm{tot}^{}} } 
 =  
\frac{
\left\langle\, {\cal U}(\beta)\,\mathcal{O} \,\right\rangle}
{\left\langle\, {\cal U}(\beta)\,\right\rangle} ,
\label{eq:average_U} 
\end{align}
where $\left\langle\, \cdots \,\right\rangle\, \equiv \,
\mathrm{Tr}\left[\, e^{-\beta\,\mathcal{H}}\,\cdots\,\right]\,/\,\Xi$, and  
$\Xi \, \equiv \,\mathrm{Tr}\, e^{-\beta\,\mathcal{H}}$.
%
For the operator 
  $\delta \mathcal{O} \equiv \mathcal{O}-\langle \mathcal{O} \rangle$   
that satisfies  $\langle \delta \mathcal{O} \rangle =0$, 
the expansion up to second order is given by  
\begin{align}
&\langle \delta \mathcal{O} \rangle_\mathrm{tot}^{} 
 =  \, 
- \int_{0}^{\beta} \!\!\! d\tau \, 
\langle 
\mathcal{H}_\mathrm{ex} (\tau) \,
\delta \mathcal{O} 
\rangle 
\nonumber \\
&  \qquad \quad
+ \frac{1}{2}
\int_{0}^{\beta} \!\!\! d\tau \!\! 
\int_{0}^{\beta} \!\!\! d\tau' \, 
\langle T_\tau 
\mathcal{H}_\mathrm{ex} (\tau) \,
\mathcal{H}_\mathrm{ex} (\tau') \,
\delta \mathcal{O} 
\rangle 
\nonumber\\
& \qquad \quad 
- 
\int_{0}^{\beta} \!\!\! d\tau \!\! 
\int_{0}^{\beta} \!\!\! d\tau'\, 
\langle \mathcal{H}_\mathrm{ex} (\tau) \,\delta \mathcal{O} \rangle \,
\langle \mathcal{H}_\mathrm{ex} (\tau') \rangle 
+ \cdots .
\end{align}
We can apply this formula to a  response of the occupation number 
  $\mathcal{O} = n_{d\sigma}$ 
against a small variation of the impurity level $\delta \epsilon_{d\sigma}$, 
for which the perturbation Hamiltonian is given by  
$\mathcal{H}_\mathrm{ex} = 
\sum_{\sigma} \delta \epsilon_{d\sigma} \, \delta n_{d\sigma}$  with 
$\delta n_{d\sigma} \equiv n_{d\sigma} - \langle n_{d\sigma}  \rangle$. 
For this case,  $\langle \mathcal{H}_\mathrm{ex} (\tau') \rangle =0$ by definition,  
and thus
\begin{align}
&\langle \delta n_{d\sigma} \rangle_\mathrm{tot}^{} 
 \, = \, 
-
\sum_{\sigma_1} 
\int_{0}^{\beta} \!\!\! d\tau \, 
\langle 
\delta n_{d\sigma_1} (\tau) \,
\delta n_{d\sigma}
\rangle 
\, \delta \epsilon_{d\sigma_1}
\nonumber \\
&
+ \frac{1}{2}
\!
\sum_{\sigma_1\sigma_2} 
\!
\int_{0}^{\beta} \!\!\!\! d\tau_1 \!\!\! 
\int_{0}^{\beta} \!\!\!\! d\tau_2 \,
\langle T_\tau 
\delta n_{d\sigma_1} (\tau_1) 
\delta n_{d\sigma_2} (\tau_2) 
\delta n_{d\sigma}
\rangle 
 \delta \epsilon_{d\sigma_1}
 \delta \epsilon_{d\sigma_2}
\nonumber \\
&   +  \cdots .
\end{align}
In this case, the impurity level is given by $\epsilon_{d\sigma}+ \delta \epsilon_{d\sigma}$. The coefficients  can also be written in terms of 
the derivative of   $\langle n_{d\sigma} \rangle$ with respect to $\epsilon_{d\sigma'}$,
\begin{align}
\chi_{\sigma\sigma'} = & \ 
- \frac{\partial \langle n_{d\sigma} \rangle }{\partial \epsilon_{d\sigma'}} 
 = 
\int^\beta \!  d \tau \, 
\left\langle  \delta n_{d\sigma'}(\tau)\,\delta  n_{d\sigma}\right\rangle  ,
\\
\chi_{\sigma\sigma_1\sigma_2}^{[3]} = &
- \frac{\partial^2 \langle n_{d\sigma} \rangle }
{\partial \epsilon_{d\sigma_1}\partial \epsilon_{d\sigma_2}} 
\nonumber \\
 =&   \ 
- 
\int_{0}^{\beta} \!\!\! d\tau_1 \!\! 
\int_{0}^{\beta} \!\!\! d\tau_2 \, 
\langle T_\tau 
\delta n_{d\sigma_1} (\tau_1) \,
\delta n_{d\sigma_2} (\tau_2) \,
\delta n_{d\sigma}
\rangle .
\end{align}

\section{
Anti-symmetrization of a homogeneous polynomial 
}
\label{sec:homogeneous polynomial}

We describe here a quite simple but an important 
property of a homogeneous polynomial of a linear form, i.e., 
it can not be anti-symmetrized in the following sense. 
We consider the homogeneous function of degree one,
 \begin{align}
\mathcal{F}(x_1,x_2;x_3, x_4)\, =\,
 a_1\, x_1 + a_2\, x_2 +a_3\, x_3  + a_4 x_4
\;. 
 \end{align}
Here,  $a_1$, $a_2$, $a_3$, and $a_4$ are constants. 
We set a requirement   $x_1+x_3=x_2+x_4$   
which corresponds to a frequency conservation, 
and thus  three variables  among four are independent.  
Introducing another variable $y$ such that 
$x_1=x_4+y$ and   $x_2=x_3+y$, 
we choose $x_3$, $x_4$, and  $y$ as three independent variables.

In order to anti-symmetrize this polynomial, 
we impose the additional conditions,
 \begin{align}
&\mathcal{F}(x_1,x_2;x_3,x_4) 
\,=\, 
\mathcal{F}(x_3,x_4;x_1,x_2) 
\nonumber \\
& \qquad 
=\, -
\mathcal{F}(x_3,x_2;x_1,x_4) 
\,=\, -
\mathcal{F}(x_1,x_4;x_3,x_2) 
\;.
\label{eq:simple_antisymmetry}
 \end{align}
These conditions can explicitly be written as,
\begin{align}
& \ (a_1+a_4)\,x_4 + (a_2+a_3)\,x_3+(a_1+a_2)\,y 
\nonumber \\
= & \  \ \ 
(a_2+a_3)\,x_4 + (a_1+a_4)\,x_3+(a_3+a_4)\,y
\nonumber \\
=& \ 
- (a_3+a_4)\,x_4 -(a_1+a_2)\,x_3  -(a_2+a_3)\,y
\nonumber \\
=& \ 
-(a_1+a_2)\,x_4 - (a_3+a_4)\,x_3-(a_1+a_4)\,y \;.
\end{align}
For these conditions to be identically  satisfied  for arbitrary  $x_3$, $x_4$, and $y$,   
\begin{align}
 a_1+a_4\,=\, a_2+a_3\,=\,-(a_1+a_2)\,=\,- (a_3+a_4)\;. 
\end{align}
The solution is $a_1=-a_2=a_3=-a_4$, 
and the anti-symmetrized function is given by 
\begin{align}
\mathcal{F}(x_1,x_2;x_3, x_4)\, =\,
 a_1\, ( x_1 - x_2 + x_3  -  x_4 )  \, \equiv \, 0 \;,
\label{eq:homogeneous_linear}
\end{align}
because  $x_1+x_3=x_2+x_4$.

We note that this simple property of the homogeneous polynomial justifies 
our observation that the vertex function 
for the parallel spins, $\Gamma_{\sigma\sigma;\sigma\sigma}
(i\omega_1,i\omega_2;i\omega_3,i\omega_4)$, 
does not have the analytic component in the $\omega$-linear terms.

\section{The $\omega^2$ contribution of 
 $\Gamma_{\sigma,-\sigma;-\sigma,\sigma}^{}(i\omega,0;0,i\omega)$  
}
\label{sec:w2_vertex_ud_coefficient}

The coefficient for the  $(i\omega)^2\, \mathrm{sgn}\, \omega$  term
of $\Gamma_{\sigma,-\sigma;-\sigma,\sigma}^{}(i\omega,0;0,i\omega)$  
shown  in Eqs.\ \eqref{eq:GammaUD_general_matsubara} 
is calculated rewriting the derivative in the following way, 
\begin{align}
\frac{\partial}{\partial \epsilon_{d,-\sigma}^{}} \left(
  \frac{\chi_{\uparrow\downarrow}^2}{\rho_{d\sigma}^{}} \right) 
&=\, 
 \frac{2\chi_{\uparrow\downarrow}^{}}{\rho_{d\sigma}^{}}
\,\frac{\partial \chi_{\uparrow\downarrow}}{\partial \epsilon_{d,-\sigma}^{}} 
 + 
2 \pi \cot \delta_{\sigma} 
\,\frac{\chi_{\uparrow\downarrow}^3}{\rho_{d\sigma}^{}}
\nonumber 
\\
 \xrightarrow{\,h\to 0\,}&  \ \ 
 \frac{\chi_{\uparrow\downarrow}^{}}{\rho_{d}^{}}\,
\left(
\frac{\partial \chi_{\uparrow\downarrow}}{\partial \epsilon_{d}^{}} 
 + 2 \pi \cot \delta\ \chi_{\uparrow\downarrow}^2
\right) 
\label{eq:vert_u_d_w4_im} 
\\
\xrightarrow{\xi_d \to 0}&  \ \ 0 \,.
\end{align}
 The   $(i\omega)^2$ part 
 of $\Gamma_{\sigma,-\sigma;-\sigma,\sigma}^{}(i\omega,0;0,i\omega)$  
involves the fourth derivative of $\Omega$, 
\begin{align}
& 
\frac{\partial}{\partial \epsilon_{d,-\sigma}^{}}
\left(
\frac{\partial  \widetilde{\chi}_{\sigma\sigma}}{\partial \epsilon_{d\sigma}^{}} 
\right)
\ = \ 
\frac{\partial^2 \widetilde{\chi}_{\sigma,-\sigma}^{}}
{\partial \epsilon_{d\sigma}^2} 
\ =  \ 
\frac{\partial^2 }{\partial \epsilon_{d\sigma}^2} 
\left(
\frac{\chi_{\uparrow\downarrow}^{}}{\rho_{d\sigma}^{}} 
\right)
\nonumber  \\
&
=   \ 
\frac{1}{\rho_{d\sigma}^{}}\,
\frac{\partial^2 \chi_{\uparrow\downarrow}^{} }{\partial \epsilon_{d\sigma}^2} 
+ 2\, \frac{\partial \chi_{\uparrow\downarrow}^{}}{\partial \epsilon_{d\sigma}^{}} 
\,\frac{\partial }{\partial \epsilon_{d\sigma}^{}} 
\left(\frac{1}{\rho_{d\sigma}^{}}\right)
+ \chi_{\uparrow\downarrow}^{}\, 
\frac{\partial^2 }{\partial \epsilon_{d\sigma}^2} 
\left(\frac{1}{\rho_{d\sigma}^{}}\right)
\nonumber \\ 
&
= \  \frac{1}{\rho_{d\sigma}^{}}\,
\frac{\partial^2 \chi_{\uparrow\downarrow}^{} }{\partial \epsilon_{d\sigma}^2} 
 +
\frac{2\pi^2 \left(2 \cos ^2\delta_{\sigma}  + 1\right) }
{\rho_{d\sigma}^{} \sin^2 \delta_{\sigma}}
\,\chi_{\uparrow\downarrow}^{}\chi_{\sigma\sigma}^2
\nonumber \\
&   
\quad 
+
\frac{2 \pi \cot \delta_{\sigma}}{\rho_{d\sigma}^{}} 
\left( 
2 \, \chi_{\sigma\sigma}
\, \frac{\partial \chi_{\uparrow\downarrow}^{}}{\partial \epsilon_{d\sigma}^{}} 
+
\chi_{\uparrow\downarrow}^{} 
\frac{\partial  \chi_{\sigma\sigma}}{\partial \epsilon_{d\sigma}^{}} 
\right). 
\label{eq:vert_u_d_w4_re1} 
\end{align}
We have used  Eq.\ \eqref{eq:Dren_to_Dsus_org2} 
 for the double derivative of the inverse density of states     
and  
\begin{align}
 \frac{\partial^2 }{\partial \epsilon_{d\sigma}^2} 
\left(\frac{1}{\rho_{d\sigma}^{}}\right)
 =\  
& 
\frac{2 \pi  \cot \delta_{\sigma}  }{\rho_{d\sigma}^{}} 
\frac{\partial  \chi_{\sigma\sigma}}{\partial \epsilon_{d\sigma}} 
+
\frac{2\pi \chi_{\sigma\sigma}^2}{\Delta \rho_{d\sigma}^{2}}
\left(
2 \cos ^2\delta_{\sigma}  
 + 1
\right) . 
\end{align}
Equation \eqref{eq:vert_u_d_w4_re1} simplifies 
in zero magnetic  field, and  in the particle-hole symmetric case:
\begin{align}
& 
\frac{\partial}{\partial \epsilon_{d,-\sigma}^{}}
\left(
\frac{\partial  \widetilde{\chi}_{\sigma\sigma}}{\partial \epsilon_{d\sigma}^{}} 
\right) 
\nonumber \\
& \xrightarrow{h\to 0} \   
  \frac{1}{4\rho_{d}^{}}\left( 
 \frac{\partial^2 }{\partial \epsilon_{d}^2} 
 + \frac{\partial^2}{\partial h^2}  
  \right)  \chi_{\uparrow\downarrow}^{} 
 +
\frac{2\pi^2 \left(2 \cos ^2\delta  + 1\right) }
{\rho_{d\sigma}^{} \sin^2 \delta}
\,\chi_{\uparrow\downarrow}^{}\chi_{\uparrow\uparrow}^2
\nonumber \\
& \qquad \quad 
+
\frac{2 \pi \cot \delta}{\rho_{d}^{}} 
\left[ 
\left(\chi_{\uparrow\uparrow} -\frac{1}{2} \chi_{\uparrow\downarrow}\right) 
\frac{\partial \chi_{\uparrow\downarrow}^{}}{\partial \epsilon_{d}^{}} 
+
\chi_{\uparrow\downarrow}^{} 
\frac{\partial  \chi_{\uparrow\uparrow}}{\partial \epsilon_{d}^{}} 
\right]
 \label{eq:vert_u_d_w4_re2} 
 \\
& \xrightarrow{h\to 0 \atop \xi_d \to 0} \ 
\pi\Delta \,\left[
\frac{1}{4}\left( 
 \frac{\partial^2 \chi_{\uparrow\downarrow}^{} }{\partial \epsilon_{d}^2} 
 + \frac{\partial^2 \chi_{\uparrow\downarrow}^{}}{\partial h^2}  
  \right)  
+  2\pi^2 
\chi_{\uparrow\downarrow}^{}\,\chi_{\uparrow\uparrow}^{2}
\right]
.
\end{align}


%



\newpage

\ 

\newpage

\ 

\newpage

\ 

\newpage

\begin{widetext}

\begin{center}
{\large \bf 
Higher-order Fermi-liquid corrections for an Anderson impurity away from half-filling 

II: equilibrium properties  (Supplemental Material)
\rule{0cm}{0.5cm}
}
\end{center}


\begin{center}
Akira Oguri$^1$  and  A.\ C.\ Hewson$^2$ 
\smallskip

{\it 
$^1$Department of Physics, Osaka City University, Sumiyoshi-ku, 
Osaka 558-8585, Japan

$^2$Department of Mathematics, Imperial College London, London SW7 2AZ, 
United Kingdom
}
\end{center}


\bigskip

\begin{center}
{\bf Order $U^4$ skeleton-diagram expansion}
\end{center}

 \addtocounter{section}{15}

In this supplemental materials, 
we consider the order $U^4$ skeleton diagrams of the vertex function for parallel spins 
to show how the $\omega$-linear {\it regular} contributions cancel each other out. 
Specifically, 
we calculate the contributions by operating $\widehat{\partial}_{i\omega}^{+}$, 
defined in Sec.\ \ref{sec:green_function_product_expansion} of the text, 
upon $\Gamma_{\uparrow\uparrow;\uparrow\uparrow}^{}(i\omega , 0; 0 , i\omega)$.
\bigskip

\begin{figure}[h]
 \leavevmode
\begin{minipage}{1\linewidth}
\includegraphics[width=0.3\linewidth]{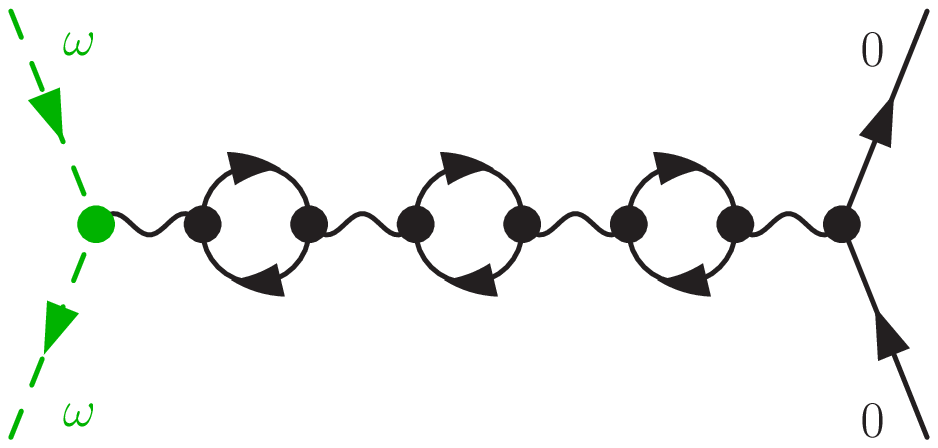}
\rule{0.08\linewidth}{0cm}
\includegraphics[width=0.3\linewidth]{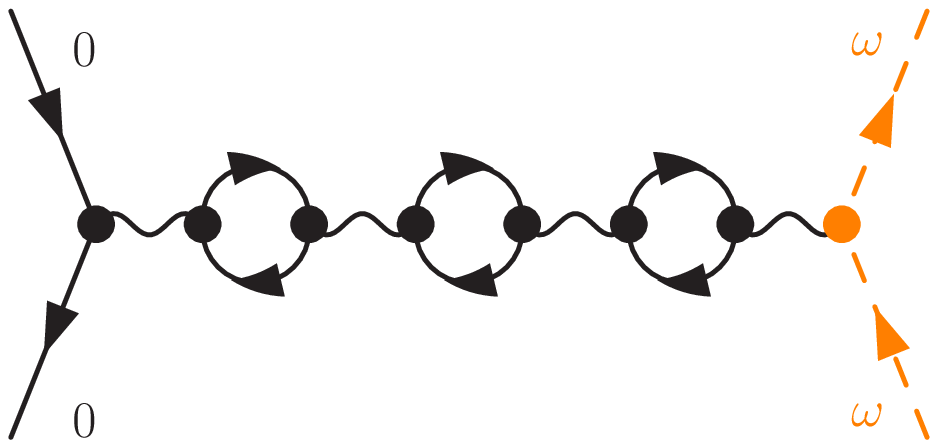}
\end{minipage}

\rule{0cm}{0.5cm}

\begin{minipage}{1\linewidth}
\includegraphics[width=0.3\linewidth]{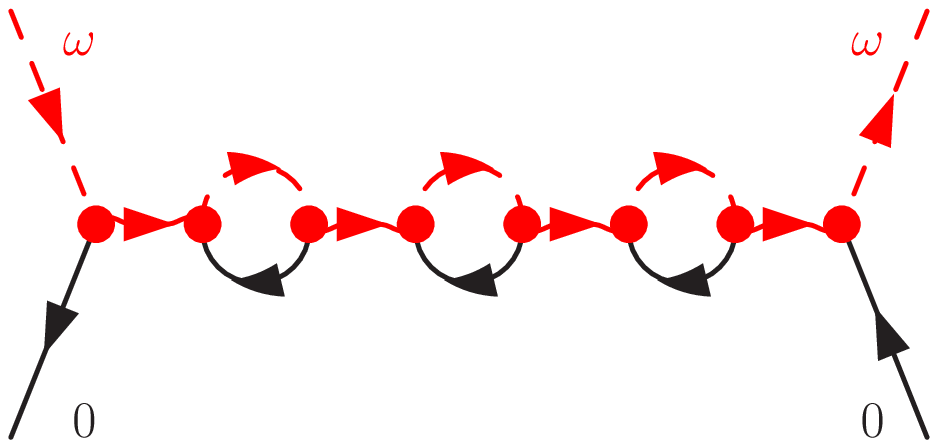}
\rule{0.08\linewidth}{0cm}
\includegraphics[width=0.3\linewidth]{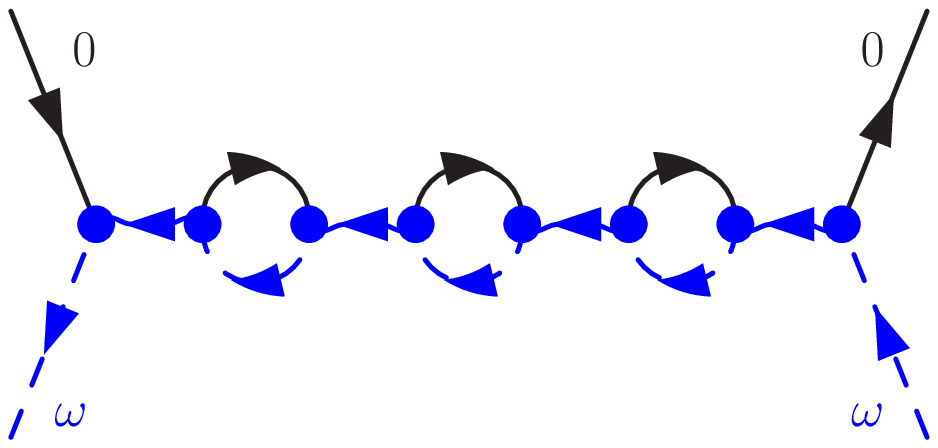}
\end{minipage}

 \caption{
(Color online) 
A set of four diagrams for 
$\Gamma_{\uparrow\uparrow;\uparrow\uparrow}^{(4A)}$, 
contribution of which is given in Eq.\ \eqref{eq:vertex_4A}.
 The dashed line represents the propagator 
 which is assigned to carry the external frequency $\omega$. 
}
 \label{fig:vertex_w00w_order_u4_1_color_each}
\end{figure}

\begin{figure}[h]
 \leavevmode
\begin{minipage}{1\linewidth}
\includegraphics[width=0.35\linewidth]{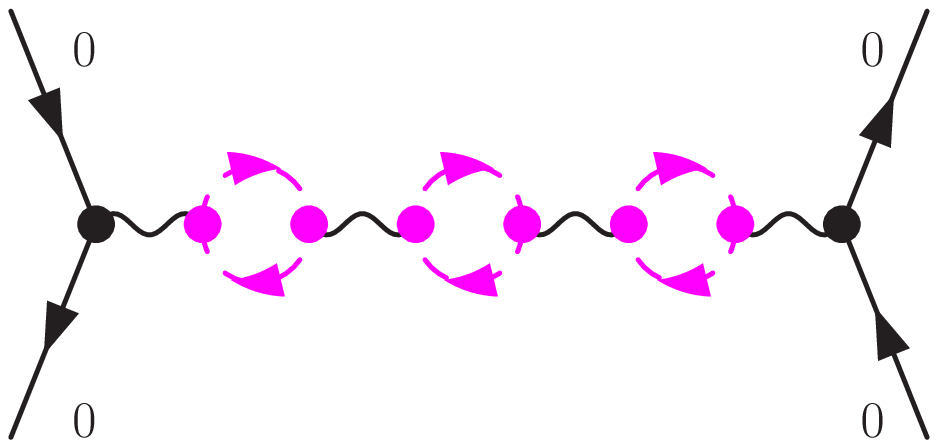}
\end{minipage}
 \caption{
(Color online) 
Schematic picture for 
the total  contribution 
 $\widehat{\partial}_{i\omega}^{+}
 \Gamma_{\uparrow\uparrow;\uparrow\uparrow}^{(4A)}$  
 of the set
shown in Fig.\ \ref{fig:vertex_w00w_order_u4_1_color_each}. 
}
 \label{fig:vertex_w00w_order_u4_1_color_sum}
\end{figure}


Total contribution of the diagrams shown 
in Fig.\ \ref{fig:vertex_w00w_order_u4_1_color_each} 
can be rewritten in a total derivative form
 (see also Fig.\ \ref{fig:vertex_w00w_order_u4_1_color_sum}) : 
\begin{align}
& 
\!\!\!\!\!\!\!\!\!\!\!\!\!
\widehat{\partial}_{i\omega}^{+}
\Gamma_{\uparrow\uparrow;\uparrow\uparrow}^{(4A)}(i\omega , 0; 0 , i\omega) 
\,= \,  
U^4 \,
\widehat{\partial}_{i\omega}^{+} 
\left[\, 
\left\{ \chi_{\downarrow\downarrow}^{qp}(i\omega) \right\}^2
\chi_{\uparrow\uparrow}^{qp}(i\omega)
- \left\{ \chi_{\downarrow\downarrow}^{qp}(0) \right\}^2
\chi_{\uparrow\uparrow}^{qp}(0)
\, \right] 
\nonumber \\
&\,= \, 
-U^4 \!
\int_{-\infty}^{\infty} \!
\int_{-\infty}^{\infty} \! 
\int_{-\infty}^{\infty} \! 
\frac{d\varepsilon_1\,d\varepsilon_2\,d\varepsilon_3}{(2\pi)^3}\  
\widehat{\partial}_{i\omega}^{+} 
\Bigl[
G_{\downarrow}^{}(i\varepsilon_1)\,
G_{\downarrow}^{}(i\varepsilon_1+i\omega)\,
G_{\uparrow}^{}(i\varepsilon_2)\,
G_{\uparrow}^{}(i\varepsilon_2+i\omega)\,
G_{\downarrow}^{}(i\varepsilon_3)\,
G_{\downarrow}^{}(i\varepsilon_3+\omega)
\Bigr] 
\nonumber \\
&\,= \, 
-U^4 \,
\frac{1}{2}\, 
{\color[rgb]{1,0,1} 
\widehat{\partial}_{i\omega}^{+} 
\left[
\int_{-\infty}^{\infty} \!
\int_{-\infty}^{\infty} \! 
\int_{-\infty}^{\infty} \! 
\frac{d\varepsilon_1\,d\varepsilon_2\,d\varepsilon_3}{(2\pi)^3}\  
\left\{G_{\downarrow}^{}(i\varepsilon_1+\omega)\right\}^2
\left\{G_{\uparrow}^{}(i\varepsilon_2+i\omega)\right\}^2
\left\{G_{\downarrow}^{}(i\varepsilon_3+i\omega)\right\}^2
\right] 
} 
\ = \ 0 .
\label{eq:vertex_4A}
\end{align}
For this set,  the external frequency $\omega$
transverses through the three intermediate closed loops.

\newpage

\begin{figure}[h]
 \leavevmode
\begin{minipage}{1\linewidth}
\includegraphics[width=0.3\linewidth]{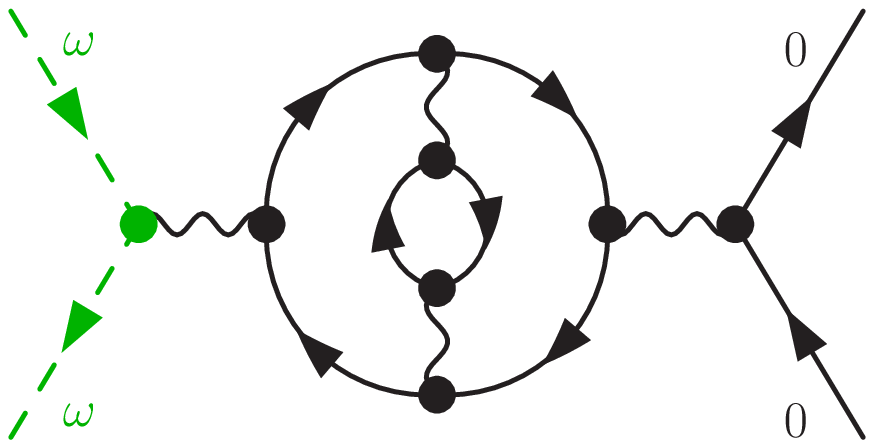}
\rule{0.08\linewidth}{0cm}
\includegraphics[width=0.3\linewidth]{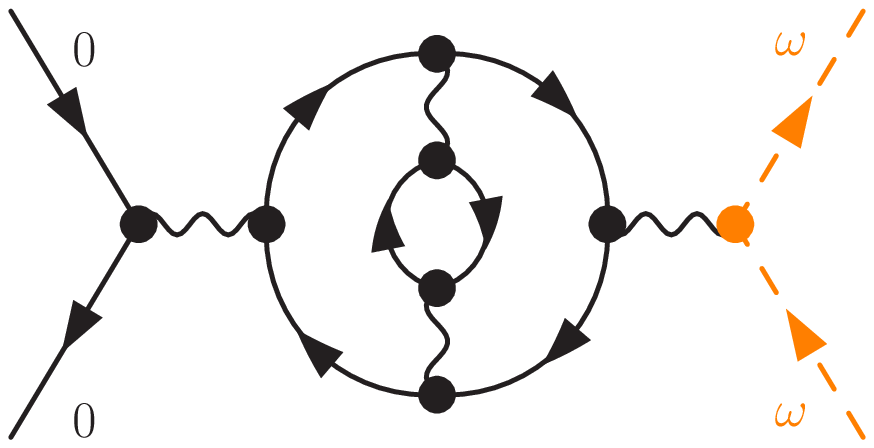}
\end{minipage}

 \rule{0cm}{0.5cm}

\begin{minipage}{1\linewidth}
\includegraphics[width=0.3\linewidth]{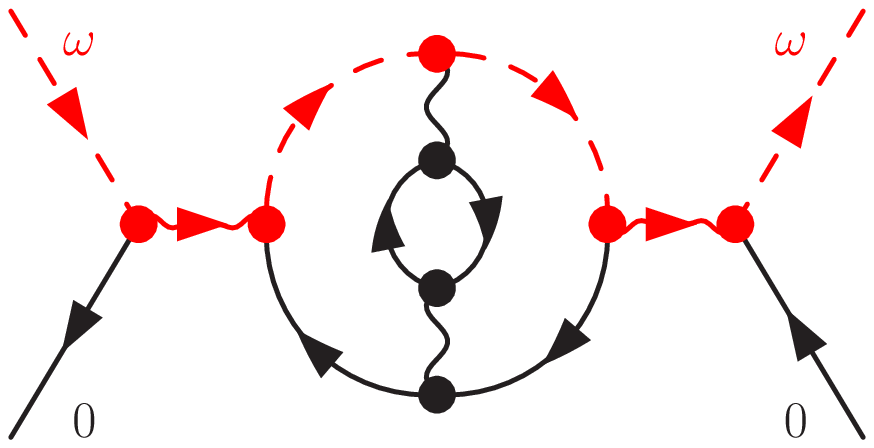}
\rule{0.08\linewidth}{0cm}
\includegraphics[width=0.3\linewidth]{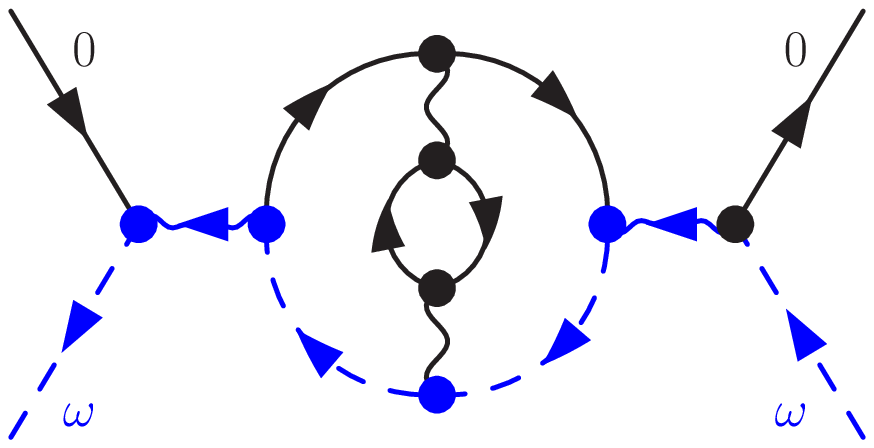}
\end{minipage}
 \caption{
(Color online) 
A set of four diagrams for 
$\Gamma_{\uparrow\uparrow;\uparrow\uparrow}^{(4B)}$, 
contribution of which is given in Eq.\ \eqref{eq:vertex_4B}.
}
 \label{fig:vertex_w00w_order_u4_4_color_each}
\end{figure}

\begin{figure}[h]
 \leavevmode
\begin{minipage}{1\linewidth}
\includegraphics[width=0.35\linewidth]{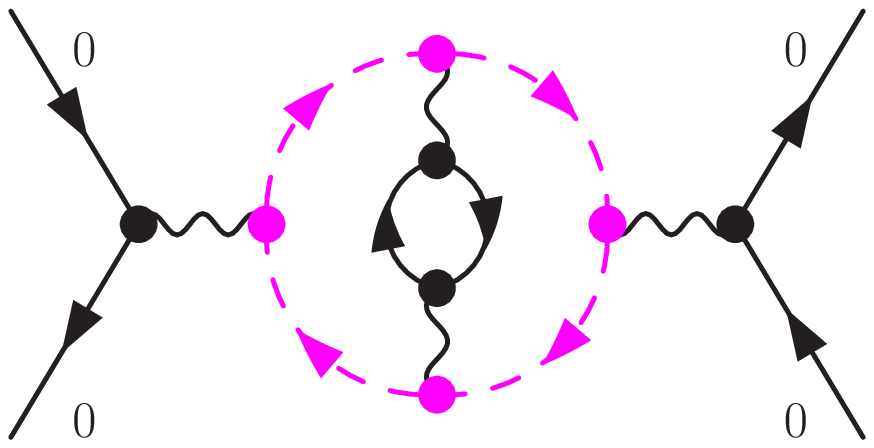}
\end{minipage}
 \caption{
(Color online) 
Schematic picture for 
the total  contribution  
 $\widehat{\partial}_{i\omega}^{+} \Gamma_{\uparrow\uparrow;\uparrow\uparrow}^{(4B)}$ 
of the diagrams 
shown in Fig.\ \ref{fig:vertex_w00w_order_u4_4_color_each}. 
%
}
 \label{fig:vertex_w00w_order_u4_4_color_sum}
\end{figure}


Total contribution of the diagrams shown in Fig.\ 
\ref{fig:vertex_w00w_order_u4_4_color_each} 
can be rewritten in a total derivative form 
 (see also  Fig.\ \ref{fig:vertex_w00w_order_u4_4_color_sum}): 
\begin{align}
& 
\!\!\!\!\!\!\!\!\!\!\!\!\!
\widehat{\partial}_{i\omega}^{+}
\Gamma_{\uparrow\uparrow;\uparrow\uparrow}^{(4B)}(i\omega , 0; 0 , i\omega) 
\,= \, 
-U^4\!
\int_{-\infty}^{\infty} \!\! 
\int_{-\infty}^{\infty} \!\! d\varepsilon'  
d\varepsilon \  
\widehat{\partial}_{i\omega}^{+} 
\Bigl[\, 
 G_{\downarrow}(i\varepsilon)\, 
G_{\downarrow}(i\varepsilon+i\omega) \, 
 G_{\downarrow}(i\varepsilon')\, 
G_{\downarrow}(i\varepsilon'+i\omega)\, 
\,-\, 
\left\{G_{\downarrow}(i\varepsilon)\right\}^2 
\left\{G_{\downarrow}(i\varepsilon')\right\}^2 
\,\Bigr] 
\chi_{\uparrow\uparrow}^{qp}(i\varepsilon-i\varepsilon')
\nonumber \\
& =  \ 
-U^4\!
\int_{-\infty}^{\infty} \!\! 
\int_{-\infty}^{\infty} \!\! d\varepsilon'  
d\varepsilon 
\,
\Biggl\{ 
\widehat{\partial}_{i\omega}^{+} 
\Bigl[\, 
 G_{\downarrow}(i\varepsilon)\, 
G_{\downarrow}(i\varepsilon+i\omega) \, 
\Bigr]
\left\{G_{\downarrow}(i\varepsilon')\right\}^2 
+ 
\left\{ G_{\downarrow}(i\varepsilon) \right\}^2
\widehat{\partial}_{i\omega}^{+} 
\Bigl[\, 
 G_{\downarrow}(i\varepsilon')\, 
G_{\downarrow}(i\varepsilon'+i\omega) \, 
\Bigr]\,
\Biggr\}\,
\chi_{\uparrow\uparrow}^{qp}(i\varepsilon-i\varepsilon')
 \nonumber \\
& =\,
-U^4\,   
\frac{1}{2}\, 
{\color[rgb]{1,0,1}
\widehat{\partial}_{i\omega}^{+} 
\left[\, 
\int_{-\infty}^{\infty} \!\! 
\int_{-\infty}^{\infty} \!\! d\varepsilon'  d\varepsilon  
\,
\left\{ G_{\downarrow}(i\varepsilon+i\omega) \right\}^2 \, 
\left\{ G_{\downarrow}(i\varepsilon'+i\omega)\right\}^2\, 
\,
\chi_{\uparrow\uparrow}^{qp}(i\varepsilon-i\varepsilon')
\right] 
}
 \ = \ 0. 
\label{eq:vertex_4B}
\end{align}
This is the simplest example of  Fig.\ \ref{fig:cancellation_loop_general} (c1): 
the intermediate closed loop consists 
of two singular Green's-function  products  carrying $\omega$ in the horizontal direction.


\newpage

\begin{figure}[h]
 \leavevmode
\begin{minipage}{1\linewidth}
\includegraphics[width=0.3\linewidth]{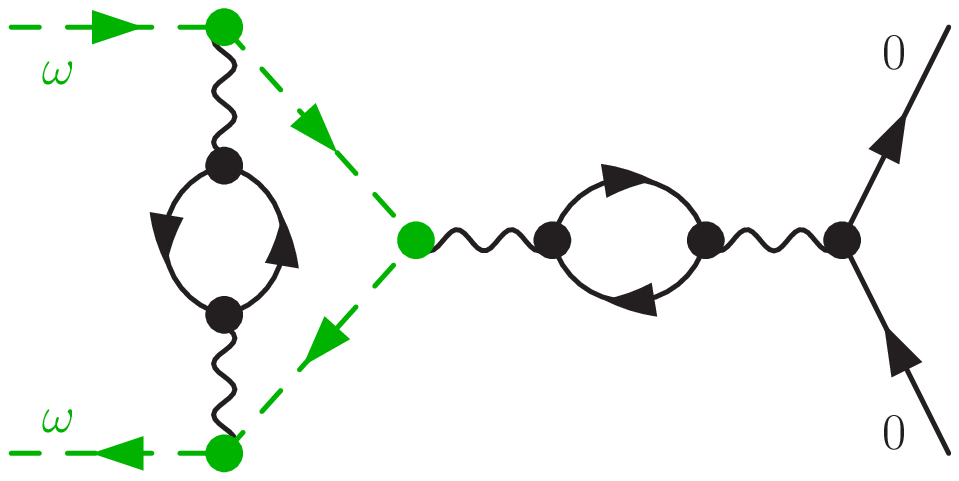}
\rule{0.08\linewidth}{0cm}
\includegraphics[width=0.3\linewidth]{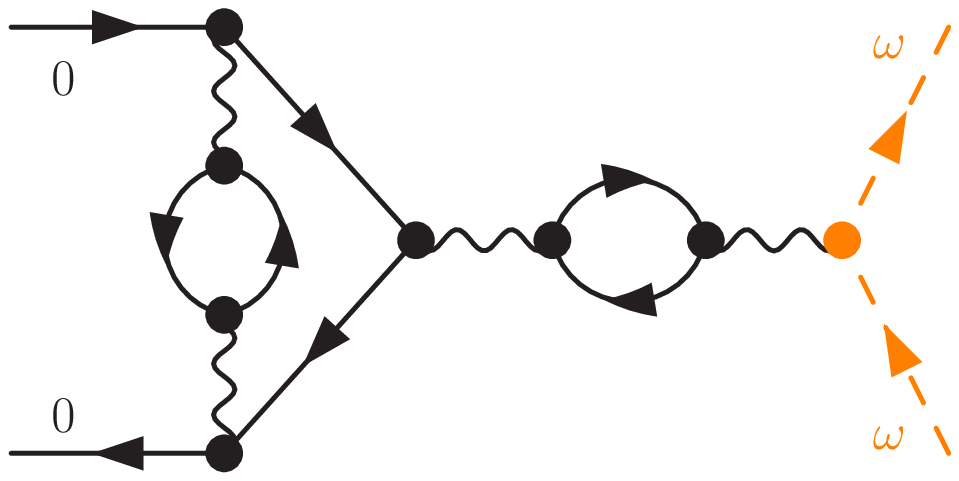}
\end{minipage}

 \rule{0cm}{0.5cm}

\begin{minipage}{1\linewidth}
\includegraphics[width=0.3\linewidth]{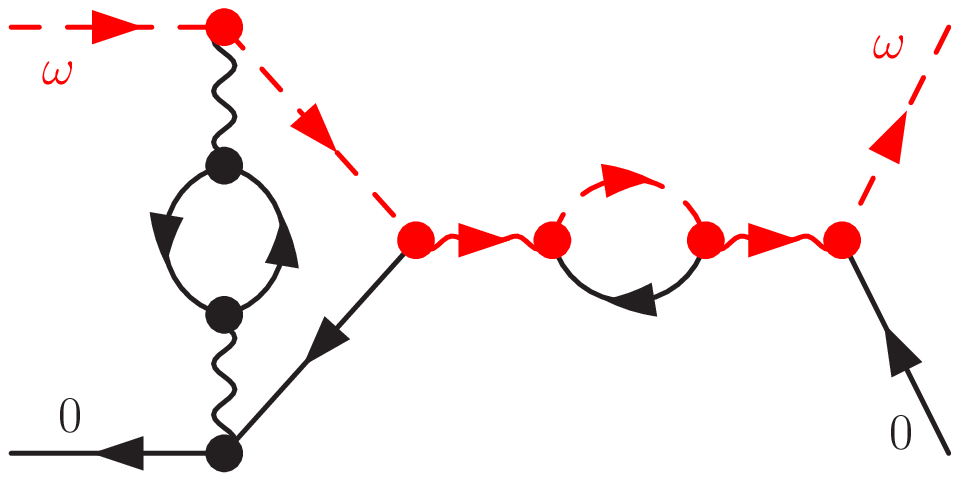}
\rule{0.08\linewidth}{0cm}
\includegraphics[width=0.3\linewidth]{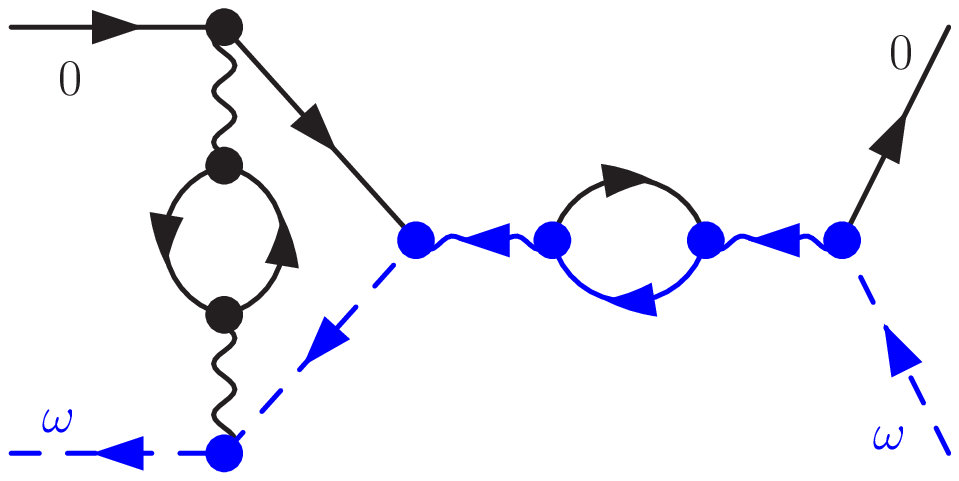}
\end{minipage}
 \caption{
(Color online) 
A set of four diagrams for 
$\Gamma_{\uparrow\uparrow;\uparrow\uparrow}^{(4C)}$ 
contribution of which is given in Eq.\ \eqref{eq:vertex_4C}.
}
 \label{fig:vertex_w00w_order_u4_2L_color_each}
\end{figure}

\begin{figure}[h]
 \leavevmode
\begin{minipage}{1\linewidth}
\includegraphics[width=0.35\linewidth]{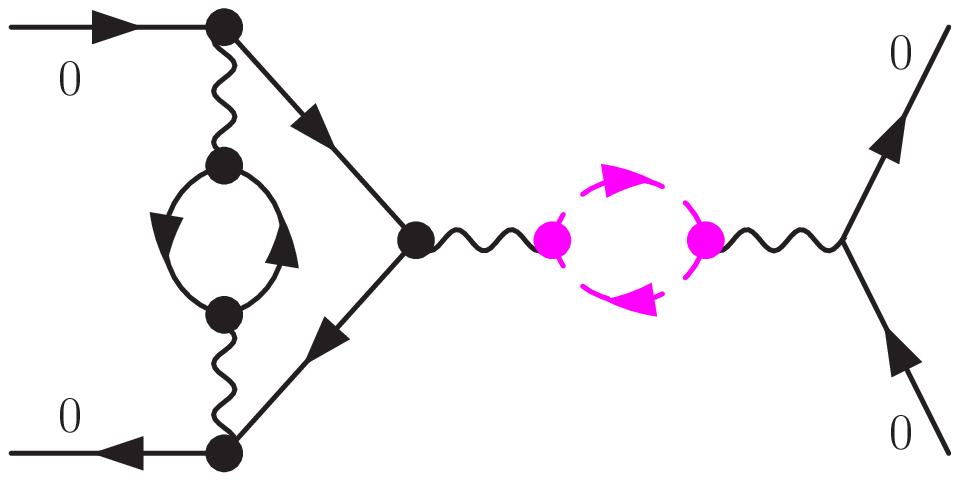}
\end{minipage}
 \caption{
(Color online) 
Schematic picture for 
the total  contribution 
 $\widehat{\partial}_{i\omega}^{+} \Gamma_{\uparrow\uparrow;\uparrow\uparrow}^{(4C)}$  of the set
shown in Fig.\ \ref{fig:vertex_w00w_order_u4_2L_color_each}. 
%
}
 \label{fig:vertex_w00w_order_u4_2L_color_sum}
\end{figure}


Total contribution of the diagrams shown in 
 Fig.\ \ref{fig:vertex_w00w_order_u4_2L_color_each} can be  
rewritten in a total derivative form with respect to the loop frequency 
(see also Fig.\ \ref{fig:vertex_w00w_order_u4_2L_color_sum}): 
\begin{align}
& \!\!\!\!\!\!
\widehat{\partial}_{i\omega}^{+}
\Gamma_{\uparrow\uparrow;\uparrow\uparrow}^{(4C)}(i\omega , 0; 0 , i\omega) 
\nonumber \\
&= 
U^4 \! 
\int_{-\infty}^{\infty} \!
\int_{-\infty}^{\infty} \! 
\frac{d\varepsilon\,d\varepsilon'}{(2\pi)^2}\  
\chi_{\downarrow\downarrow}^{qp}(i\varepsilon) \  
\widehat{\partial}_{i\omega}^{+} \, 
\biggl[\, 
-
\, 
{\color{red}
G_{\uparrow}^{}(i\varepsilon+i\omega)\,
}\,
G_{\uparrow}^{}(i\varepsilon)\, 
{\color{red}
G_{\downarrow}^{}(i\varepsilon'+i\omega)\, 
}
G_{\downarrow}^{}(i\varepsilon')\, 
\, -
\, 
G_{\uparrow}^{}(i\varepsilon)\, 
{\color{blue}
G_{\uparrow}^{}(i\varepsilon+i\omega)
}
\,
G_{\downarrow}^{}(i\varepsilon') 
{\color{blue} 
G_{\downarrow}^{}(i\varepsilon'+i\omega)\, 
}\,
 \nonumber \\
&  
\qquad \qquad \qquad \qquad  \qquad \qquad \qquad \qquad 
+ 
{\color[rgb]{0,0.5,0}
\left\{G_{\uparrow}^{}(i\varepsilon+i\omega)\right\}^2\, 
}
\left\{G_{\downarrow}^{}(i\varepsilon')\right\}^2
\, +
\left\{G_{\uparrow}^{}(i\varepsilon)\right\}^2
\, 
\left\{G_{\downarrow}^{}(i\varepsilon')\right\}^2
\,\biggr] 
\nonumber \\
=& \ 
-U^4 \! 
\int_{-\infty}^{\infty} \!
\frac{d\varepsilon}{2\pi}\  
\chi_{\downarrow\downarrow}^{qp}(i\varepsilon) \  
\left\{G_{\uparrow}^{}(i\varepsilon)\right\}^2
\int_{-\infty}^{\infty} \! 
\frac{d\varepsilon'}{2\pi}\  
\widehat{\partial}_{i\omega}^{+} \, 
\biggl[\, 
{\color{red} 
G_{\downarrow}^{}(i\varepsilon'+i\omega)\,
} 
G_{\downarrow}^{}(i\varepsilon')\, 
+
G_{\downarrow}^{}(i\varepsilon') \,
{\color{blue} 
G_{\downarrow}^{}(i\varepsilon'+i\omega)\,
} 
\,\biggr]
\nonumber \\
=& \ 
-U^4 \! 
\int_{-\infty}^{\infty} \!
\frac{d\varepsilon}{2\pi}\  
\chi_{\downarrow\downarrow}^{qp}(i\varepsilon) \  
\left\{G_{\uparrow}^{}(i\varepsilon)\right\}^2
{\color[rgb]{1,0,1} 
\widehat{\partial}_{i\omega}^{+} \, 
\left[\, 
\int_{-\infty}^{\infty} \! 
\frac{d\varepsilon'}{2\pi}\  
\left\{G_{\downarrow}^{}(i\varepsilon'+i\omega)\right\}^2
\,
\,\right]} \ = \ 0 .
\label{eq:vertex_4C}
\end{align}
Here, $U^2 \chi_{\downarrow\downarrow}^{qp}(i\varepsilon)$ 
due to the particle-hole pair in the vertical direction 
can be regarded as  a vertex correction  
between the external lines on the left side.

\newpage

\begin{figure}[h]
 \leavevmode
\begin{minipage}{1\linewidth}
\includegraphics[width=0.3\linewidth]{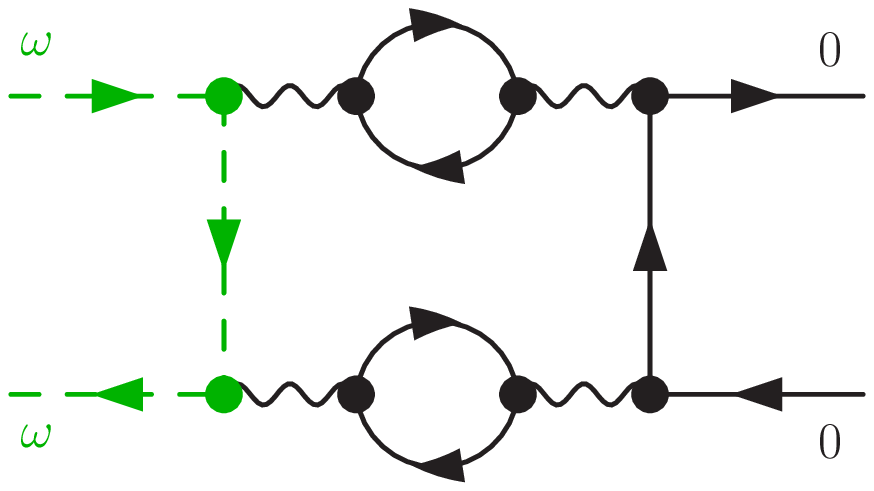}
\rule{0.08\linewidth}{0cm}
\includegraphics[width=0.3\linewidth]{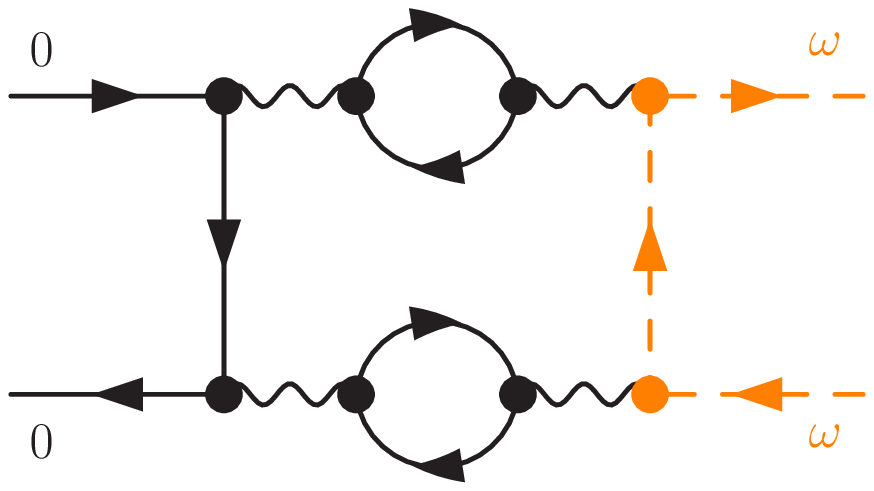}
\end{minipage}

\rule{0cm}{0.5cm}

\begin{minipage}{1\linewidth}
\includegraphics[width=0.3\linewidth]{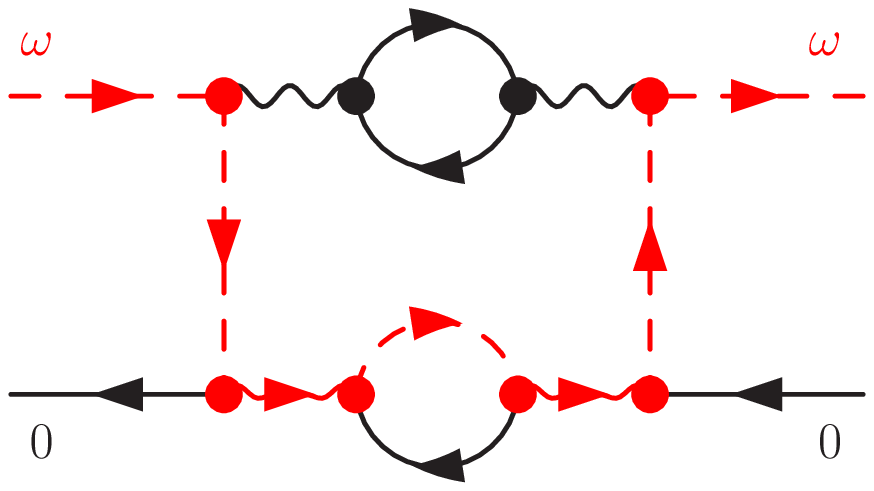}
\rule{0.08\linewidth}{0cm}
\includegraphics[width=0.3\linewidth]{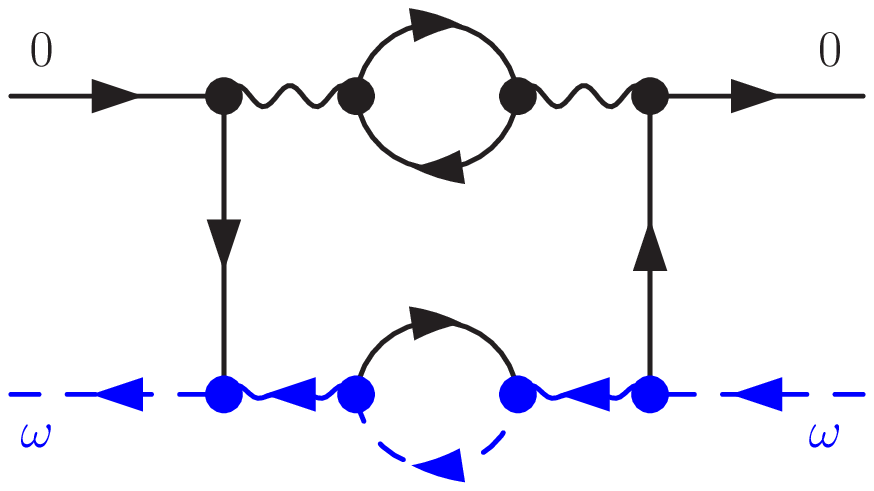}
\end{minipage}
 \caption{
(Color online) 
A set of four diagrams for 
$\Gamma_{\uparrow\uparrow;\uparrow\uparrow}^{(4D)}$,  
contribution of which is given in Eq.\ \eqref{eq:vertex_4D}.
}
 \label{fig:vertex_w00w_order_u4_17ph_color_each}
\end{figure}

\begin{figure}[h]
 \leavevmode
\begin{minipage}{1\linewidth}
\includegraphics[width=0.35\linewidth]{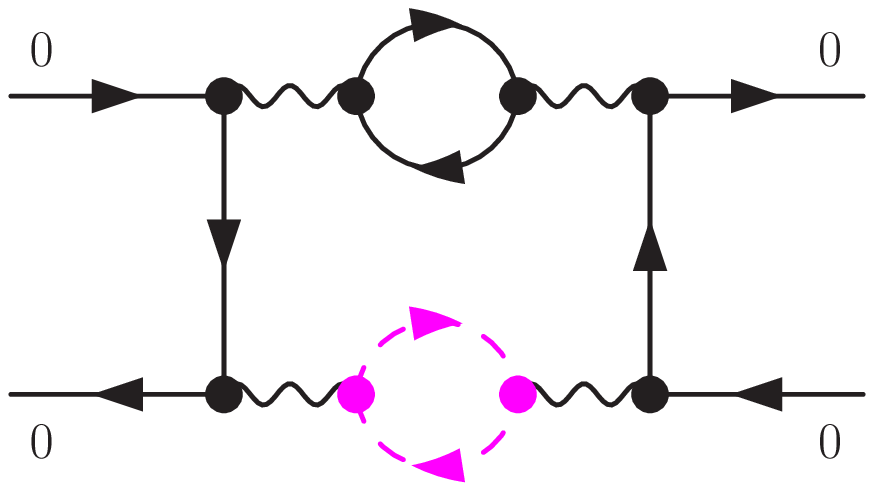}
\end{minipage}
 \caption{
(Color online) 
Schematic picture for 
the total  contribution 
 $\widehat{\partial}_{i\omega}^{+} 
\Gamma_{\uparrow\uparrow;\uparrow\uparrow}^{(4D)}$  
 of the set
shown in Fig.\ \ref{fig:vertex_w00w_order_u4_17ph_color_each}. 
The dashed propagators 
carrying the external frequency  $\omega$ 
form a closed loop.}
 \label{fig:vertex_w00w_order_u4_17ph_color_sum}
\end{figure}


Total contribution of the diagrams shown in Fig.\ 
\ref{fig:vertex_w00w_order_u4_17ph_color_each} 
can be rewritten in a total derivative form 
 (see also  Fig.\ \ref{fig:vertex_w00w_order_u4_17ph_color_sum}):  
\begin{align}
& 
 \!\!\!\!\!\!\!\!\!
\widehat{\partial}_{i\omega}^{+}
\Gamma_{\uparrow\uparrow;\uparrow\uparrow}^{(4D)}(i\omega , 0; 0 , i\omega) 
\nonumber \\
& = \,  \frac{U^4}{2}   \int_{-\infty}^{\infty}
 \frac{d\varepsilon}{2\pi}\  \widehat{\partial}_{i\omega}^{+} 
 \biggl[\,
{\color{red}
\chi_{\downarrow\downarrow}^{qp}(i\varepsilon + i\omega)
}\,  
 \chi_{\downarrow\downarrow}^{qp}(i\varepsilon)\,  
{\color{red}
\left\{ G_{\uparrow}^{}(i\varepsilon+i\omega) \right\}^2
} 
+
{\color{blue}
 \chi_{\downarrow\downarrow}^{qp}(i\varepsilon- i\omega)
}\,  
 \chi_{\downarrow\downarrow}^{qp}(i\varepsilon)\,  
{\color{blue}
\left\{ G_{\uparrow}^{}(i\varepsilon) \right\}^2  
}
\nonumber \\
& \qquad  \qquad \qquad \qquad \quad 
 - \left\{\chi_{\downarrow\downarrow}^{qp}(i\varepsilon) \right\}^2 
{\color[rgb]{0,0.5,0}
G_{\uparrow}^{}(i\varepsilon+i\omega)\,
} 
G_{\uparrow}^{}(i\varepsilon)\,
 - 
\left\{\chi_{\downarrow\downarrow}^{qp}(i\varepsilon) \right\}^2 
G_{\uparrow}^{}(i\varepsilon)\,
{\color[rgb]{1,0.5,0}
G_{\uparrow}^{}(i\varepsilon+i\omega)\, 
}
 \biggr] 
\nonumber \\
& = \,  \frac{U^4}{2}   \int_{-\infty}^{\infty}
 \frac{d\varepsilon}{2\pi}\, 
\left\{ G_{\uparrow}^{}(i\varepsilon) \right\}^2  
\widehat{\partial}_{i\omega}^{+}
\left[\,
{\color{red}
\chi_{\downarrow\downarrow}^{qp}(i\varepsilon + i\omega)
}\,  
 \chi_{\downarrow\downarrow}^{qp}(i\varepsilon)\,  
+
{\color{blue}
 \chi_{\downarrow\downarrow}^{qp}(i\varepsilon- i\omega)
}\,  
 \chi_{\downarrow\downarrow}^{qp}(i\varepsilon)\,  
-2  \left\{\chi_{\downarrow\downarrow}^{qp}(i\varepsilon) \right\}^2 
\right]
\nonumber \\
& = \, -\, \frac{U^4}{2}   
\int_{-\infty}^{\infty}
 \frac{d\varepsilon}{2\pi}\, 
\left\{ G_{\uparrow}^{}(i\varepsilon) \right\}^2  
 \chi_{\downarrow\downarrow}^{qp}(i\varepsilon)\,  
\int_{-\infty}^{\infty}
 \frac{d\varepsilon'}{2\pi}\ \widehat{\partial}_{i\omega}^{+} 
\biggl[\,
{\color{red}
G_{\downarrow}^{}(i\varepsilon'+i\varepsilon+i\omega)\,
}
G_{\downarrow}^{}(i\varepsilon')\,
+
G_{\downarrow}^{}(i\varepsilon'+i\varepsilon)\,
{\color{blue}
G_{\downarrow}^{}(i\varepsilon'+i\omega)\,
}\,
\biggr]
\nonumber \\
& = \, -\, \frac{U^4}{2}   
\int_{-\infty}^{\infty}
 \frac{d\varepsilon}{2\pi}\, 
\left\{ G_{\uparrow}^{}(i\varepsilon) \right\}^2  
 \chi_{\downarrow\downarrow}^{qp}(i\varepsilon)\   
{\color[rgb]{1,0,1}
\widehat{\partial}_{i\omega}^{+} 
\left[\,
\int_{-\infty}^{\infty}
 \frac{d\varepsilon'}{2\pi}\, 
G_{\downarrow}^{}(i\varepsilon'+i\varepsilon+i\omega)\,
G_{\downarrow}^{}(i\varepsilon'+i\omega)\,
\,\right]
}
  \ = \ 0. 
\label{eq:vertex_4D}
\end{align}
This set contains one singular particle-hole product 
$G_{\uparrow}^{}(i\varepsilon+i\omega) G_{\uparrow}^{}(i\varepsilon)$ 
carrying the same spin $\uparrow$ as that of the external one.   
The contribution of this product  
in the upper two diagrams of  
Fig.\ \ref{fig:vertex_w00w_order_u4_17ph_color_each}
and the contribution of the corresponding internal lines in 
 the lower panel cancel each other out 
in the second line of Eq.\ \eqref{eq:vertex_4D},   
using the {\it generalized\/}  chain rule 
for $\widehat{\partial}_{i\omega}^{+}$.

\newpage

\begin{figure}[h]
 \leavevmode
\begin{minipage}{1\linewidth}
\includegraphics[width=0.3\linewidth]{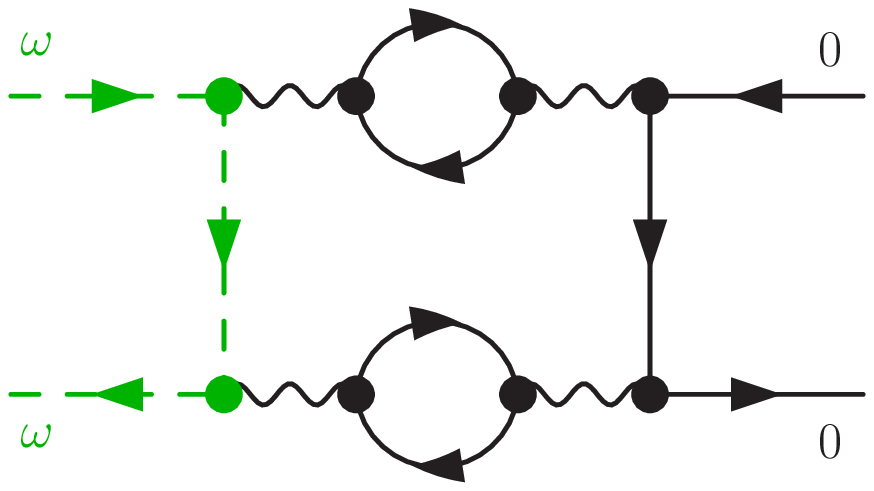}
\rule{0.08\linewidth}{0cm}
\includegraphics[width=0.3\linewidth]{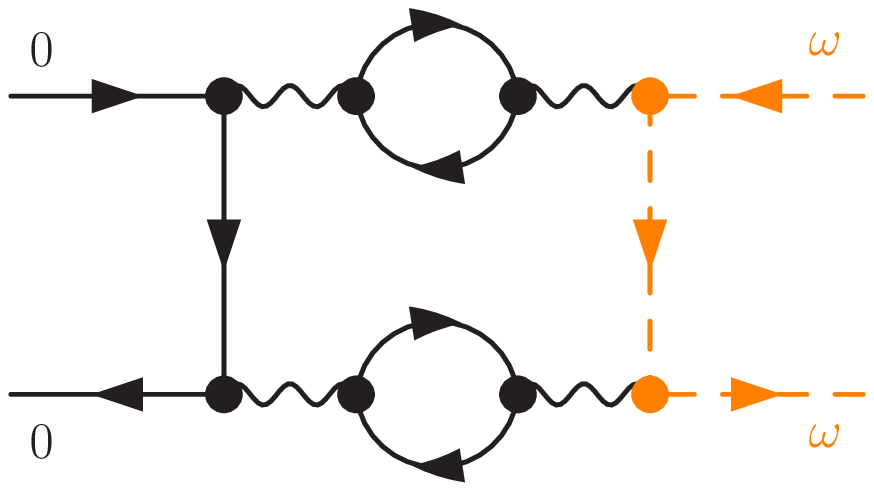}
\end{minipage}

\rule{0cm}{0.5cm}

\begin{minipage}{1\linewidth}
\includegraphics[width=0.3\linewidth]{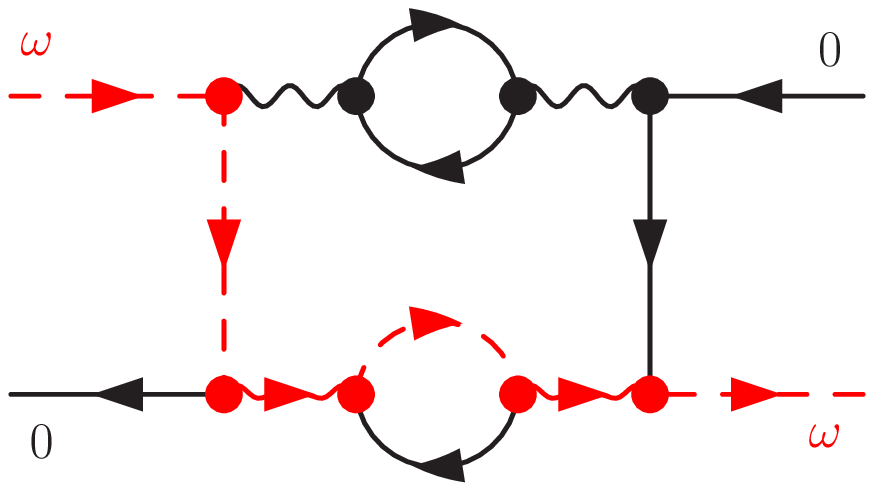}
\rule{0.08\linewidth}{0cm}
\includegraphics[width=0.3\linewidth]{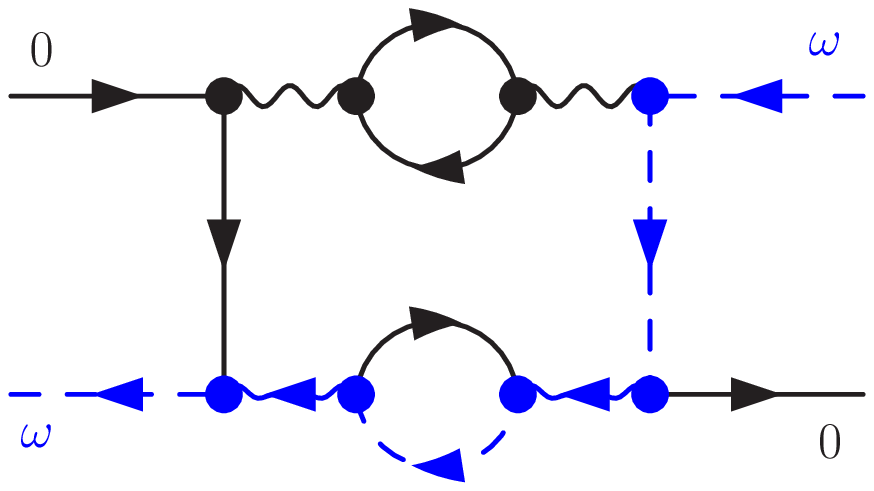}
\end{minipage}
 \caption{
(Color online) 
A set of four diagrams for 
$\Gamma_{\uparrow\uparrow;\uparrow\uparrow}^{(4E)}$,
contribution of which is given in Eq.\ \eqref{eq:vertex_4E}.
}
 \label{fig:vertex_w00w_order_u4_17pp_color_each}
\end{figure}

\begin{figure}[h]
 \leavevmode
\begin{minipage}{1\linewidth}
\includegraphics[width=0.35\linewidth]{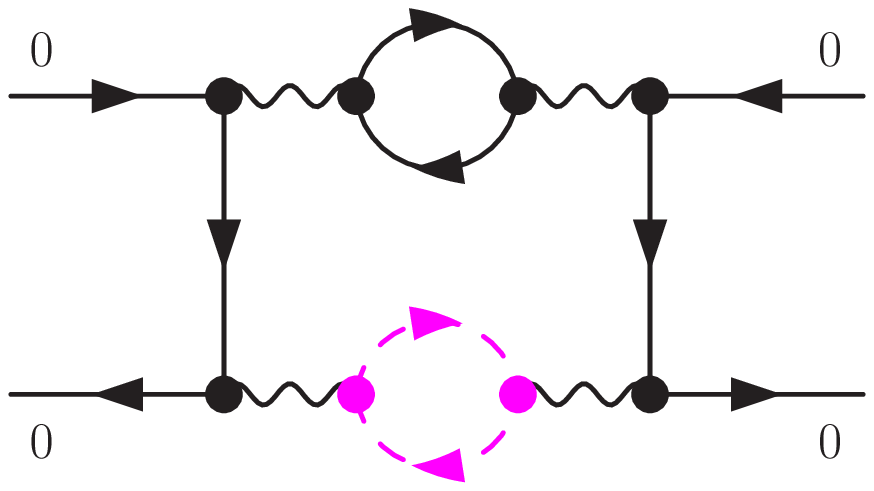}
\end{minipage}
 \caption{
(Color online) 
Schematic picture for 
the total  contribution 
$\widehat{\partial}_{i\omega}^{+} 
\Gamma_{\uparrow\uparrow;\uparrow\uparrow}^{(4E)}$ 
 of the set
shown in Fig.\ \ref{fig:vertex_w00w_order_u4_17pp_color_each}. 
%
}
 \label{fig:vertex_w00w_order_u4_17pp_color_sum}
\end{figure}


Next one  is a set of diagrams 
that include the particle-particle pair. 
Specifically, this is the simplest example 
that the crossing symmetry cancels the singularity 
caused by the particle-particle pair excitation, 
described in Eq.\ \eqref{eq:d+_for_pp_pair_example}. 
Total contribution of the diagrams shown in Fig.\ 
\ref{fig:vertex_w00w_order_u4_17pp_color_each} 
can be rewritten in a total derivative form 
 (see also  Fig.\ \ref{fig:vertex_w00w_order_u4_17pp_color_sum}): 
\begin{align}
& 
\!\!\!\!\!\!\!\!\!\!
\widehat{\partial}_{i\omega}^{+}
\Gamma_{\uparrow\uparrow;\uparrow\uparrow}^{(4E)}(i\omega , 0; 0 , i\omega) 
\nonumber \\
& = \,  \frac{U^4}{2}   \int_{-\infty}^{\infty}
 \frac{d\varepsilon}{2\pi}\  \widehat{\partial}_{i\omega}^{+}
 \biggl[\,
{\color{red}
\chi_{\downarrow\downarrow}^{qp}(i\varepsilon + i\omega)
}\,  
 \chi_{\downarrow\downarrow}^{qp}(i\varepsilon)\,  
{\color{red}
 G_{\uparrow}^{}(i\varepsilon+i\omega) 
} 
\,G_{\uparrow}^{}(-i\varepsilon)
+
{\color{blue} 
\chi_{\downarrow\downarrow}^{qp}(i\varepsilon- i\omega)
}\,  
 \chi_{\downarrow\downarrow}^{qp}(i\varepsilon)\,  
G_{\uparrow}^{}(i\varepsilon)  \,  
{\color{blue}
G_{\uparrow}^{}(i\omega-i\varepsilon)
}
\nonumber \\
& \qquad  \qquad \qquad \qquad \quad 
 - \left\{\chi_{\downarrow\downarrow}^{qp}(i\varepsilon) \right\}^2 
{\color[rgb]{0,0.5,0}
G_{\uparrow}^{}(i\varepsilon+i\omega)\,
} 
G_{\uparrow}^{}(-i\varepsilon)\,
 - \left\{\chi_{\downarrow\downarrow}^{qp}(i\varepsilon) \right\}^2 
G_{\uparrow}^{}(i\varepsilon)\,
{\color[rgb]{1,0.5,0}
G_{\uparrow}^{}(i\omega-i\varepsilon)\, 
}
 \biggr] 
\nonumber \\
& = \,  \frac{U^4}{2}   \int_{-\infty}^{\infty}
 \frac{d\varepsilon}{2\pi}\, 
 G_{\uparrow}^{}(i\varepsilon) \, G_{\uparrow}^{}(-i\varepsilon)\   
\widehat{\partial}_{i\omega}^{+} 
\left[\, 
{\color{red}
\chi_{\downarrow\downarrow}^{qp}(i\varepsilon + i\omega
)}\,  
 \chi_{\downarrow\downarrow}^{qp}(i\varepsilon)\,  
+
{\color{blue}
 \chi_{\downarrow\downarrow}^{qp}(i\varepsilon- i\omega)
}\,  
 \chi_{\downarrow\downarrow}^{qp}(i\varepsilon)\,  
-2  \left\{\chi_{\downarrow\downarrow}^{qp}(i\varepsilon) \right\}^2 
\right]
\nonumber \\
& = \, -\, \frac{U^4}{2}   
\int_{-\infty}^{\infty}
 \frac{d\varepsilon}{2\pi}\, 
 G_{\uparrow}^{}(i\varepsilon) \, G_{\uparrow}^{}(-i\varepsilon)\, 
 \chi_{\downarrow\downarrow}^{qp}(i\varepsilon)\,  
\int_{-\infty}^{\infty}
 \frac{d\varepsilon'}{2\pi}\ 
\widehat{\partial}_{i\omega}^{+} 
\biggl[\,
{\color{red}
G_{\downarrow}^{}(i\varepsilon'+i\varepsilon+i\omega)\,
}
G_{\downarrow}^{}(i\varepsilon')\,
+
G_{\downarrow}^{}(i\varepsilon'+i\varepsilon)\,
{\color{blue}
G_{\downarrow}^{}(i\varepsilon'+i\omega)\,
}\,
\biggr]
\nonumber \\
& = \, -\, \frac{U^4}{2}   
\int_{-\infty}^{\infty}
 \frac{d\varepsilon}{2\pi}\, 
 G_{\uparrow}^{}(i\varepsilon) \, G_{\uparrow}^{}(-i\varepsilon) \, 
 \chi_{\downarrow\downarrow}^{qp}(i\varepsilon)\   
{\color[rgb]{1,0,1}
\widehat{\partial}_{i\omega}^{+} 
\left[\,
\int_{-\infty}^{\infty}
 \frac{d\varepsilon'}{2\pi}\, 
G_{\downarrow}^{}(i\varepsilon'+i\varepsilon+i\omega)\,
G_{\downarrow}^{}(i\varepsilon'+i\omega)\,
\,\right] 
}
 \ = \ 0. 
\label{eq:vertex_4E}
\end{align}
This set contains one singular particle-particle product, 
$ G_{\uparrow}^{}(i\varepsilon+i\omega) G_{\uparrow}^{}(-i\varepsilon)$, or 
$ G_{\uparrow}^{}(i\varepsilon) G_{\uparrow}^{}(i\omega-i\varepsilon)$,
carrying the same spin $\uparrow$ as that of the external one. 
 The contribution of this product  
in the upper two diagrams of  
Fig.\ \ref{fig:vertex_w00w_order_u4_17pp_color_each}
and those in the lower panel cancel each other out 
in the second line of Eq.\ \eqref{eq:vertex_4E},  
using the {\it generalized\/}  chain rule 
for $\widehat{\partial}_{i\omega}^{+}$.



\newpage

\begin{figure}[h]
 \leavevmode
\begin{minipage}{1\linewidth}
\includegraphics[width=0.3\linewidth]{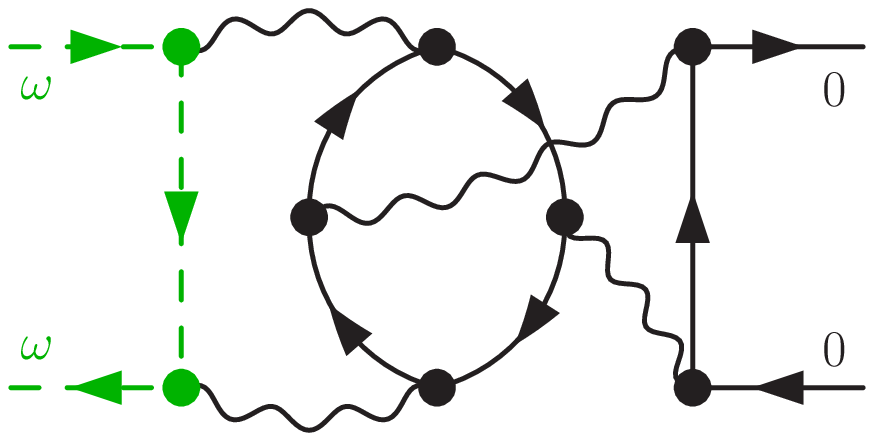}
\rule{0.08\linewidth}{0cm}
\includegraphics[width=0.3\linewidth]{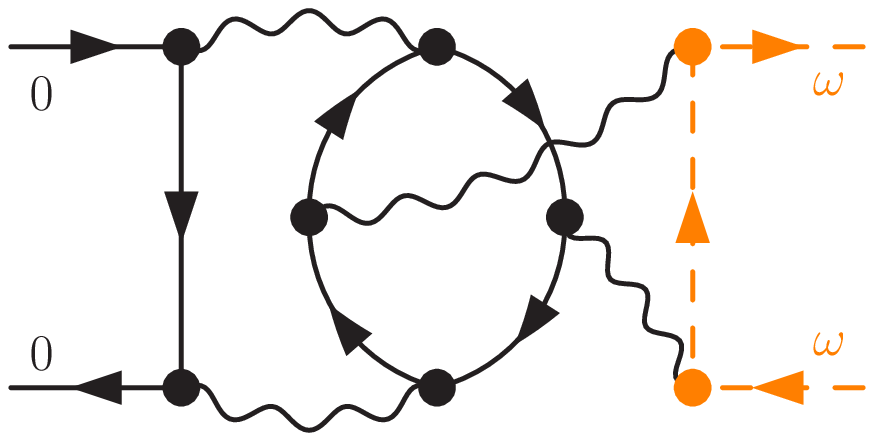}
\end{minipage}

 \rule{0cm}{0.5cm}

\begin{minipage}{1\linewidth}
\includegraphics[width=0.3\linewidth]{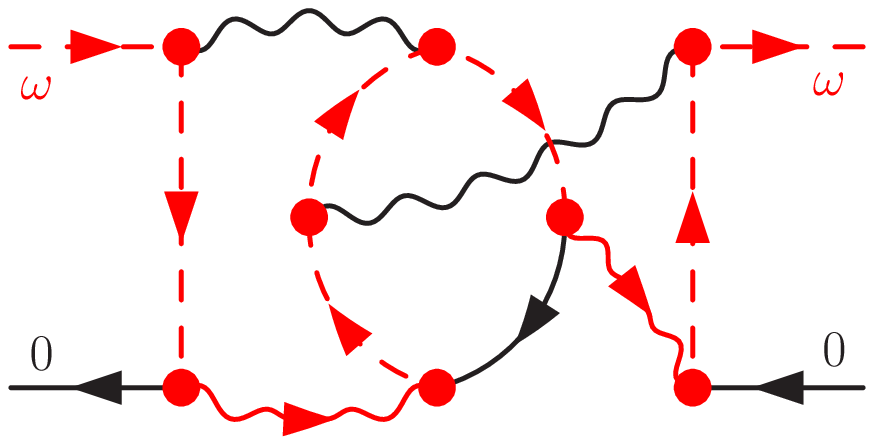}
\rule{0.08\linewidth}{0cm}
\includegraphics[width=0.3\linewidth]{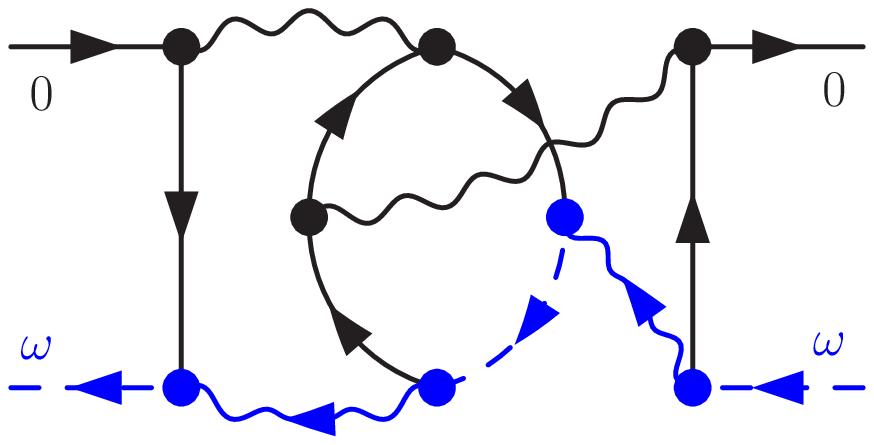}
\end{minipage}
 \caption{
(Color online) 
A set of four diagrams for  
$\Gamma_{\uparrow\uparrow;\uparrow\uparrow}^{(4F)}$,  
contribution of which is given in Eq.\ \eqref{eq:vertex_4F}.
}
 \label{fig:vertex_w00w_order_u4_0a_color_each}
\end{figure}

\begin{figure}[h]
 \leavevmode
\begin{minipage}{1\linewidth}
\includegraphics[width=0.35\linewidth]{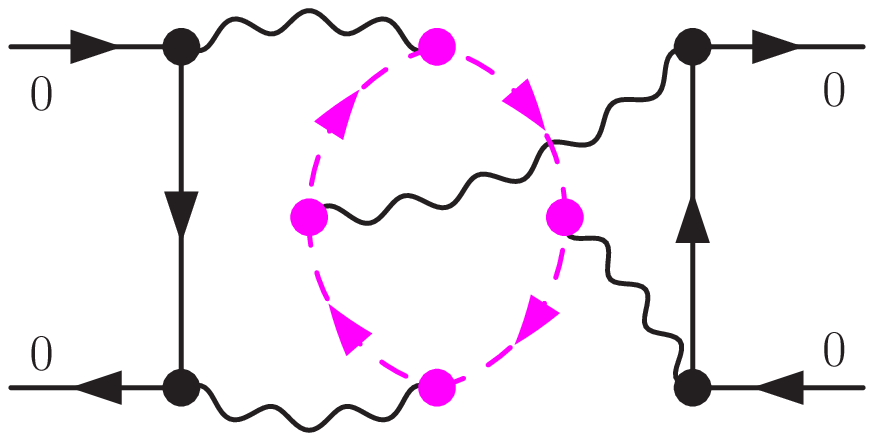}
\end{minipage}
 \caption{
(Color online) 
Schematic picture for 
the total  contribution  
 $\widehat{\partial}_{i\omega}^{+} 
\Gamma_{\uparrow\uparrow;\uparrow\uparrow}^{(4F)}$ 
 of the set
shown in Fig.\ \ref{fig:vertex_w00w_order_u4_0a_color_each}. 
%
}
 \label{fig:vertex_w00w_order_u4_0a_color_sum}
\end{figure}

Total contribution of the diagrams shown in Fig.\ 
\ref{fig:vertex_w00w_order_u4_0a_color_each} 
can be rewritten in a total derivative form 
 (see also  Fig.\ \ref{fig:vertex_w00w_order_u4_0a_color_sum}):  
\begin{align}
& 
\!\!\!\!\!
\widehat{\partial}_{i\omega}^{+}
\Gamma_{\uparrow\uparrow;\uparrow\uparrow}^{(4F)}(i\omega , 0; 0 , i\omega) 
\nonumber \\
& = \,
 - U^4   
\int_{-\infty}^{\infty}\!
\int_{-\infty}^{\infty}\!
\int_{-\infty}^{\infty}\!
 \frac{d\varepsilon\,d\varepsilon_1\,d\varepsilon_2}{(2\pi)^3}\  
\widehat{\partial}_{i\omega}^{+}
 \biggl[\,
G_{\uparrow}^{}(i\varepsilon_1) \,
G_{\downarrow}^{}(i\varepsilon_1+i\varepsilon) \,
G_{\downarrow}^{}(i\varepsilon_1+i\varepsilon_2+i\varepsilon) \,
{\color{blue}
G_{\downarrow}^{}(i\varepsilon+i\omega) \,
}
G_{\downarrow}^{}(i\varepsilon_2+i\varepsilon) \,
G_{\uparrow}^{}(i\varepsilon_2) \,
\nonumber \\
& \qquad \qquad \qquad 
+
{\color{red}
G_{\uparrow}^{}(i\varepsilon_1+i\omega) \,
G_{\downarrow}^{}(i\varepsilon_1+i\varepsilon+i\omega) \,
G_{\downarrow}^{}(i\varepsilon_1+i\varepsilon_2
+i\varepsilon+i\omega) \,
}
G_{\downarrow}^{}(i\varepsilon) \,
{\color{red}
G_{\downarrow}^{}(i\varepsilon_2+i\varepsilon+i\omega) \,
G_{\uparrow}^{}(i\varepsilon_2+i\omega) \,
}
 \nonumber \\
& \qquad \qquad \qquad
- 
{\color[rgb]{0,0.5,0}
G_{\uparrow}^{}(i\varepsilon_1+i\omega) \,
}
G_{\downarrow}^{}(i\varepsilon_1 +i\varepsilon) \,
G_{\downarrow}^{}(i\varepsilon+i\varepsilon_1 +i\varepsilon_2) \,
G_{\downarrow}^{}(i\varepsilon) \,
G_{\downarrow}^{}(i\varepsilon_2+i\varepsilon) \,
G_{\uparrow}^{}(i\varepsilon_2) \,
\nonumber \\
& \qquad \qquad \qquad
- 
G_{\uparrow}^{}(i\varepsilon_1) \,
G_{\downarrow}^{}(i\varepsilon_1 +i\varepsilon) \,
G_{\downarrow}^{}(i\varepsilon+i\varepsilon_1 +i\varepsilon_2) \,
G_{\downarrow}^{}(i\varepsilon) \,
G_{\downarrow}^{}(i\varepsilon_2+i\varepsilon) \,
{\color[rgb]{1,0.5,0}
G_{\uparrow}^{}(i\varepsilon_2+i\omega) \,
}
\,\biggr]
\nonumber \\
& = \ 
 - U^4   
\int_{-\infty}^{\infty}\!
\int_{-\infty}^{\infty}\!
 \frac{d\varepsilon_1\,d\varepsilon_2}{(2\pi)^2} \  
G_{\uparrow}^{}(i\varepsilon_1) \,
G_{\uparrow}^{}(i\varepsilon_2) 
\int_{-\infty}^{\infty}\!
 \frac{d\varepsilon}{2\pi}\  
\widehat{\partial}_{i\omega}^{+}
 \biggl[\,
G_{\downarrow}^{}(i\varepsilon_1+i\varepsilon) \,
G_{\downarrow}^{}(i\varepsilon+i\varepsilon_1 +i\varepsilon_2) \,
{\color{blue}
G_{\downarrow}^{}(i\varepsilon+i\omega) \,
}
G_{\downarrow}^{}(i\varepsilon_2+i\varepsilon) \,
\nonumber \\
& 
+
{\color{red}
G_{\downarrow}^{}(i\varepsilon_1+i\varepsilon+i\omega) \,
G_{\downarrow}^{}(i\varepsilon_1 +i\varepsilon_2+i\varepsilon+i\omega) \,
}
G_{\downarrow}^{}(i\varepsilon) \,
{\color{red}
G_{\downarrow}^{}(i\varepsilon_2+i\varepsilon+i\omega) \,
}
- 2
G_{\downarrow}^{}(i\varepsilon_1 +i\varepsilon) \,
G_{\downarrow}^{}(i\varepsilon+i\varepsilon_1 +i\varepsilon_2) \,
G_{\downarrow}^{}(i\varepsilon) \,
G_{\downarrow}^{}(i\varepsilon_2+i\varepsilon) \,
\,\biggr]
\nonumber \\
& = 
 - U^4\!\!   
\int_{-\infty}^{\infty}\!\!
\int_{-\infty}^{\infty}\!\!
 \frac{d\varepsilon_1 d\varepsilon_2}{(2\pi)^2}  \, 
G_{\uparrow}^{}(i\varepsilon_1) \,
G_{\uparrow}^{}(i\varepsilon_2 ) \ 
{\color[rgb]{1,0,1}
\widehat{\partial}_{i\omega}^{+}
 \left[
\int_{-\infty}^{\infty}\!\!
 \frac{d\varepsilon}{2\pi}\  
G_{\downarrow}^{}(i\varepsilon_1+i\varepsilon+i\omega) \,
G_{\downarrow}^{}(i\varepsilon+i\varepsilon_1 +i\varepsilon_2+i\omega) \,
G_{\downarrow}^{}(i\varepsilon+i\omega) \,
G_{\downarrow}^{}(i\varepsilon_2+i\varepsilon+i\omega) 
\,\right] 
}
\nonumber \\
  &  = \ 0 \;. 
\label{eq:vertex_4F}
\end{align}
The diagram of this set cannot be separated into two parts 
by cutting two internal lines, and thus it has no singular Green's-function product.
To obtain the second line,  the derivative 
with respect to  $\widehat{\partial}_{i\omega}^{+}$ 
is taken for $\omega$'s which are assigned for  the $\uparrow$ spin 
propagators in the vertical direction.
Then,  the remaining contribution  arising from the 
two diagrams in the lower panel of 
Fig.\ \ref{fig:vertex_w00w_order_u4_0a_color_each} 
 is extracted to obtain the third line of  
Eq.\ \eqref{eq:vertex_4F}.

\begin{figure}[h]
 \leavevmode
\begin{minipage}{1\linewidth}
\includegraphics[width=0.3\linewidth]{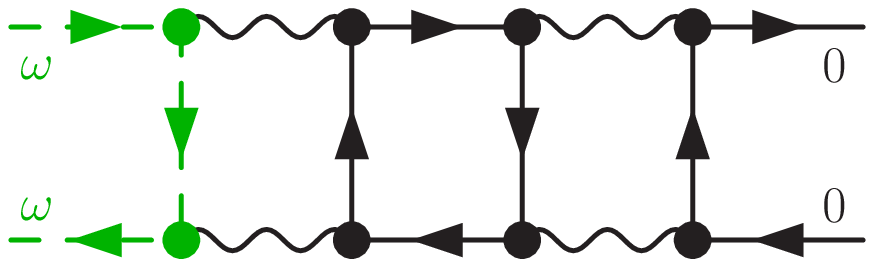}
\rule{0.08\linewidth}{0cm}
\includegraphics[width=0.3\linewidth]{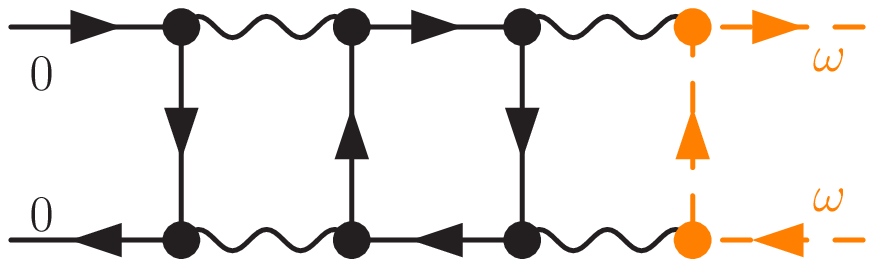}
\end{minipage}

\rule{0cm}{0.5cm}

\begin{minipage}{1\linewidth}
\includegraphics[width=0.3\linewidth]{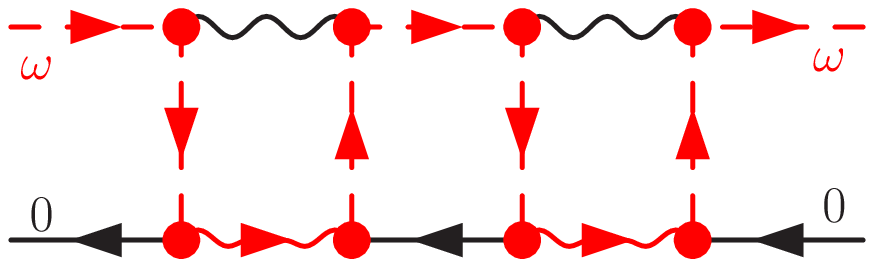}
\rule{0.08\linewidth}{0cm}
\includegraphics[width=0.3\linewidth]{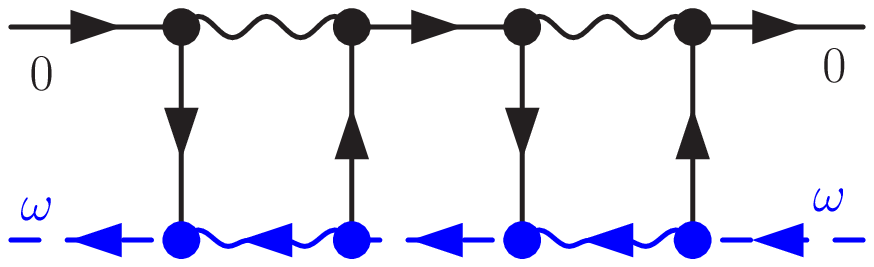}
\end{minipage}
 \caption{
(Color online) 
A set of four diagrams for 
$\Gamma_{\uparrow\uparrow;\uparrow\uparrow}^{(4G)}$,  
contribution of which is given in Eq.\ \eqref{eq:vertex_4G}.
}
 \label{fig:vertex_w00w_order_u4_18ph_color_each}
\end{figure}

\begin{figure}[h]
 \leavevmode
\begin{minipage}{1\linewidth}
\includegraphics[width=0.35\linewidth]{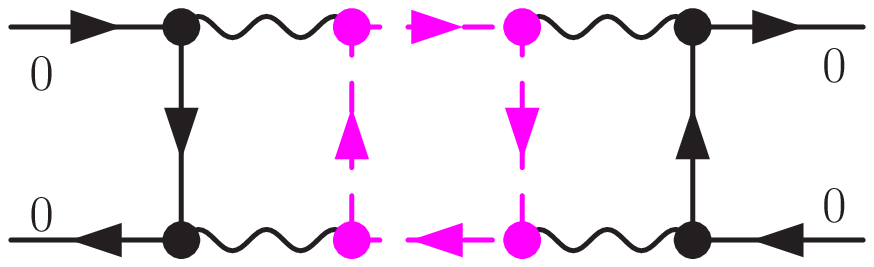}
\end{minipage}
 \caption{
(Color online) 
Schematic picture for 
the total  contribution 
 $\widehat{\partial}_{i\omega}^{+} 
\Gamma_{\uparrow\uparrow;\uparrow\uparrow}^{(4G)}$. 
 of the set
shown in Fig.\ \ref{fig:vertex_w00w_order_u4_18ph_color_each}. 
%
}
 \label{fig:vertex_w00w_order_u4_18ph_color_sum}
\end{figure}


Total contribution of the diagrams shown in Fig.\ 
\ref{fig:vertex_w00w_order_u4_18ph_color_each} 
can be rewritten in a total derivative form 
 (see also  Fig.\ \ref{fig:vertex_w00w_order_u4_18ph_color_sum}): 
\begin{align}
& 
\!\!\!\!\!
\widehat{\partial}_{i\omega}^{+}
\Gamma_{\uparrow\uparrow;\uparrow\uparrow}^{(4G)}(i\omega , 0; 0 , i\omega) 
\nonumber \\
& = \,
 - \frac{U^4}{2}   
\int_{-\infty}^{\infty}\!
\int_{-\infty}^{\infty}\!
\int_{-\infty}^{\infty}\!
 \frac{d\varepsilon\,d\varepsilon_1\,d\varepsilon_2}{(2\pi)^3}\  
\widehat{\partial}_{i\omega}^{+}
 \biggl[\,
G_{\uparrow}^{}(i\varepsilon_1) \,
G_{\downarrow}^{}(i\varepsilon_1+i\varepsilon) \,
G_{\downarrow}^{}(i\varepsilon) \,
{\color{blue}
G_{\downarrow}^{}(i\varepsilon+i\omega) \,
}
G_{\downarrow}^{}(i\varepsilon_2+i\varepsilon) \,
G_{\uparrow}^{}(i\varepsilon_2) \,
\nonumber \\
& \qquad \qquad \qquad 
+
{\color{red}
G_{\uparrow}^{}(i\varepsilon_1+i\omega) \,
G_{\downarrow}^{}(i\varepsilon_1+i\varepsilon+i\omega) \,
G_{\downarrow}^{}(i\varepsilon+i\omega) \,
}
G_{\downarrow}^{}(i\varepsilon) \,
{\color{red}
G_{\downarrow}^{}(i\varepsilon_2+i\varepsilon+i\omega) \,
G_{\uparrow}^{}(i\varepsilon_2+i\omega) \,
}
 \nonumber \\
& \qquad \qquad \qquad
- 
{\color[rgb]{0,0.5,0}
G_{\uparrow}^{}(i\varepsilon_1+i\omega) \,
}
G_{\downarrow}^{}(i\varepsilon_1 +i\varepsilon) \,
\left\{G_{\downarrow}^{}(i\varepsilon) \right\}^2 \,
G_{\downarrow}^{}(i\varepsilon_2+i\varepsilon) \,
G_{\uparrow}^{}(i\varepsilon_2) \,
\nonumber \\
& \qquad \qquad \qquad
- 
G_{\uparrow}^{}(i\varepsilon_1) \,
G_{\downarrow}^{}(i\varepsilon_1 +i\varepsilon) \,
\left\{G_{\downarrow}^{}(i\varepsilon) \right\}^2 \,
G_{\downarrow}^{}(i\varepsilon_2+i\varepsilon) \,
{\color[rgb]{1,0.5,0}
G_{\uparrow}^{}(i\varepsilon_2+i\omega) \,
}
\,\biggr]
\nonumber \\
& = \ 
 - \frac{U^4}{2}   
\int_{-\infty}^{\infty}\!
\int_{-\infty}^{\infty}\!
 \frac{d\varepsilon_1\,d\varepsilon_2}{(2\pi)^2} \  
G_{\uparrow}^{}(i\varepsilon_1) \,
G_{\uparrow}^{}(i\varepsilon_2) 
\int_{-\infty}^{\infty}\!
 \frac{d\varepsilon}{2\pi}\  
\widehat{\partial}_{i\omega}^{+}
 \biggl[\,
G_{\downarrow}^{}(i\varepsilon_1+i\varepsilon) \,
G_{\downarrow}^{}(i\varepsilon) \,
{\color{blue}
G_{\downarrow}^{}(i\varepsilon+i\omega) \,
}
G_{\downarrow}^{}(i\varepsilon_2+i\varepsilon) \,
\nonumber \\
& \qquad \qquad
+
{\color{red}
G_{\downarrow}^{}(i\varepsilon_1+i\varepsilon+i\omega) \,
G_{\downarrow}^{}(i\varepsilon+i\omega) \,
}
G_{\downarrow}^{}(i\varepsilon) \,
{\color{red}
G_{\downarrow}^{}(i\varepsilon_2+i\varepsilon+i\omega) \,
}
- 2
G_{\downarrow}^{}(i\varepsilon_1 +i\varepsilon) \,
\left\{G_{\downarrow}^{}(i\varepsilon) \right\}^2 \,
G_{\downarrow}^{}(i\varepsilon_2+i\varepsilon) \,
\,\biggr]
\nonumber \\
& = \ 
 - \frac{U^4}{2}   
\int_{-\infty}^{\infty}\!
\int_{-\infty}^{\infty}\!
 \frac{d\varepsilon_1\,d\varepsilon_2}{(2\pi)^2} \  
G_{\uparrow}^{}(i\varepsilon_1) \,
G_{\uparrow}^{}(i\varepsilon_2 ) \ 
{\color[rgb]{1,0,1}
\widehat{\partial}_{i\omega}^{+}
 \left[\,
\int_{-\infty}^{\infty}\!
 \frac{d\varepsilon}{2\pi}\  
G_{\downarrow}^{}(i\varepsilon_1+i\varepsilon+i\omega) \,
\left\{G_{\downarrow}^{}(i\varepsilon+i\omega) \right\}^2 \,
G_{\downarrow}^{}(i\varepsilon_2+i\varepsilon+i\omega) 
\,\right] 
}
\nonumber \\
  &  = \ 0 \;. 
\label{eq:vertex_4G}
\end{align}
This set contains one  singular particle-hole pair 
$G_{\downarrow}^{}(i\varepsilon) G_{\downarrow}^{}(i\varepsilon+i\omega)$
 carrying $\omega$ in the horizontal direction.
To obtain the second line,  the derivative 
with respect to  $\widehat{\partial}_{i\omega}^{+}$ 
is taken for $\omega$'s which are assigned for  the $\uparrow$ spin 
 propagators in the vertical direction.
Then,  the remaining contribution  arising from the 
two diagrams in the lower panel of 
Fig.\ \ref{fig:vertex_w00w_order_u4_18ph_color_each}  
 is extracted to obtain the third line of  
Eq.\ \eqref{eq:vertex_4G}.

\newpage

\begin{figure}[h]
 \leavevmode
\begin{minipage}{1\linewidth}
\includegraphics[width=0.3\linewidth]{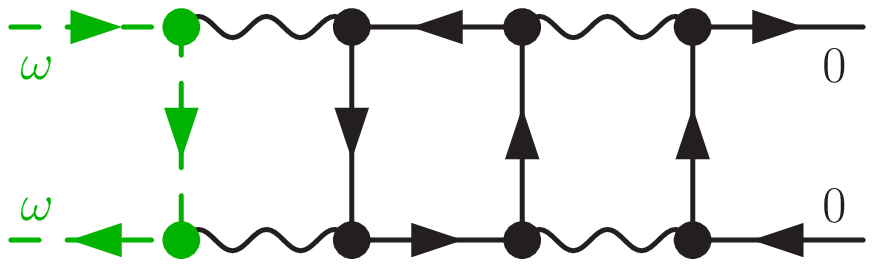}
\rule{0.08\linewidth}{0cm}
\includegraphics[width=0.3\linewidth]{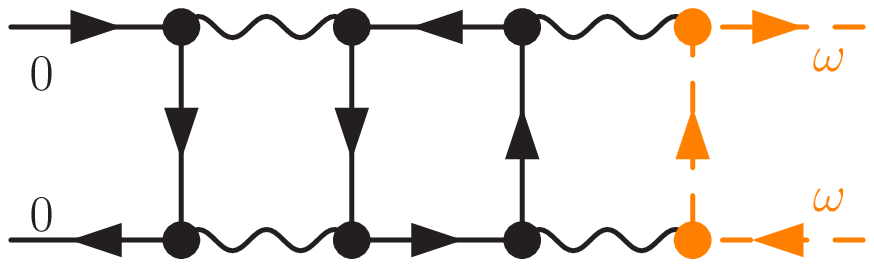}
\end{minipage}

 \rule{0cm}{0.5cm}

\begin{minipage}{1\linewidth}
\includegraphics[width=0.3\linewidth]{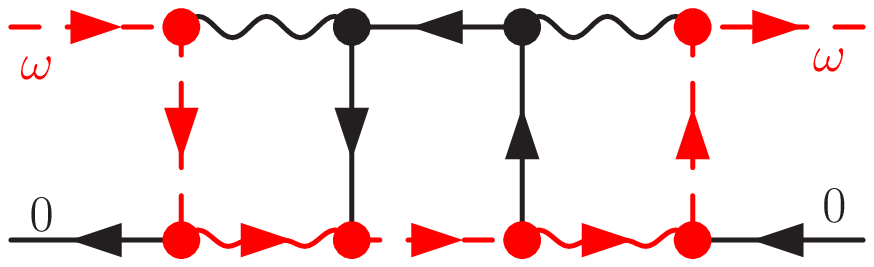}
\rule{0.08\linewidth}{0cm}
\includegraphics[width=0.3\linewidth]{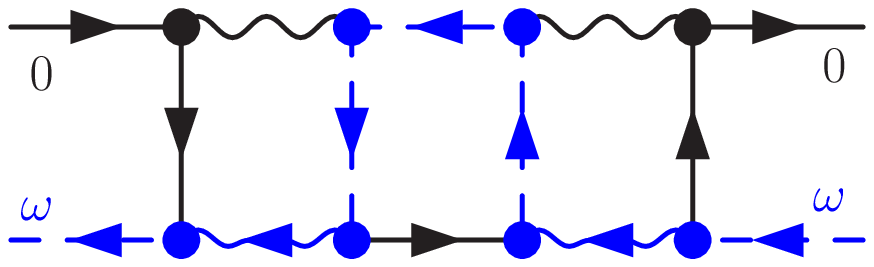}
\end{minipage}
 \caption{
(Color online) 
A set of four diagrams for 
$\Gamma_{\uparrow\uparrow;\uparrow\uparrow}^{(4H)}$,  
contribution of which is given in Eq.\ \eqref{eq:vertex_4H}.
}
 \label{fig:vertex_w00w_order_u4_18pp_color_each}
\end{figure}

\begin{figure}[h]
 \leavevmode
\begin{minipage}{1\linewidth}
\includegraphics[width=0.35\linewidth]{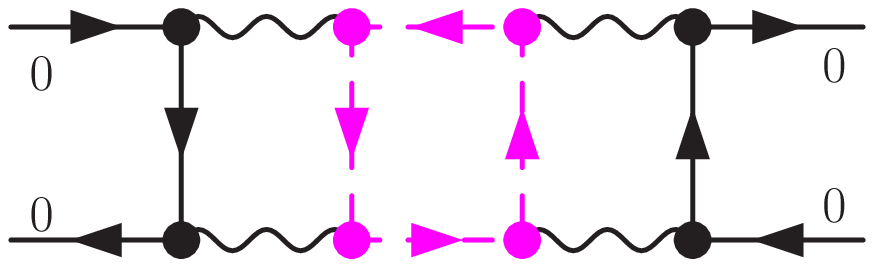}
\end{minipage}

 \caption{
(Color online) 
Sum of the derivative 
$\widehat{\partial}_{i\omega}^{+} 
\Gamma_{\uparrow\uparrow;\uparrow\uparrow}^{(4H)}$ 
and the related other three.
Schematic picture for 
the total  contribution 
 $\widehat{\partial}_{i\omega}^{+} \Gamma_{\uparrow\uparrow;\uparrow\uparrow}^{(4H)}$ 
 of the set
shown in Fig.\ \ref{fig:vertex_w00w_order_u4_18pp_color_each}. 
%
}
 \label{fig:vertex_w00w_order_u4_18pp_color_sum}
\end{figure}

Total contribution of the diagrams shown in Fig.\ 
\ref{fig:vertex_w00w_order_u4_18pp_color_each} 
can be rewritten in a total derivative form 
 (see also  Fig.\ \ref{fig:vertex_w00w_order_u4_18pp_color_sum}): 
\begin{align}
& 
\!\!\!\!\!
\widehat{\partial}_{i\omega}^{+}
\Gamma_{\uparrow\uparrow;\uparrow\uparrow}^{(4H)}(i\omega , 0; 0 , i\omega) 
\nonumber \\
& = \,
 - \frac{U^4}{2}   
\int_{-\infty}^{\infty}\!
\int_{-\infty}^{\infty}\!
\int_{-\infty}^{\infty}\!
 \frac{d\varepsilon\,d\varepsilon_1\,d\varepsilon_2}{(2\pi)^3}\  
\widehat{\partial}_{i\omega}^{+}
 \biggl[\,
G_{\uparrow}^{}(i\varepsilon_1) \,
G_{\downarrow}^{}(i\varepsilon) \,
{\color{blue}
G_{\downarrow}^{}(i\varepsilon -i\varepsilon_1+i\omega) \,
G_{\downarrow}^{}(i\varepsilon +i\omega) \,
G_{\downarrow}^{}(i\varepsilon -i\varepsilon_2 +i\omega) \,
}
G_{\uparrow}^{}(i\varepsilon_2) \,
\nonumber \\
& \qquad \qquad \qquad 
+
{\color{red}
G_{\uparrow}^{}(i\varepsilon_1+i\omega) \,
G_{\downarrow}^{}(i\varepsilon +i\omega) \,
}
G_{\downarrow}^{}(i\varepsilon -i\varepsilon_1) \,
G_{\downarrow}^{}(i\varepsilon) \,
G_{\downarrow}^{}(i\varepsilon -\varepsilon_2) \,
{\color{red}
G_{\uparrow}^{}(i\varepsilon_2 +i\omega) \,
}
 \nonumber \\
& \qquad \qquad \qquad
- 
{\color[rgb]{0,0.5,0}
G_{\uparrow}^{}(i\varepsilon_1+i\omega) \,
}
G_{\downarrow}^{}(i\varepsilon -i\varepsilon_1) \,
\left\{G_{\downarrow}^{}(i\varepsilon) \right\}^2 \,
G_{\downarrow}^{}(i\varepsilon - i\varepsilon_2) \,
G_{\uparrow}^{}(i\varepsilon_2) \,
\nonumber \\
& \qquad \qquad \qquad
- 
G_{\uparrow}^{}(i\varepsilon_1) \,
G_{\downarrow}^{}(i\varepsilon -i\varepsilon_1) \,
\left\{G_{\downarrow}^{}(i\varepsilon) \right\}^2 \,
G_{\downarrow}^{}(i\varepsilon - i\varepsilon_2) \,
{\color[rgb]{1,0.5,0}
G_{\uparrow}^{}(i\varepsilon_2 +i\omega) \,
}
\,\biggr]
\nonumber \\
& = \ 
 - \frac{U^4}{2}   
\int_{-\infty}^{\infty}\!
\int_{-\infty}^{\infty}\!
 \frac{d\varepsilon_1\,d\varepsilon_2}{(2\pi)^2} \  
G_{\uparrow}^{}(i\varepsilon_1) \,
G_{\uparrow}^{}(i\varepsilon_2) 
\int_{-\infty}^{\infty}\!
 \frac{d\varepsilon}{2\pi}\  
\widehat{\partial}_{i\omega}^{+}
 \biggl[\,
G_{\downarrow}^{}(i\varepsilon) \,
{\color{blue}
G_{\downarrow}^{}(i\varepsilon-i\varepsilon_1+i\omega) \,
G_{\downarrow}^{}(i\varepsilon+i\omega) \,
G_{\downarrow}^{}(i\varepsilon-i\varepsilon_2+i\omega) \,
}
\nonumber \\
& \qquad \qquad
+
{\color{red}
G_{\downarrow}^{}(i\varepsilon+i\omega) \,
}
G_{\downarrow}^{}(i\varepsilon-i\varepsilon_1) \,
G_{\downarrow}^{}(i\varepsilon) \,
G_{\downarrow}^{}(i\varepsilon-i\varepsilon_2) \,
- 2
G_{\downarrow}^{}(i\varepsilon- i\varepsilon_1 ) \,
\left\{G_{\downarrow}^{}(i\varepsilon) \right\}^2 \,
G_{\downarrow}^{}(i\varepsilon - i\varepsilon_2) \,
\,\biggr]
\nonumber \\
& = \ 
 - \frac{U^4}{2}   
\int_{-\infty}^{\infty}\!
\int_{-\infty}^{\infty}\!
 \frac{d\varepsilon_1\,d\varepsilon_2}{(2\pi)^2} \  
G_{\uparrow}^{}(i\varepsilon_1) \,
G_{\uparrow}^{}(i\varepsilon_2 ) \ 
{\color[rgb]{1,0,1}
\widehat{\partial}_{i\omega}^{+}
 \left[\,
\int_{-\infty}^{\infty}\!
 \frac{d\varepsilon}{2\pi}\  
G_{\downarrow}^{}(i\varepsilon - i\varepsilon_1 +i\omega) \,
\left\{G_{\downarrow}^{}(i\varepsilon+i\omega) \right\}^2 \,
G_{\downarrow}^{}(i\varepsilon- i\varepsilon_2+i\omega) 
\,\right] 
}
\nonumber \\
  &  = \ 0 \;. 
\label{eq:vertex_4H}
\end{align}
This set also contains one  singular particle-hole pair 
$G_{\downarrow}^{}(i\varepsilon) G_{\downarrow}^{}(i\varepsilon+i\omega)$
 carrying $\omega$ in the horizontal direction.
To obtain the second line,  the derivative 
with respect to  $\widehat{\partial}_{i\omega}^{+}$ 
is taken for $\omega$'s which are assigned for  the $\uparrow$ spin 
propagators in the vertical direction.
Then,  the remaining contribution  arising from the 
two diagrams in the lower panel of 
Fig.\ \ref{fig:vertex_w00w_order_u4_18pp_color_each} 
 is extracted to obtain the third line of  Eq.\ \eqref{eq:vertex_4H}.


\newpage

\begin{figure}[h]
 \leavevmode
\begin{minipage}{1\linewidth}
\includegraphics[width=0.3\linewidth]{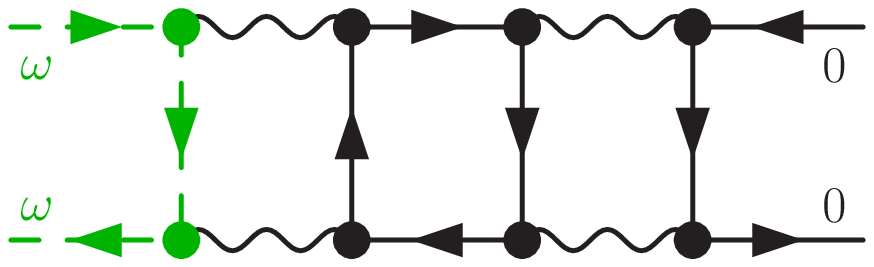}
\rule{0.08\linewidth}{0cm}
\includegraphics[width=0.3\linewidth]{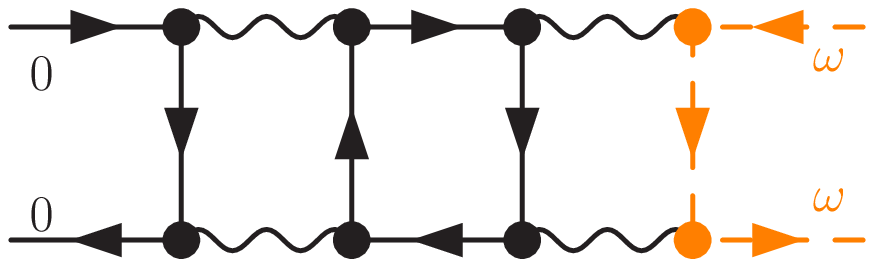}
\end{minipage}

 \rule{0cm}{0.5cm}

\begin{minipage}{1\linewidth}
\includegraphics[width=0.3\linewidth]{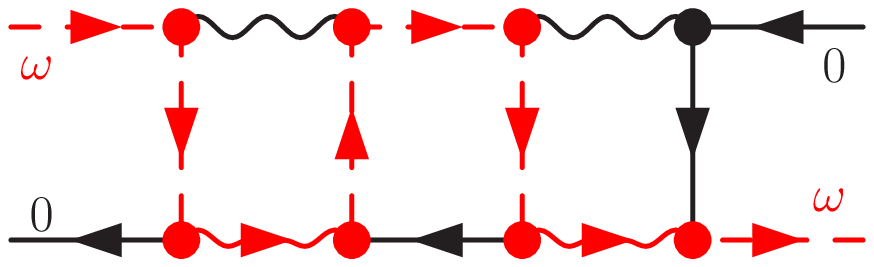}
\rule{0.08\linewidth}{0cm}
\includegraphics[width=0.3\linewidth]{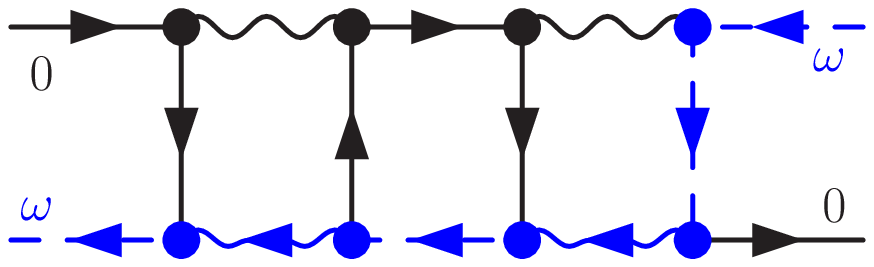}
\end{minipage}
 \caption{
(Color online) 
A set of four diagrams for 
$\Gamma_{\uparrow\uparrow;\uparrow\uparrow}^{(4I)}$, 
contribution of which is given in Eq.\ \eqref{eq:vertex_4I}.
}
 \label{fig:vertex_w00w_order_u4_19_color_each}
\end{figure}

\begin{figure}[h]
 \leavevmode
\begin{minipage}{1\linewidth}
\includegraphics[width=0.35\linewidth]{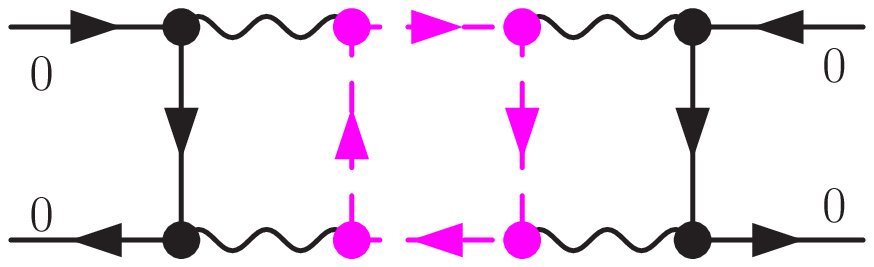}
\end{minipage}
 \caption{
(Color online) 
Schematic picture for 
the total  contribution  
 $\widehat{\partial}_{i\omega}^{+} \Gamma_{\uparrow\uparrow;\uparrow\uparrow}^{(4I)}$ 
 of the set
shown in Fig.\ \ref{fig:vertex_w00w_order_u4_19_color_each} 
%
}
 \label{fig:vertex_w00w_order_u4_19_color_sum}
\end{figure}

Total contribution of the diagrams shown in Fig.\ 
\ref{fig:vertex_w00w_order_u4_19_color_each} 
can be rewritten in a total derivative form 
 (see also  Fig.\ \ref{fig:vertex_w00w_order_u4_19_color_sum}):  
\begin{align}
& 
\!\!\!\!\!
\widehat{\partial}_{i\omega}^{+}
\Gamma_{\uparrow\uparrow;\uparrow\uparrow}^{(4I)}(i\omega , 0; 0 , i\omega) 
\nonumber \\
& = \,
 - U^4 
\int_{-\infty}^{\infty}\!
\int_{-\infty}^{\infty}\!
\int_{-\infty}^{\infty}\!
 \frac{d\varepsilon\,d\varepsilon_1\,d\varepsilon_2}{(2\pi)^3}\  
\widehat{\partial}_{i\omega}^{+}
 \biggl[\,
G_{\uparrow}^{}(i\varepsilon_1) \,
G_{\downarrow}^{}(i\varepsilon) \,
G_{\downarrow}^{}(i\varepsilon +i\varepsilon_1) \,
G_{\downarrow}^{}(i\varepsilon -i\varepsilon_2) \,
{\color{blue}
G_{\downarrow}^{}(i\varepsilon+i\omega) \,
G_{\uparrow}^{}(i\varepsilon_2+i\omega) \,
}
\nonumber \\
& \qquad \qquad \qquad 
+
{\color{red}
G_{\uparrow}^{}(i\varepsilon_1+i\omega) \,
G_{\downarrow}^{}(i\varepsilon +i\omega) \,
G_{\downarrow}^{}(i\varepsilon +i\varepsilon_1+i\omega) \,
G_{\downarrow}^{}(i\varepsilon -\varepsilon_2+i\omega) \,
}
G_{\downarrow}^{}(i\varepsilon) \,
G_{\uparrow}^{}(i\varepsilon_2) \,
 \nonumber \\
& \qquad \qquad \qquad
- 
{\color[rgb]{0,0.5,0}
G_{\uparrow}^{}(i\varepsilon_1+i\omega) \,
}
G_{\downarrow}^{}(i\varepsilon +i\varepsilon_1) \,
G_{\downarrow}^{}(i\varepsilon - i\varepsilon_2) \,
\left\{G_{\downarrow}^{}(i\varepsilon) \right\}^2 \,
G_{\uparrow}^{}(i\varepsilon_2) \,
\nonumber \\
& \qquad \qquad \qquad
- 
G_{\uparrow}^{}(i\varepsilon_1) \,
G_{\downarrow}^{}(i\varepsilon +i\varepsilon_1) \,
G_{\downarrow}^{}(i\varepsilon - i\varepsilon_2) \,
\left\{G_{\downarrow}^{}(i\varepsilon) \right\}^2 \,
{\color[rgb]{1,0.5,0}
G_{\uparrow}^{}(i\varepsilon_2 +i\omega) \,
}
\,\biggr]
\nonumber \\
& = \ 
 - U^4   
\int_{-\infty}^{\infty}\!
\int_{-\infty}^{\infty}\!
 \frac{d\varepsilon_1\,d\varepsilon_2}{(2\pi)^2} \  
G_{\uparrow}^{}(i\varepsilon_1) \,
G_{\uparrow}^{}(i\varepsilon_2) 
\int_{-\infty}^{\infty}\!
 \frac{d\varepsilon}{2\pi}\  
\widehat{\partial}_{i\omega}^{+}
 \biggl[\,
G_{\downarrow}^{}(i\varepsilon) \,
G_{\downarrow}^{}(i\varepsilon+i\varepsilon_1) \,
G_{\downarrow}^{}(i\varepsilon-i\varepsilon_2) \,
{\color{blue}
G_{\downarrow}^{}(i\varepsilon+i\omega) \,
}
\nonumber \\
& \qquad \qquad
+
{\color{red}
G_{\downarrow}^{}(i\varepsilon+i\omega) \,
G_{\downarrow}^{}(i\varepsilon+i\varepsilon_1+i\omega) \,
G_{\downarrow}^{}(i\varepsilon-i\varepsilon_2+i\omega) \,
}
G_{\downarrow}^{}(i\varepsilon) \,
- 2
G_{\downarrow}^{}(i\varepsilon + i\varepsilon_1 ) \,
\left\{G_{\downarrow}^{}(i\varepsilon) \right\}^2 \,
G_{\downarrow}^{}(i\varepsilon - i\varepsilon_2) \,
\,\biggr]
\nonumber \\
& = \ 
 - U^4   
\int_{-\infty}^{\infty}\!
\int_{-\infty}^{\infty}\!
 \frac{d\varepsilon_1\,d\varepsilon_2}{(2\pi)^2} \  
G_{\uparrow}^{}(i\varepsilon_1) \,
G_{\uparrow}^{}(i\varepsilon_2 ) \ 
{\color[rgb]{1,0,1}
\widehat{\partial}_{i\omega}^{+}
 \left[\,
\int_{-\infty}^{\infty}\!
 \frac{d\varepsilon}{2\pi}\  
G_{\downarrow}^{}(i\varepsilon + i\varepsilon_1 +i\omega) \,
\left\{G_{\downarrow}^{}(i\varepsilon+i\omega) \right\}^2 \,
G_{\downarrow}^{}(i\varepsilon- i\varepsilon_2+i\omega) 
\,\right] 
}
\nonumber \\
  &  = \ 0 \;. 
\label{eq:vertex_4I}
\end{align}
This set also contains one  singular particle-hole pair 
$G_{\downarrow}^{}(i\varepsilon) G_{\downarrow}^{}(i\varepsilon+i\omega)$
 carrying $\omega$ in the horizontal direction.
To obtain the second line,  the derivative 
with respect to  $\widehat{\partial}_{i\omega}^{+}$ 
is taken for $\omega$'s which are assigned for  the $\uparrow$ spin 
propagators in the vertical direction.
Then,  the remaining contribution  arising from the 
two diagrams in the lower panel of 
Fig.\ \ref{fig:vertex_w00w_order_u4_19_color_each} 
 is extracted to obtain the third line of  
Eq.\ \eqref{eq:vertex_4I}.


\newpage


\begin{figure}[h]
 \leavevmode
\begin{minipage}{1\linewidth}
\includegraphics[width=0.3\linewidth]{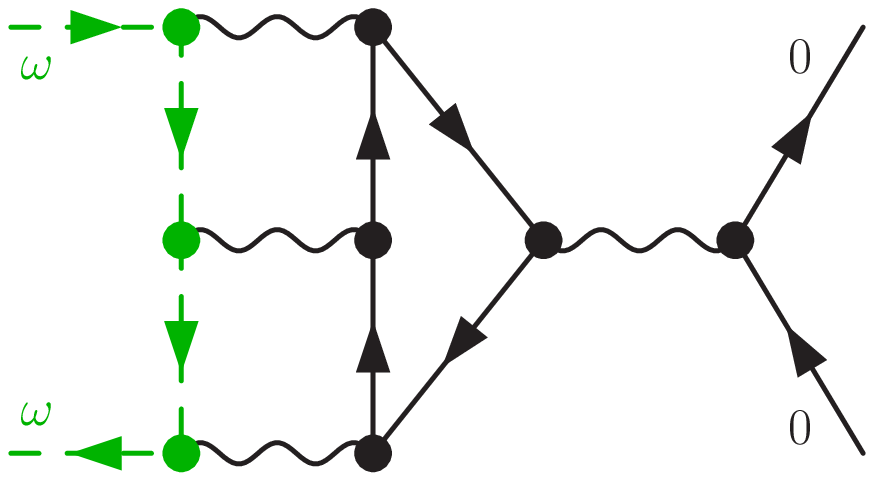}
\rule{0.08\linewidth}{0cm}
\includegraphics[width=0.3\linewidth]{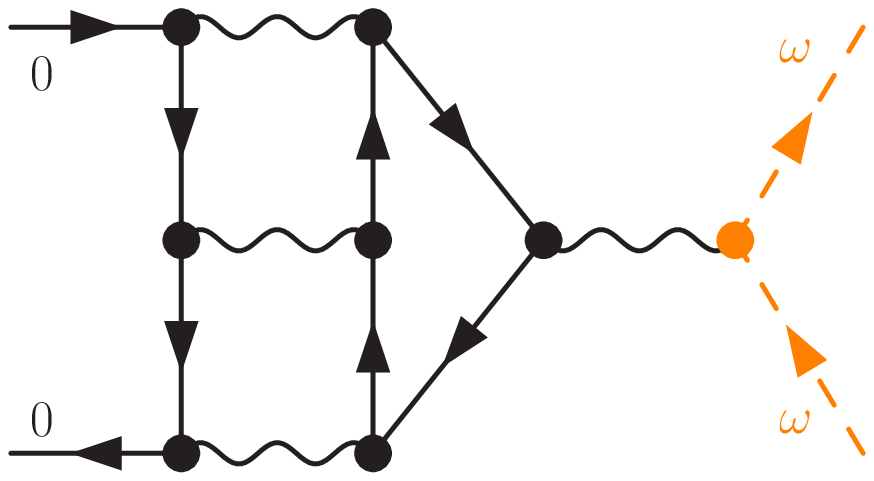}
\end{minipage}

 \rule{0cm}{0.5cm}

\begin{minipage}{1\linewidth}
\includegraphics[width=0.3\linewidth]{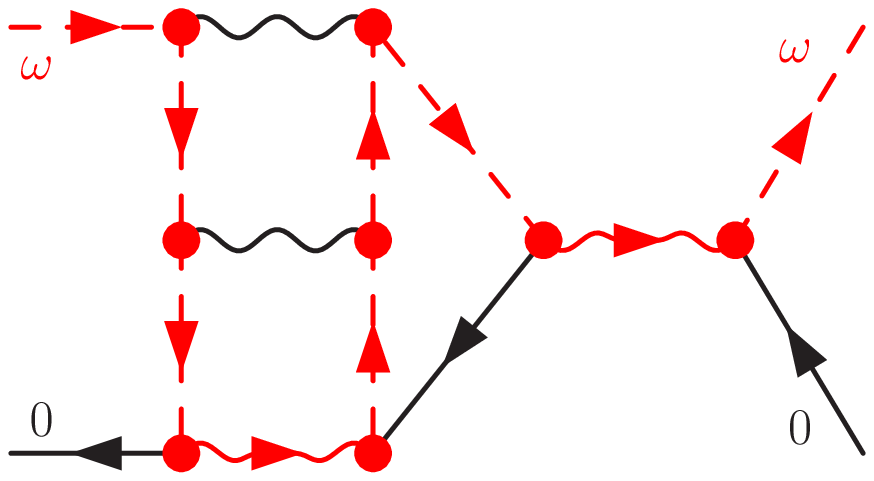}
\rule{0.08\linewidth}{0cm}
\includegraphics[width=0.3\linewidth]{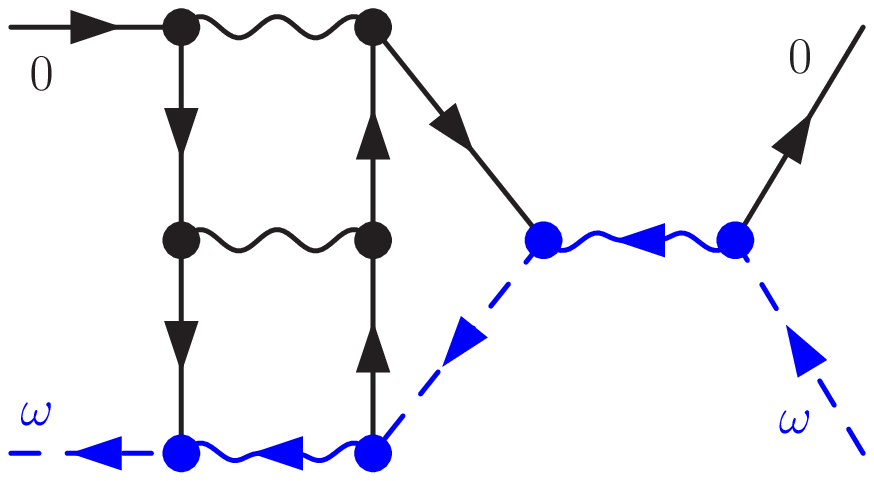}
\end{minipage}
 \caption{
(Color online) 
A set of four diagrams for 
$\Gamma_{\uparrow\uparrow;\uparrow\uparrow}^{(4J)}$,
contribution of which is given in Eq.\ \eqref{eq:vertex_4J}.
}
 \label{fig:vertex_w00w_order_u4_6phL_color_each}
\end{figure}

\begin{figure}[h]
 \leavevmode
\begin{minipage}{1\linewidth}
\includegraphics[width=0.35\linewidth]{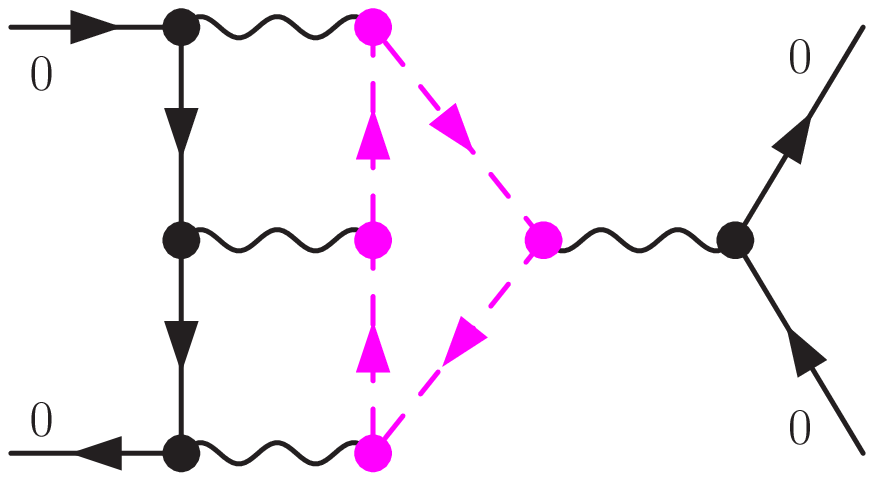}
\end{minipage}
 \caption{
(Color online) 
Schematic picture for 
the total  contribution 
 $\widehat{\partial}_{i\omega}^{+} 
\Gamma_{\uparrow\uparrow;\uparrow\uparrow}^{(4J)}$ 
 of the set
shown in Fig.\ \ref{fig:vertex_w00w_order_u4_6phL_color_each}.
%
}
 \label{fig:vertex_w00w_order_u4_6phL_color_sum}
\end{figure}


Total contribution of the diagrams shown in Fig.\ 
\ref{fig:vertex_w00w_order_u4_6phL_color_each} 
can be rewritten in a total derivative form 
 (see also  Fig.\ \ref{fig:vertex_w00w_order_u4_6phL_color_sum}):  
\begin{align}
& 
\!\!\!\!\!
\widehat{\partial}_{i\omega}^{+}
\Gamma_{\uparrow\uparrow;\uparrow\uparrow}^{(4J)}(i\omega , 0; 0 , i\omega) 
\nonumber \\
& = \,
 - U^4   
\int_{-\infty}^{\infty}\!
\int_{-\infty}^{\infty}\!
\int_{-\infty}^{\infty}\!
 \frac{d\varepsilon\,d\varepsilon_1\,d\varepsilon_2}{(2\pi)^3}\  
\widehat{\partial}_{i\omega}^{+}
 \biggl[\,
G_{\uparrow}^{}(i\varepsilon_1) \,
G_{\uparrow}^{}(i\varepsilon_2) \,
G_{\downarrow}^{}(i\varepsilon+i\varepsilon_1) \,
G_{\downarrow}^{}(i\varepsilon+i\varepsilon_2) \,
G_{\downarrow}^{}(i\varepsilon) \,
{\color[rgb]{0,0,1}
G_{\downarrow}^{}(i\varepsilon+i\omega) \,
}
\nonumber \\
& \qquad \qquad \qquad  \qquad \qquad
+
{\color[rgb]{1,0,0}
G_{\uparrow}^{}(i\varepsilon_1+i\omega) \,
G_{\uparrow}^{}(i\varepsilon_2+i\omega) \,
G_{\downarrow}^{}(i\varepsilon+i\varepsilon_1+i\omega) \,
G_{\downarrow}^{}(i\varepsilon+i\varepsilon_2+i\omega) \,
G_{\downarrow}^{}(i\varepsilon+i\omega) 
}\,
G_{\downarrow}^{}(i\varepsilon) \,
 \nonumber \\
& \qquad \qquad \qquad \qquad \qquad
- 
{\color[rgb]{0,0.5,0}
G_{\uparrow}^{}(i\varepsilon_1+i\omega) \,
G_{\uparrow}^{}(i\varepsilon_2+i\omega) \,
}
G_{\downarrow}^{}(i\varepsilon +i\varepsilon_1) \,
G_{\downarrow}^{}(i\varepsilon +i\varepsilon_2) \,
G_{\downarrow}^{}(i\varepsilon) \,
G_{\downarrow}^{}(i\varepsilon) \,
\nonumber \\
& \qquad \qquad \qquad  \qquad \qquad
- 
G_{\uparrow}^{}(i\varepsilon_1) \,
G_{\uparrow}^{}(i\varepsilon_2) \,
G_{\downarrow}^{}(i\varepsilon +i\varepsilon_1) \,
G_{\downarrow}^{}(i\varepsilon +i\varepsilon_2) \,
G_{\downarrow}^{}(i\varepsilon) \,
G_{\downarrow}^{}(i\varepsilon) \,
\,\biggr]
\nonumber \\
& = \ 
 - U^4   
\int_{-\infty}^{\infty}\!
\int_{-\infty}^{\infty}\!
 \frac{d\varepsilon_1\,d\varepsilon_2}{(2\pi)^2} \  
G_{\uparrow}^{}(i\varepsilon_1) \,
G_{\uparrow}^{}(i\varepsilon_2) 
\int_{-\infty}^{\infty}\!
 \frac{d\varepsilon}{2\pi}\  
\widehat{\partial}_{i\omega}^{+}
 \biggl[\,
G_{\downarrow}^{}(i\varepsilon+i\varepsilon_1) \,
G_{\downarrow}^{}(i\varepsilon+i\varepsilon_2) \,
G_{\downarrow}^{}(i\varepsilon) \,
{\color[rgb]{0,0,1}
G_{\downarrow}^{}(i\varepsilon+i\omega) \,
}
\nonumber \\
& 
 \qquad \qquad \qquad
+
{\color[rgb]{1,0,0}
G_{\downarrow}^{}(i\varepsilon+i\varepsilon_1+i\omega) \,
G_{\downarrow}^{}(i\varepsilon+i\varepsilon_2+i\omega) \,
G_{\downarrow}^{}(i\varepsilon+i\omega) 
}\,
G_{\downarrow}^{}(i\varepsilon) \,
- 2
G_{\downarrow}^{}(i\varepsilon +i\varepsilon_1) \,
G_{\downarrow}^{}(i\varepsilon +i\varepsilon_2) \,
G_{\downarrow}^{}(i\varepsilon) \,
G_{\downarrow}^{}(i\varepsilon) \,
 \,
\,\biggr]
\nonumber \\
& = 
 - U^4\!\!   
\int_{-\infty}^{\infty}\!\!
\int_{-\infty}^{\infty}\!\!
 \frac{d\varepsilon_1 d\varepsilon_2}{(2\pi)^2}  \, 
G_{\uparrow}^{}(i\varepsilon_1) \,
G_{\uparrow}^{}(i\varepsilon_2 ) \ 
{\color[rgb]{1,0,1}
\widehat{\partial}_{i\omega}^{+}
 \left[
\int_{-\infty}^{\infty}\!\!
 \frac{d\varepsilon}{2\pi}\  
G_{\downarrow}^{}(i\varepsilon+i\varepsilon_1+i\omega) \,
G_{\downarrow}^{}(i\varepsilon+i\varepsilon_2+i\omega) \,
\left\{G_{\downarrow}^{}(i\varepsilon+i\omega) \right\}^2
\,\right] 
}
\nonumber \\
  &  = \ 0 \;. 
\label{eq:vertex_4J}
\end{align}
To obtain the second line,  the derivative 
with respect to  $\widehat{\partial}_{i\omega}^{+}$ 
is taken for $\omega$'s which are assigned for  the $\uparrow$ spin 
propagators in the vertical direction.
The remaining contribution  arising from the two diagrams 
in the lower panel of Fig.\ \ref{fig:vertex_w00w_order_u4_6phL_color_each}  
 is extracted by applying the {\it generalized\/} chain rule for the product 
$G_{\downarrow}^{}(i\varepsilon) G_{\downarrow}^{}(i\varepsilon+i\omega)$.


\newpage

\begin{figure}[h]
 \leavevmode
\begin{minipage}{1\linewidth}
\includegraphics[width=0.3\linewidth]{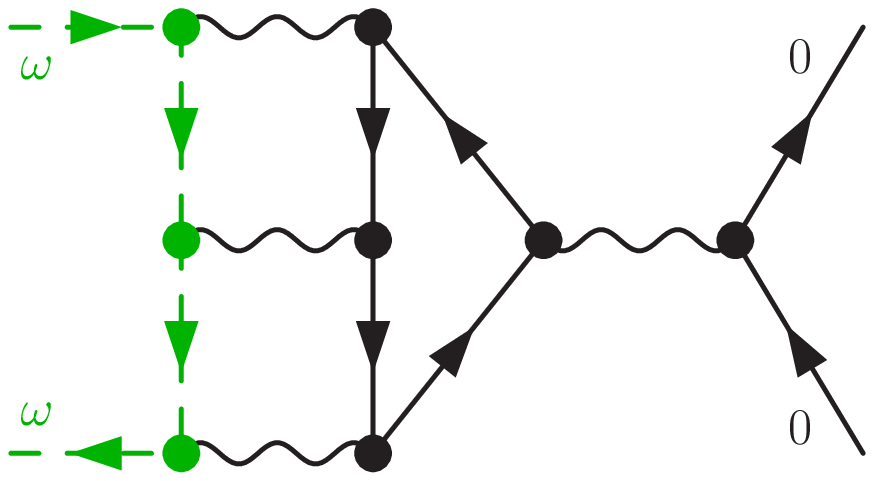}
\rule{0.08\linewidth}{0cm}
\includegraphics[width=0.3\linewidth]{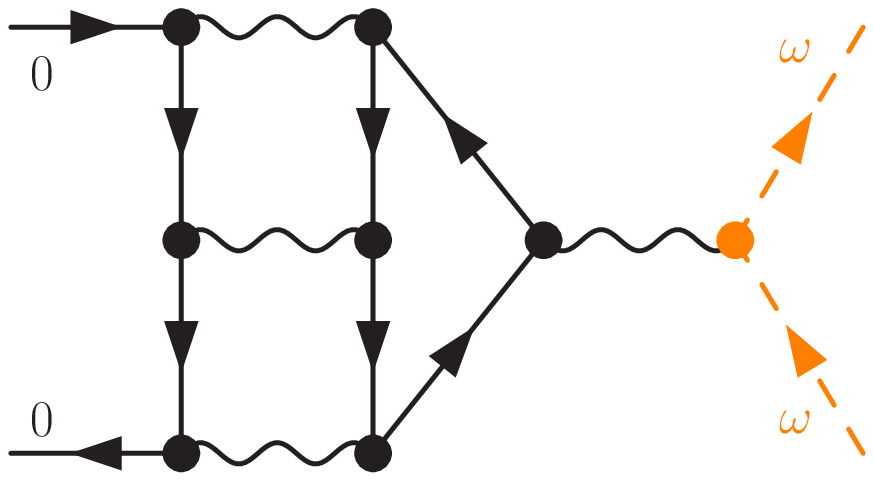}
\end{minipage}

 \rule{0cm}{0.5cm}

\begin{minipage}{1\linewidth}
\includegraphics[width=0.3\linewidth]{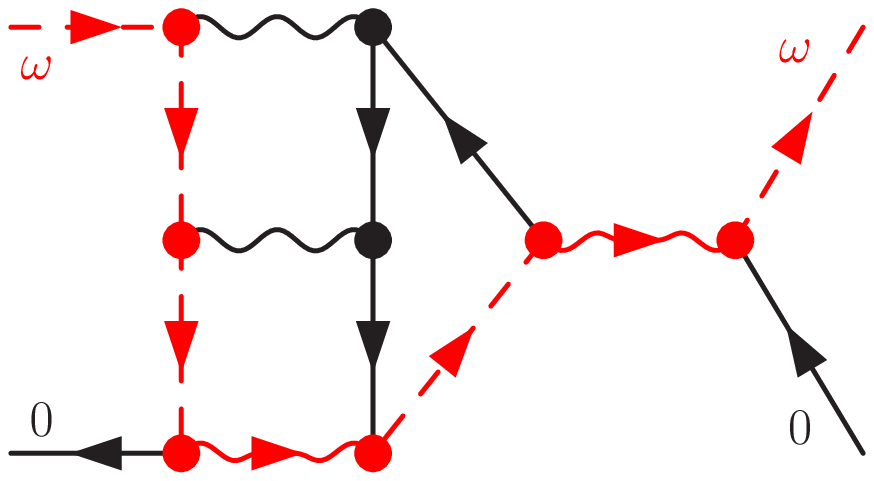}
\rule{0.08\linewidth}{0cm}
\includegraphics[width=0.3\linewidth]{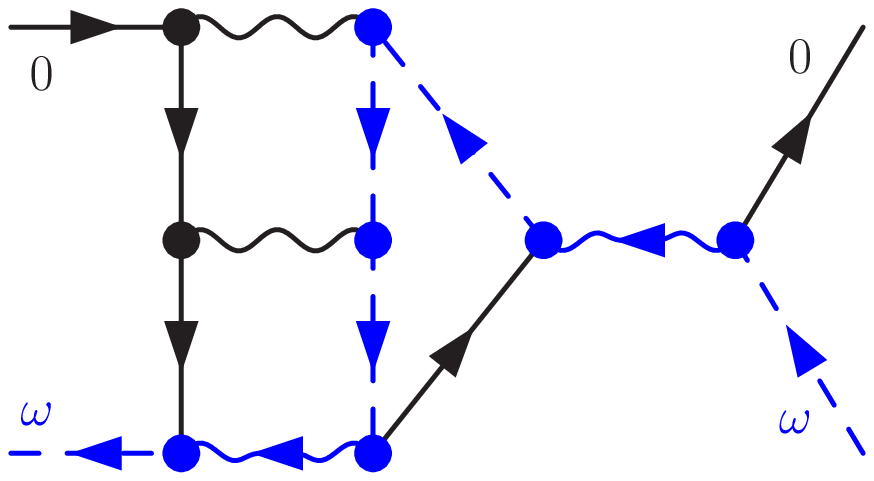}
\end{minipage}
 \caption{
(Color online) 
A set of four diagrams for 
$\Gamma_{\uparrow\uparrow;\uparrow\uparrow}^{(4K)}$,
contribution of which is given in Eq.\ \eqref{eq:vertex_4K}.
}
 \label{fig:vertex_w00w_order_u4_6ppL_color_each}
\end{figure}

\begin{figure}[h]
 \leavevmode
\begin{minipage}{1\linewidth}
\includegraphics[width=0.35\linewidth]{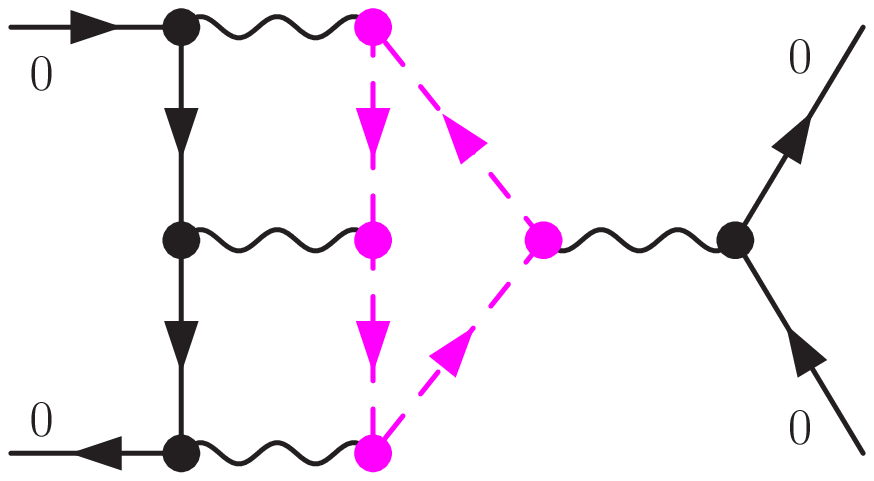}
\end{minipage}

 \caption{
(Color online) 
Schematic picture for 
the total  contribution 
 $\widehat{\partial}_{i\omega}^{+} \Gamma_{\uparrow\uparrow;\uparrow\uparrow}^{(4K)}$ 
 of the set
shown in Fig.\ \ref{fig:vertex_w00w_order_u4_6ppL_color_each}. 
%
}
 \label{fig:vertex_w00w_order_u4_6ppL_color_sum}
\end{figure}

Total contribution of the diagrams shown in Fig.\ 
\ref{fig:vertex_w00w_order_u4_6ppL_color_each} 
can be rewritten in a total derivative form 
 (see also  Fig.\ \ref{fig:vertex_w00w_order_u4_6ppL_color_sum}):  
\begin{align}
& 
\!\!\!\!\!
\widehat{\partial}_{i\omega}^{+}
\Gamma_{\uparrow\uparrow;\uparrow\uparrow}^{(4K)}(i\omega , 0; 0 , i\omega) 
\nonumber \\
& = \,
 - U^4   
\int_{-\infty}^{\infty}\!
\int_{-\infty}^{\infty}\!
\int_{-\infty}^{\infty}\!
 \frac{d\varepsilon\,d\varepsilon_1\,d\varepsilon_2}{(2\pi)^3}\  
\widehat{\partial}_{i\omega}^{+}
 \biggl[\,
G_{\uparrow}^{}(i\varepsilon_1) \,
G_{\uparrow}^{}(i\varepsilon_2) \,
{\color[rgb]{0,0,1}
G_{\downarrow}^{}(i\varepsilon-i\varepsilon_1+i\omega) \,
G_{\downarrow}^{}(i\varepsilon-i\varepsilon_2+i\omega) \,
G_{\downarrow}^{}(i\varepsilon+i\omega) \,
}
G_{\downarrow}^{}(i\varepsilon) \,
\nonumber \\
& \qquad \qquad \qquad  \qquad \qquad
+
{\color[rgb]{1,0,0}
G_{\uparrow}^{}(i\varepsilon_1+i\omega) \,
G_{\uparrow}^{}(i\varepsilon_2+i\omega) \,
}
G_{\downarrow}^{}(i\varepsilon-i\varepsilon_1) \,
G_{\downarrow}^{}(i\varepsilon-i\varepsilon_2) \,
G_{\downarrow}^{}(i\varepsilon) \,
{\color[rgb]{1,0,0}
G_{\downarrow}^{}(i\varepsilon+i\omega) \,
}
 \nonumber \\
& \qquad \qquad \qquad \qquad \qquad
- 
{\color[rgb]{0,0.5,0}
G_{\uparrow}^{}(i\varepsilon_1+i\omega) \,
G_{\uparrow}^{}(i\varepsilon_2+i\omega) \,
}
G_{\downarrow}^{}(i\varepsilon -i\varepsilon_1) \,
G_{\downarrow}^{}(i\varepsilon -i\varepsilon_2) \,
G_{\downarrow}^{}(i\varepsilon) \,
G_{\downarrow}^{}(i\varepsilon) \,
\nonumber \\
& \qquad \qquad \qquad  \qquad \qquad
- 
G_{\uparrow}^{}(i\varepsilon_1) \,
G_{\uparrow}^{}(i\varepsilon_2) \,
G_{\downarrow}^{}(i\varepsilon -i\varepsilon_1) \,
G_{\downarrow}^{}(i\varepsilon -i\varepsilon_2) \,
G_{\downarrow}^{}(i\varepsilon) \,
G_{\downarrow}^{}(i\varepsilon) \,
\,\biggr]
\nonumber \\
& = \ 
 - U^4   
\int_{-\infty}^{\infty}\!
\int_{-\infty}^{\infty}\!
 \frac{d\varepsilon_1\,d\varepsilon_2}{(2\pi)^2} \  
G_{\uparrow}^{}(i\varepsilon_1) \,
G_{\uparrow}^{}(i\varepsilon_2) 
\int_{-\infty}^{\infty}\!
 \frac{d\varepsilon}{2\pi}\  
\widehat{\partial}_{i\omega}^{+}
 \biggl[\,
{\color[rgb]{0,0,1}
G_{\downarrow}^{}(i\varepsilon-i\varepsilon_1+i\omega) \,
G_{\downarrow}^{}(i\varepsilon-i\varepsilon_2+i\omega) \,
G_{\downarrow}^{}(i\varepsilon+i\omega) \,
}
G_{\downarrow}^{}(i\varepsilon) \,
\nonumber \\
& 
 \qquad \qquad \qquad
+
G_{\downarrow}^{}(i\varepsilon-i\varepsilon_1) \,
G_{\downarrow}^{}(i\varepsilon-i\varepsilon_2) \,
G_{\downarrow}^{}(i\varepsilon) \,
{\color[rgb]{1,0,0}
G_{\downarrow}^{}(i\varepsilon+i\omega) \,
}
- 2
G_{\downarrow}^{}(i\varepsilon -i\varepsilon_1) \,
G_{\downarrow}^{}(i\varepsilon -i\varepsilon_2) \,
G_{\downarrow}^{}(i\varepsilon) \,
G_{\downarrow}^{}(i\varepsilon) \,
\,\biggr]
\nonumber \\
& = 
 - U^4\!\!   
\int_{-\infty}^{\infty}\!\!
\int_{-\infty}^{\infty}\!\!
 \frac{d\varepsilon_1 d\varepsilon_2}{(2\pi)^2}  \, 
G_{\uparrow}^{}(i\varepsilon_1) \,
G_{\uparrow}^{}(i\varepsilon_2 ) \ 
{\color[rgb]{1,0,1}
\widehat{\partial}_{i\omega}^{+}
 \left[
\int_{-\infty}^{\infty}\!\!
 \frac{d\varepsilon}{2\pi}\  
G_{\downarrow}^{}(i\varepsilon-i\varepsilon_1+i\omega) \,
G_{\downarrow}^{}(i\varepsilon-i\varepsilon_2+i\omega) \,
\left\{G_{\downarrow}^{}(i\varepsilon+i\omega) \right\}^2
\,\right] 
}
\nonumber \\
  &  = \ 0 \;. 
\label{eq:vertex_4K}
\end{align}
To obtain the second line,  the derivative 
with respect to  $\widehat{\partial}_{i\omega}^{+}$ 
is taken for $\omega$'s which are assigned for  the $\uparrow$ spin 
propagators in the vertical direction.
The remaining contribution  arising from the two diagrams 
in the lower panel of Fig.\ \ref{fig:vertex_w00w_order_u4_6ppL_color_each}
 is extracted by applying the {\it generalized\/} chain rule for the product 
$G_{\downarrow}^{}(i\varepsilon) G_{\downarrow}^{}(i\varepsilon+i\omega)$.

\newpage

\begin{figure}[h]
 \leavevmode
\begin{minipage}{1\linewidth}
\includegraphics[width=0.3\linewidth]{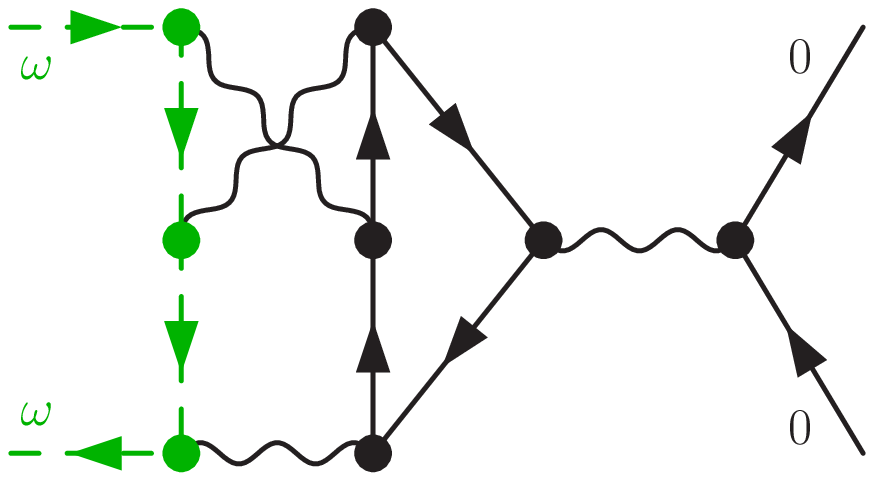}
\rule{0.08\linewidth}{0cm}
\includegraphics[width=0.3\linewidth]{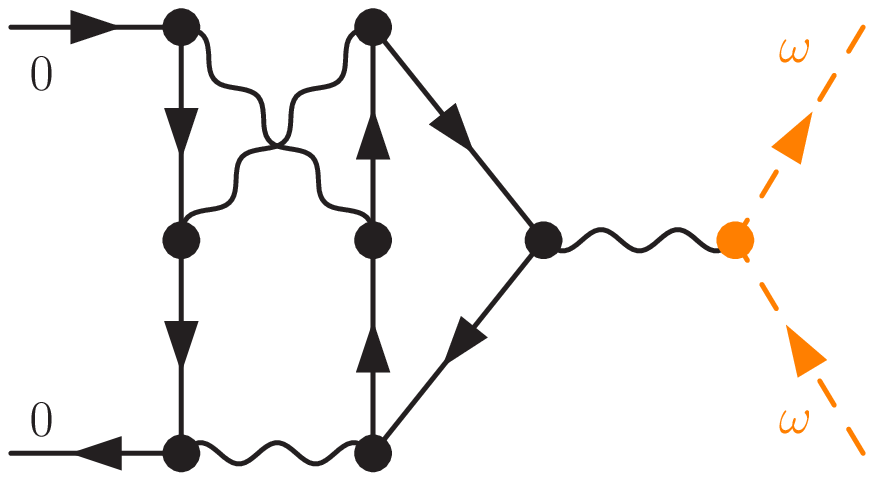}
\end{minipage}

 \rule{0cm}{0.5cm}

\begin{minipage}{1\linewidth}
\includegraphics[width=0.3\linewidth]{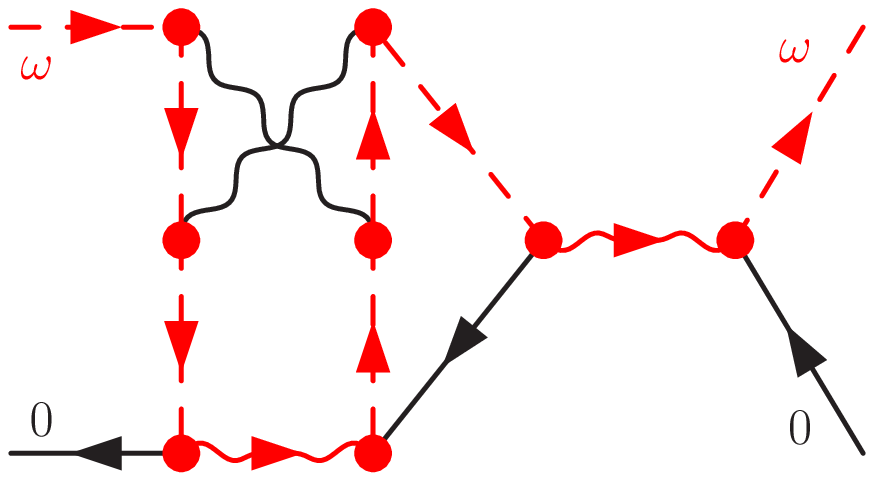}
\rule{0.08\linewidth}{0cm}
\includegraphics[width=0.3\linewidth]{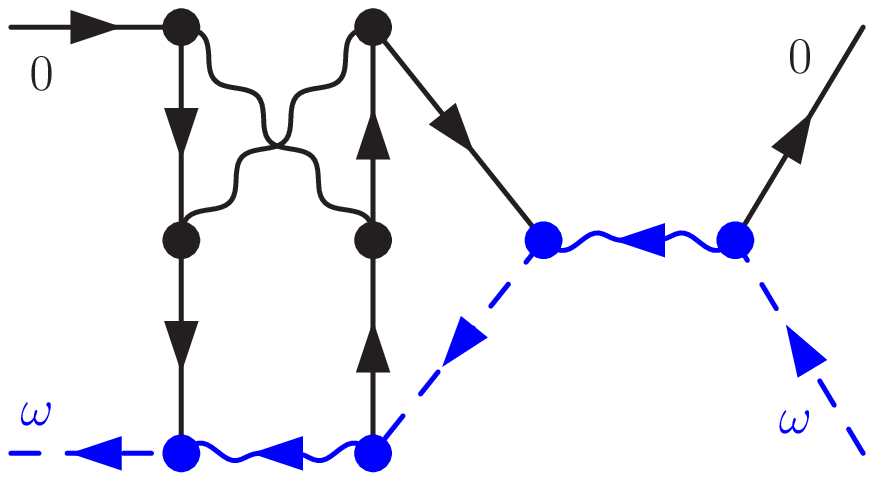}
\end{minipage}
 \caption{
(Color online) 
A set of four diagrams for $\Gamma_{\uparrow\uparrow;\uparrow\uparrow}^{(4L)}$, 
contribution of which is given in Eq.\ \eqref{eq:vertex_4L}.
}
 \label{fig:vertex_w00w_order_u4_7phL_color_each}
\end{figure}

\begin{figure}[h]
 \leavevmode
\begin{minipage}{1\linewidth}
\includegraphics[width=0.35\linewidth]{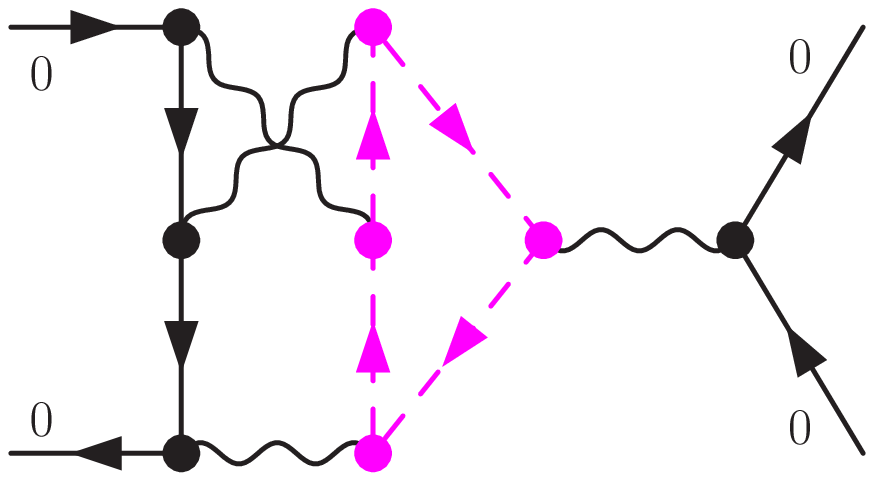}
\end{minipage}

 \caption{
(Color online) 
Schematic picture for 
the total  contribution  
 $\widehat{\partial}_{i\omega}^{+} 
\Gamma_{\uparrow\uparrow;\uparrow\uparrow}^{(4L)}$ 
of the diagrams 
shown in Fig.\ \ref{fig:vertex_w00w_order_u4_7phL_color_each}. 
%
}
 \label{fig:vertex_w00w_order_u4_7phL_color_sum}
\end{figure}

Total contribution of the diagrams shown in Fig.\ 
\ref{fig:vertex_w00w_order_u4_7phL_color_each} 
can be rewritten in a total derivative form 
 (see also  Fig.\ \ref{fig:vertex_w00w_order_u4_7phL_color_sum}):
\begin{align}
& 
\!\!\!\!\!
\widehat{\partial}_{i\omega}^{+}
\Gamma_{\uparrow\uparrow;\uparrow\uparrow}^{(4L)}(i\omega , 0; 0 , i\omega) 
\nonumber \\
& = \,
 - U^4   
\int_{-\infty}^{\infty}\!
\int_{-\infty}^{\infty}\!
\int_{-\infty}^{\infty}\!
 \frac{d\varepsilon\,d\varepsilon_1\,d\varepsilon_2}{(2\pi)^3}\  
\widehat{\partial}_{i\omega}^{+}
 \biggl[\,
G_{\uparrow}^{}(i\varepsilon_1) \,
G_{\uparrow}^{}(i\varepsilon_2) \,
G_{\downarrow}^{}(i\varepsilon+i\varepsilon_2 -i\varepsilon_1) \,
G_{\downarrow}^{}(i\varepsilon +i\varepsilon_2) \,
G_{\downarrow}^{}(i\varepsilon) \,
{\color[rgb]{0,0,1}
G_{\downarrow}^{}(i\varepsilon+i\omega) \,
}
\nonumber \\
& \qquad \qquad \qquad  \qquad \qquad
+
{\color[rgb]{1,0,0}
G_{\uparrow}^{}(i\varepsilon_1+i\omega) \,
G_{\uparrow}^{}(i\varepsilon_2+i\omega) \,
G_{\downarrow}^{}(i\varepsilon+i\varepsilon_2 -i\varepsilon_1+i\omega) \,
G_{\downarrow}^{}(i\varepsilon +i\varepsilon_2+\omega) \,
G_{\downarrow}^{}(i\varepsilon+i\omega) \,
}
G_{\downarrow}^{}(i\varepsilon) \,
 \nonumber \\
& \qquad \qquad \qquad \qquad \qquad
- 
{\color[rgb]{0,0.5,0}
G_{\uparrow}^{}(i\varepsilon_1+i\omega) \,
G_{\uparrow}^{}(i\varepsilon_2+i\omega) \,
}
G_{\downarrow}^{}(i\varepsilon+i\varepsilon_2 -i\varepsilon_1) \,
G_{\downarrow}^{}(i\varepsilon +i\varepsilon_2) \,
G_{\downarrow}^{}(i\varepsilon) \,
G_{\downarrow}^{}(i\varepsilon) \,
\nonumber \\
& \qquad \qquad \qquad  \qquad \qquad
- 
G_{\uparrow}^{}(i\varepsilon_1) \,
G_{\uparrow}^{}(i\varepsilon_2) \,
G_{\downarrow}^{}(i\varepsilon+i\varepsilon_2 -i\varepsilon_1) \,
G_{\downarrow}^{}(i\varepsilon +i\varepsilon_2) \,
G_{\downarrow}^{}(i\varepsilon) \,
G_{\downarrow}^{}(i\varepsilon) \,
\,\biggr]
\nonumber \\
& = \ 
 - U^4   
\int_{-\infty}^{\infty}\!
\int_{-\infty}^{\infty}\!
 \frac{d\varepsilon_1\,d\varepsilon_2}{(2\pi)^2} \  
G_{\uparrow}^{}(i\varepsilon_1) \,
G_{\uparrow}^{}(i\varepsilon_2) 
\int_{-\infty}^{\infty}\!
 \frac{d\varepsilon}{2\pi}\  
\widehat{\partial}_{i\omega}^{+}
 \biggl[\,
G_{\downarrow}^{}(i\varepsilon+i\varepsilon_2-i\varepsilon_1) \,
G_{\downarrow}^{}(i\varepsilon+i\varepsilon_2) \,
G_{\downarrow}^{}(i\varepsilon) \,
{\color[rgb]{0,0,1}
G_{\downarrow}^{}(i\varepsilon+i\omega) \,
}
\nonumber \\
& 
 \qquad \qquad 
+
{\color[rgb]{1,0,0}
G_{\downarrow}^{}(i\varepsilon+i\varepsilon_2-i\varepsilon_1+i\omega) \,
G_{\downarrow}^{}(i\varepsilon+i\varepsilon_2+i\omega) \,
G_{\downarrow}^{}(i\varepsilon+i\omega) 
}\,
G_{\downarrow}^{}(i\varepsilon) \,
- 2
G_{\downarrow}^{}(i\varepsilon +i\varepsilon_2-i\varepsilon_1) \,
G_{\downarrow}^{}(i\varepsilon +i\varepsilon_2) \,
G_{\downarrow}^{}(i\varepsilon) \,
G_{\downarrow}^{}(i\varepsilon) \,
 \,
\,\biggr]
\nonumber \\
& = 
 - U^4\!\!   
\int_{-\infty}^{\infty}\!\!
\int_{-\infty}^{\infty}\!\!
 \frac{d\varepsilon_1 d\varepsilon_2}{(2\pi)^2}  \, 
G_{\uparrow}^{}(i\varepsilon_1) \,
G_{\uparrow}^{}(i\varepsilon_2 ) \ 
{\color[rgb]{1,0,1}
\widehat{\partial}_{i\omega}^{+}
 \left[
\int_{-\infty}^{\infty}\!\!
 \frac{d\varepsilon}{2\pi}\  
G_{\downarrow}^{}(i\varepsilon+i\varepsilon_2-i\varepsilon_1+i\omega) \,
G_{\downarrow}^{}(i\varepsilon+i\varepsilon_2+i\omega) \,
\left\{G_{\downarrow}^{}(i\varepsilon+i\omega) \right\}^2
\,\right] 
}
\nonumber \\
  &  = \ 0 \;. 
\label{eq:vertex_4L}
\end{align}
To obtain the second line,  the derivative 
with respect to  $\widehat{\partial}_{i\omega}^{+}$ 
is taken for $\omega$'s which are assigned for  the $\uparrow$ spin 
propagators in the vertical direction.
The remaining contribution  arising from the two diagrams 
in the lower panel of Fig.\ \ref{fig:vertex_w00w_order_u4_7phL_color_each}  
 is extracted by applying the {\it generalized\/} chain rule for the product 
$G_{\downarrow}^{}(i\varepsilon) G_{\downarrow}^{}(i\varepsilon+i\omega)$.


\newpage

\begin{figure}[h]
 \leavevmode
\begin{minipage}{1\linewidth}
\includegraphics[width=0.3\linewidth]{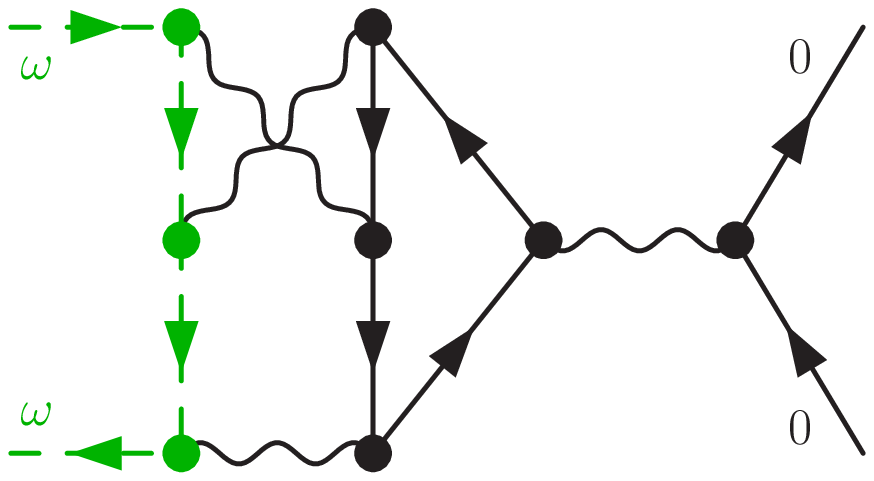}
\rule{0.08\linewidth}{0cm}
\includegraphics[width=0.3\linewidth]{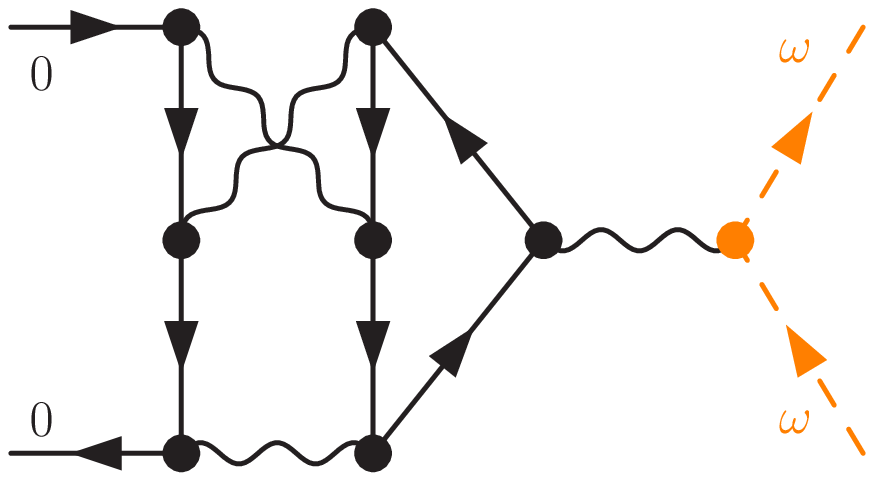}
\end{minipage}

 \rule{0cm}{0.5cm}

\begin{minipage}{1\linewidth}
\includegraphics[width=0.3\linewidth]{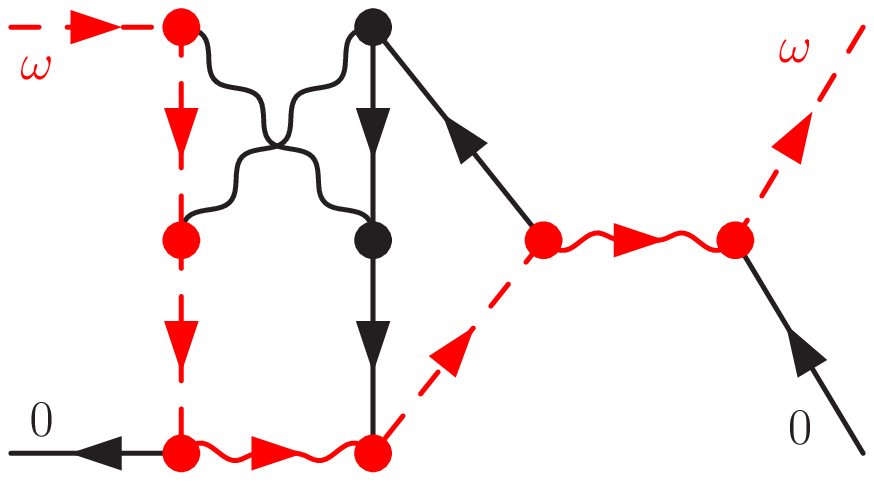}
\rule{0.08\linewidth}{0cm}
\includegraphics[width=0.3\linewidth]{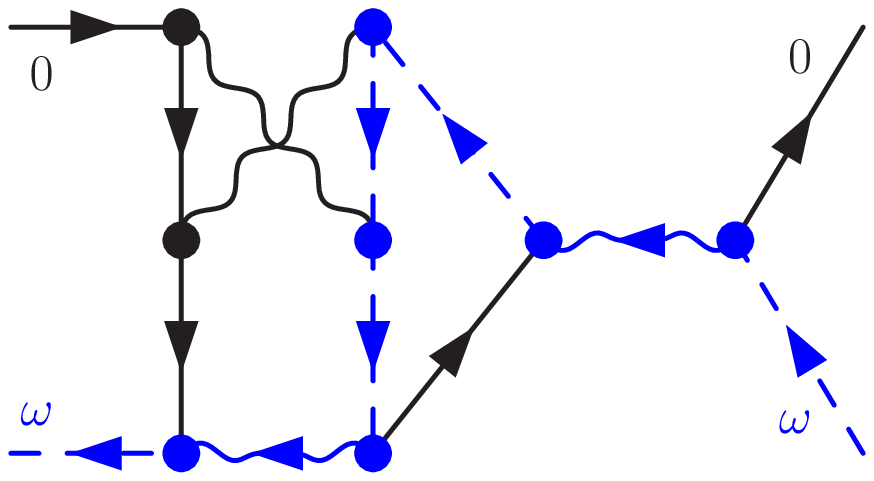}
\end{minipage}
 \caption{
(Color online) 
A set of four diagrams for $\Gamma_{\uparrow\uparrow;\uparrow\uparrow}^{(4M)}$, 
contribution of which is given in Eq.\ \eqref{eq:vertex_4M}.
}
 \label{fig:vertex_w00w_order_u4_7ppL_color_each}
\end{figure}

\begin{figure}[h]
 \leavevmode
\begin{minipage}{1\linewidth}
\includegraphics[width=0.35\linewidth]{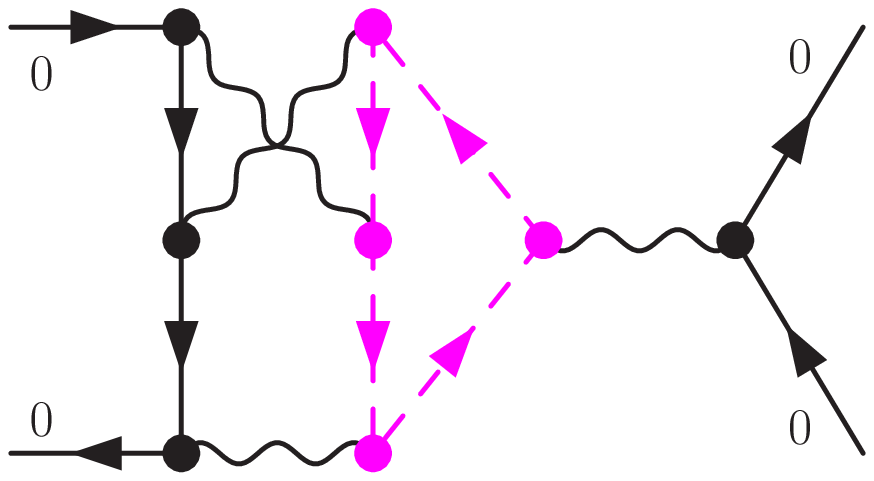}
\end{minipage}
 \caption{
(Color online) 
Schematic picture for 
the total  contribution 
 $\widehat{\partial}_{i\omega}^{+} \Gamma_{\uparrow\uparrow;\uparrow\uparrow}^{(4M)}$ 
  of the set
shown in Fig.\ \ref{fig:vertex_w00w_order_u4_7ppL_color_each}. 
%
}
 \label{fig:vertex_w00w_order_u4_7ppL_color_sum}
\end{figure}

Total contribution of the diagrams shown in Fig.\ 
\ref{fig:vertex_w00w_order_u4_7ppL_color_each} 
can be rewritten in a total derivative form 
 (see also  Fig.\ \ref{fig:vertex_w00w_order_u4_7ppL_color_sum}): 
\begin{align}
& 
\!\!\!\!\!
\widehat{\partial}_{i\omega}^{+}
\Gamma_{\uparrow\uparrow;\uparrow\uparrow}^{(4M)}(i\omega , 0; 0 , i\omega) 
\nonumber \\
& = \,
 - U^4   
\int_{-\infty}^{\infty}\!
\int_{-\infty}^{\infty}\!
\int_{-\infty}^{\infty}\!
 \frac{d\varepsilon\,d\varepsilon_1\,d\varepsilon_2}{(2\pi)^3}\  
\widehat{\partial}_{i\omega}^{+}
 \biggl[\,
G_{\uparrow}^{}(i\varepsilon_1) \,
G_{\uparrow}^{}(i\varepsilon_2) \,
{\color[rgb]{0,0,1}
G_{\downarrow}^{}(i\varepsilon+i\varepsilon_1 -i\varepsilon_2+i\omega) \,
G_{\downarrow}^{}(i\varepsilon -i\varepsilon_2+i\omega) \,
G_{\downarrow}^{}(i\varepsilon+i\omega) \,
}
G_{\downarrow}^{}(i\varepsilon) \,
\nonumber \\
& \qquad \qquad \qquad  \qquad \qquad
+
{\color[rgb]{1,0,0}
G_{\uparrow}^{}(i\varepsilon_1+i\omega) \,
G_{\uparrow}^{}(i\varepsilon_2+i\omega) \,
}
G_{\downarrow}^{}(i\varepsilon+i\varepsilon_1 -i\varepsilon_2) \,
G_{\downarrow}^{}(i\varepsilon -i\varepsilon_2) \,
G_{\downarrow}^{}(i\varepsilon) \,
{\color{red}
G_{\downarrow}^{}(i\varepsilon+i\omega) \,
}
 \nonumber \\
& \qquad \qquad \qquad \qquad \qquad
- 
{\color[rgb]{0,0.5,0}
G_{\uparrow}^{}(i\varepsilon_1+i\omega) \,
G_{\uparrow}^{}(i\varepsilon_2+i\omega) \,
}
G_{\downarrow}^{}(i\varepsilon+i\varepsilon_1 -i\varepsilon_2) \,
G_{\downarrow}^{}(i\varepsilon -i\varepsilon_2) \,
G_{\downarrow}^{}(i\varepsilon) \,
G_{\downarrow}^{}(i\varepsilon) \,
\nonumber \\
& \qquad \qquad \qquad  \qquad \qquad
- 
G_{\uparrow}^{}(i\varepsilon_1) \,
G_{\uparrow}^{}(i\varepsilon_2) \,
G_{\downarrow}^{}(i\varepsilon+i\varepsilon_1 -i\varepsilon_2) \,
G_{\downarrow}^{}(i\varepsilon -i\varepsilon_2) \,
G_{\downarrow}^{}(i\varepsilon) \,
G_{\downarrow}^{}(i\varepsilon) \,
\,\biggr]
\nonumber \\
& = \ 
 - U^4   
\int_{-\infty}^{\infty}\!
\int_{-\infty}^{\infty}\!
 \frac{d\varepsilon_1\,d\varepsilon_2}{(2\pi)^2} \  
G_{\uparrow}^{}(i\varepsilon_1) \,
G_{\uparrow}^{}(i\varepsilon_2) 
\int_{-\infty}^{\infty}\!
 \frac{d\varepsilon}{2\pi}\  
\widehat{\partial}_{i\omega}^{+}
 \biggl[\,
{\color[rgb]{0,0,1}
G_{\downarrow}^{}(i\varepsilon+i\varepsilon_1 -i\varepsilon_2 +i\omega) \,
G_{\downarrow}^{}(i\varepsilon -i\varepsilon_2 +i\omega) \,
G_{\downarrow}^{}(i\varepsilon+i\omega) \,
}
G_{\downarrow}^{}(i\varepsilon) \,
\nonumber \\
& 
 \qquad \qquad 
+
G_{\downarrow}^{}(i\varepsilon+i\varepsilon_1 -i\varepsilon_2) \,
G_{\downarrow}^{}(i\varepsilon -i\varepsilon_2) \,
G_{\downarrow}^{}(i\varepsilon) \,
{\color{red}
G_{\downarrow}^{}(i\varepsilon+i\omega) \,
}
- 2
G_{\downarrow}^{}(i\varepsilon+i\varepsilon_1 -i\varepsilon_2) \,
G_{\downarrow}^{}(i\varepsilon -i\varepsilon_2) \,
G_{\downarrow}^{}(i\varepsilon) \,
G_{\downarrow}^{}(i\varepsilon) \,
 \,
\,\biggr]
\nonumber \\
& = 
 - U^4\!\!   
\int_{-\infty}^{\infty}\!\!
\int_{-\infty}^{\infty}\!\!
 \frac{d\varepsilon_1 d\varepsilon_2}{(2\pi)^2}  \, 
G_{\uparrow}^{}(i\varepsilon_1) \,
G_{\uparrow}^{}(i\varepsilon_2 ) \ 
{\color[rgb]{1,0,1}
\widehat{\partial}_{i\omega}^{+}
 \left[
\int_{-\infty}^{\infty}\!\!
 \frac{d\varepsilon}{2\pi}\  
G_{\downarrow}^{}(i\varepsilon+i\varepsilon_1-i\varepsilon_2+i\omega) \,
G_{\downarrow}^{}(i\varepsilon-i\varepsilon_2+i\omega) \,
\left\{G_{\downarrow}^{}(i\varepsilon+i\omega) \right\}^2
\,\right] 
}
\nonumber \\
  &  = \ 0 \;. 
\label{eq:vertex_4M}
\end{align}
To obtain the second line,  the derivative 
with respect to  $\widehat{\partial}_{i\omega}^{+}$ 
is taken for $\omega$'s which are assigned for  the $\uparrow$ spin 
propagators in the vertical direction.
The remaining contribution  arising from the two diagrams 
in the lower panel of Fig.\ \ref{fig:vertex_w00w_order_u4_7ppL_color_each}  
 is extracted by applying the {\it generalized\/} chain rule for the product 
$G_{\downarrow}^{}(i\varepsilon) G_{\downarrow}^{}(i\varepsilon+i\omega)$.


\newpage

\begin{figure}[h]
 \leavevmode
\begin{minipage}{1\linewidth}
\includegraphics[width=0.3\linewidth]{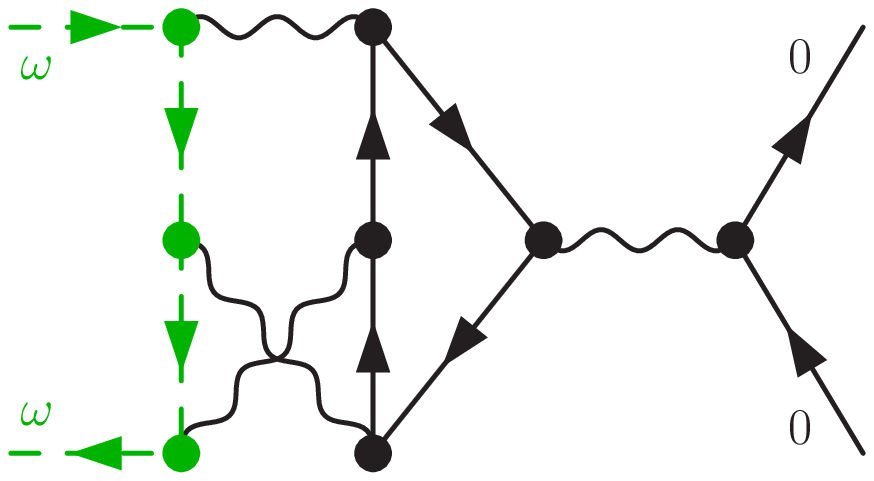}
\rule{0.08\linewidth}{0cm}
\includegraphics[width=0.3\linewidth]{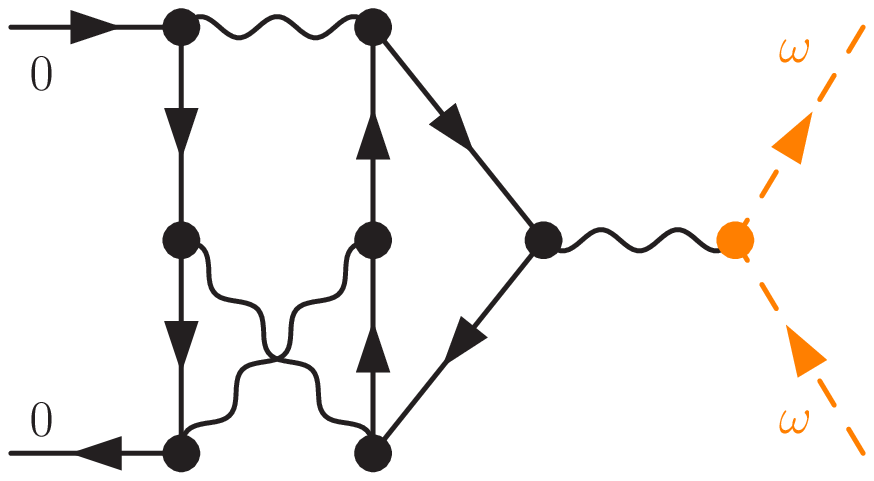}
\end{minipage}

 \rule{0cm}{0.5cm}

\begin{minipage}{1\linewidth}
\includegraphics[width=0.3\linewidth]{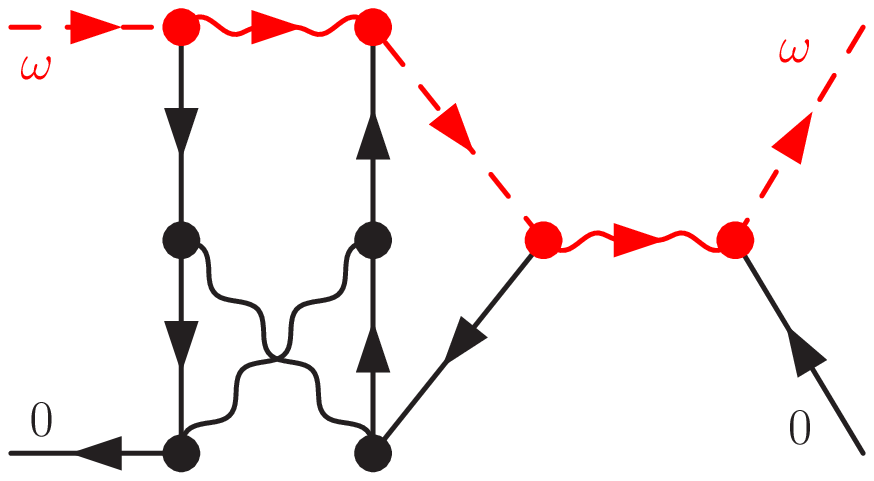}
\rule{0.08\linewidth}{0cm}
\includegraphics[width=0.3\linewidth]{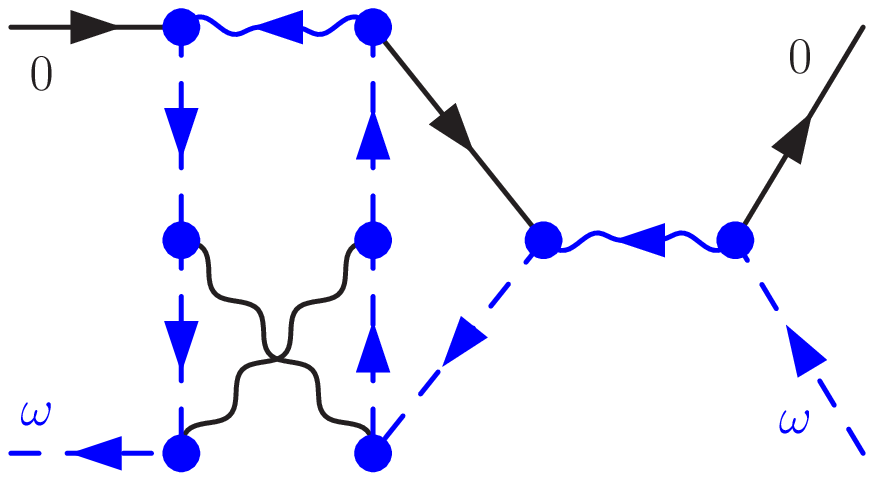}
\end{minipage}
 \caption{
(Color online) 
A set of four diagrams for $\Gamma_{\uparrow\uparrow;\uparrow\uparrow}^{(4L')}$. 
 The contribution of which the same as that of 
 $\Gamma_{\uparrow\uparrow;\uparrow\uparrow}^{(4L)}$
 in Eq.\ \eqref{eq:vertex_4L}.
}
 \label{fig:vertex_w00w_order_u4_8phL_color_each}
\end{figure}

\begin{figure}[h]
 \leavevmode
\begin{minipage}{1\linewidth}
\includegraphics[width=0.35\linewidth]{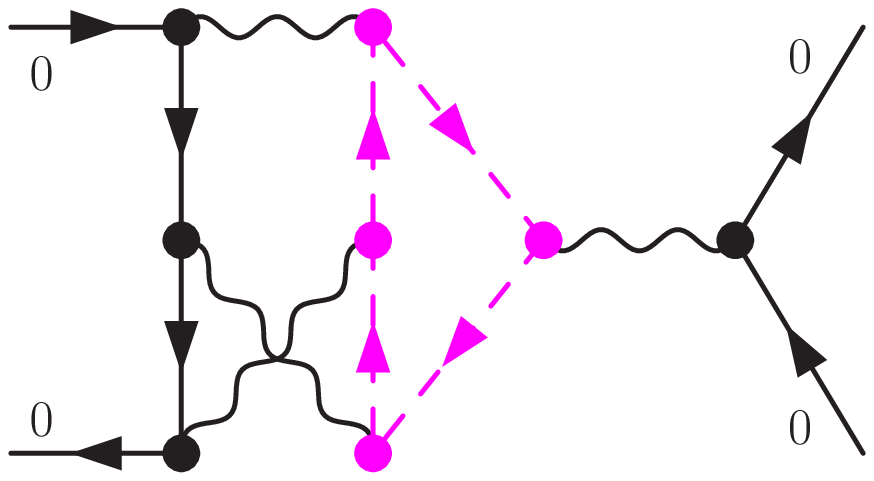}
\end{minipage}

 \caption{
(Color online) 
Schematic picture for 
the total  contribution  
 $\widehat{\partial}_{i\omega}^{+} \Gamma_{\uparrow\uparrow;\uparrow\uparrow}^{(4L')}$ 
of the diagrams 
shown in Fig.\ \ref{fig:vertex_w00w_order_u4_8phL_color_each}. 
%
Note that the contribution of this set $(4L')$ is the same the contribution of   $(4L)$. 
}
 \label{fig:vertex_w00w_order_u4_8phL_color_sum}
\end{figure}

Total contribution of the diagrams shown in Fig.\ 
\ref{fig:vertex_w00w_order_u4_8phL_color_each} 
can be rewritten in a total derivative form 
as illustrated in Fig.\ \ref{fig:vertex_w00w_order_u4_8phL_color_sum}.  
This set ($4L'$) gives the same contribution as 
that of the set ($4L$) described in in Fig.\ 
\ref{fig:vertex_w00w_order_u4_7phL_color_each}, 
namely  it also vanishes 
$\widehat{\partial}_{i\omega}^{+} \Gamma_{\uparrow\uparrow;\uparrow\uparrow}^{(4L')}=0$. 
It can be confirmed, for instance, by 
interchanging the internal frequencies 
$\varepsilon_1$ and $\varepsilon_2$ in 
Eq.\ \eqref{eq:vertex_4L}: 
then one get the corresponding expression for ($4L'$) 
in our way of the frequency assignment. 


 \newpage

\begin{figure}[h]
 \leavevmode
\begin{minipage}{1\linewidth}
\includegraphics[width=0.3\linewidth]{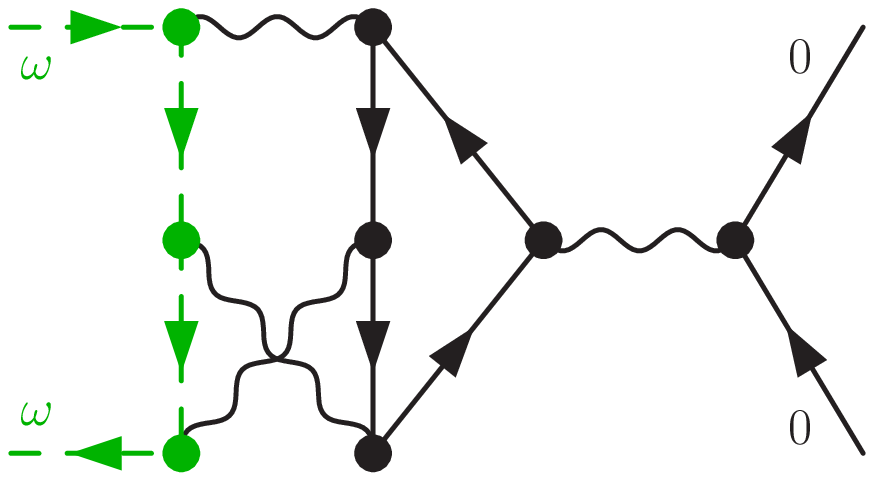}
\rule{0.08\linewidth}{0cm}
\includegraphics[width=0.3\linewidth]{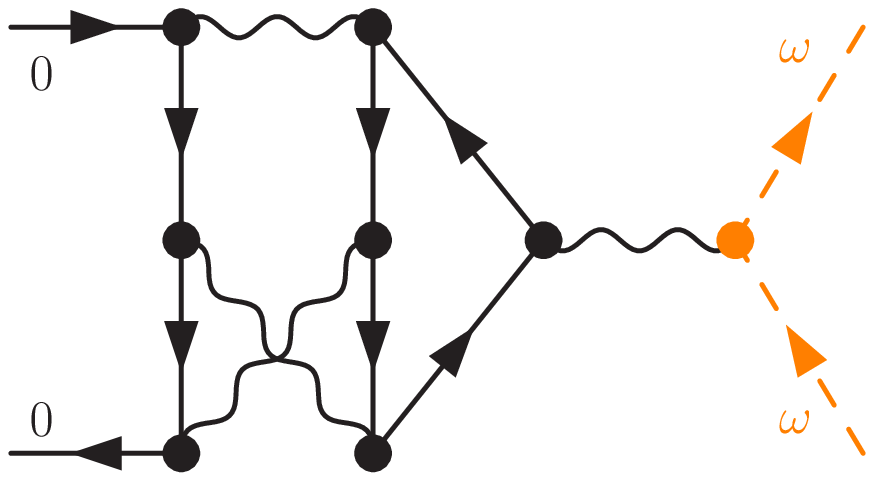}
\end{minipage}

 \rule{0cm}{0.5cm}

\begin{minipage}{1\linewidth}
\includegraphics[width=0.3\linewidth]{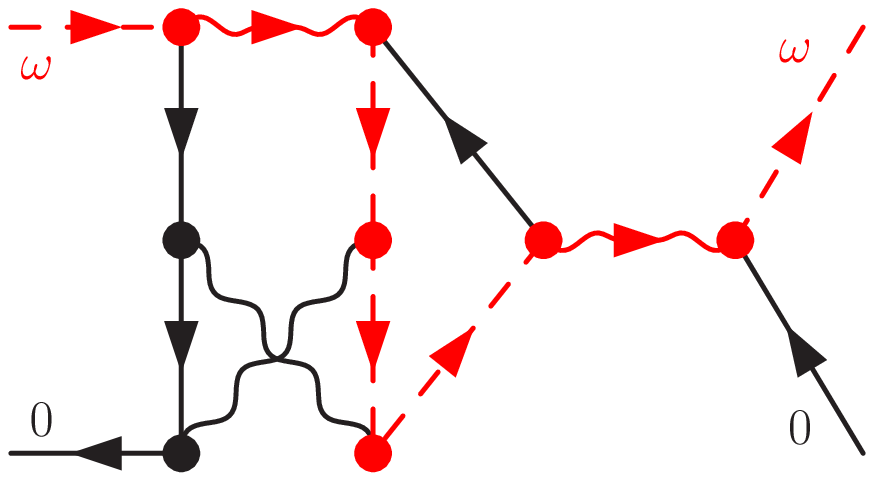}
\rule{0.08\linewidth}{0cm}
\includegraphics[width=0.3\linewidth]{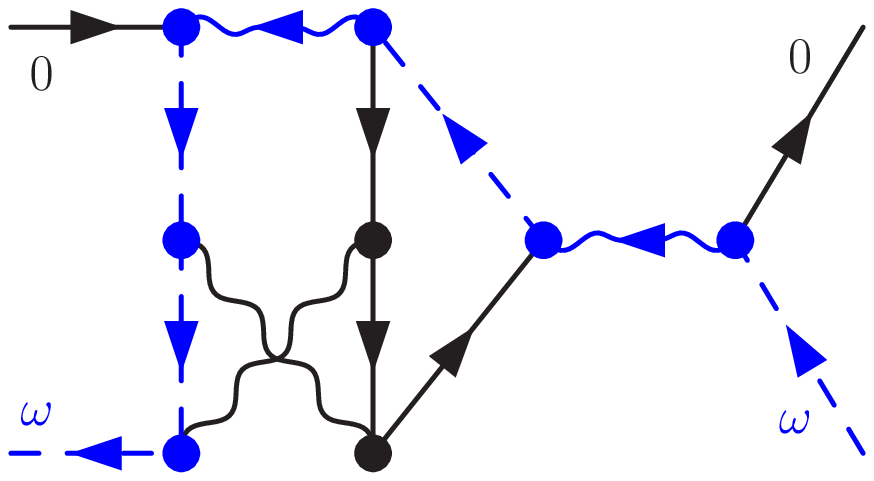}
\end{minipage}

 \caption{
(Color online) 
A set of four diagrams for $\Gamma_{\uparrow\uparrow;\uparrow\uparrow}^{(4M')}$ 
 The contribution of which the same as that of 
 $\Gamma_{\uparrow\uparrow;\uparrow\uparrow}^{(4M)}$
 in Eq.\ \eqref{eq:vertex_4M}.
}
 \label{fig:vertex_w00w_order_u4_8ppL_color_each}
\end{figure}

\begin{figure}[h]
 \leavevmode
\begin{minipage}{1\linewidth}
\includegraphics[width=0.35\linewidth]{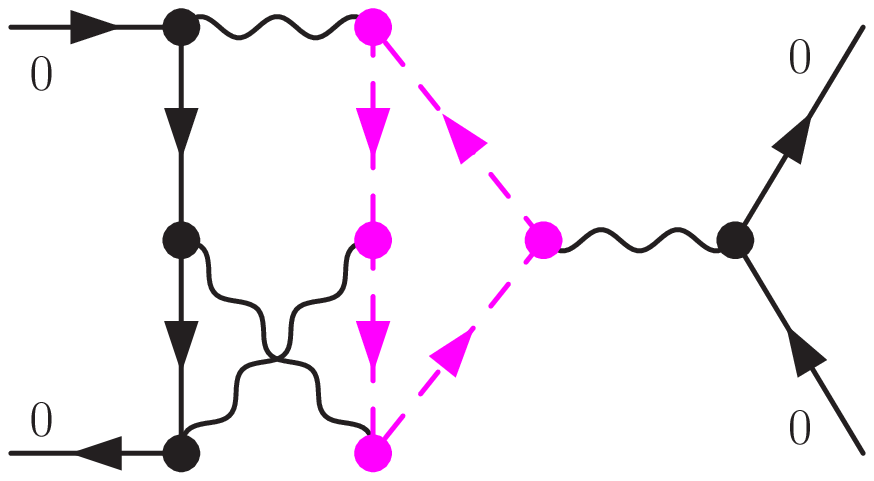}
\end{minipage}

 \caption{
(Color online) 
Schematic picture for 
the total  contribution  
 $\widehat{\partial}_{i\omega}^{+} \Gamma_{\uparrow\uparrow;\uparrow\uparrow}^{(4M')}$ 
  of the set
shown in Fig.\ \ref{fig:vertex_w00w_order_u4_8ppL_color_each}.  
%
Note that the contribution of this set $(4M')$ is the same the contribution of   $(4M)$. 
}
 \label{fig:vertex_w00w_order_u4_8ppL_color_sum}
\end{figure}

Total contribution of the diagrams shown in Fig.\ 
\ref{fig:vertex_w00w_order_u4_8ppL_color_each} 
can be rewritten in a total derivative form 
as illustrated in Fig.\ \ref{fig:vertex_w00w_order_u4_8ppL_color_sum}.  
This set ($4M'$) gives the same contribution as 
that of the set ($4M$) described in in Fig.\ 
\ref{fig:vertex_w00w_order_u4_7ppL_color_each}, 
namely  it also vanishes 
$\widehat{\partial}_{i\omega}^{+} 
\Gamma_{\uparrow\uparrow;\uparrow\uparrow}^{(4M')}=0$. 
It can be confirmed, for instance, by 
interchanging the internal frequencies 
$\varepsilon_1$ and $\varepsilon_2$ in 
Eq.\ \eqref{eq:vertex_4M}: 
then one get the corresponding expression for ($4M'$) 
in our way of the frequency assignment. 

\newpage

\end{widetext}

\end{document}